\documentclass[
aps,%
12pt,%
final,%
notitlepage,%
oneside,%
twocolumn,%
nobibnotes,%
nofootinbib,%
superscriptaddress,%
noshowpacs,%
centertags]%
{revtex4}

\begin{document}
\selectlanguage{russian}

\title{Search for Variable Sources Using Data of ``Cold'' Surveys}

\author{\firstname{E.~K.}~\surname{Majorova}}
\email{len@sao.ru} \affiliation{\saoname}

\author{\firstname{O.~P.}~\surname{Zhelenkova}}
\email{zhe@sao.ru} \affiliation{\saoname}
\affiliation{St.~Petersburg National Research University of
Information Technologies, Mechanics, and Optics, St.~Petersburg,
197101 Russia}


\begin{abstract}
We search for variable sources, using the data of the surveys
conducted on the \mbox{RATAN-600} radio telescope in 1980--1994 at
3.94~GHz. To test the radio sources of the  RCR (RATAN Cold
Refined) catalog for variability, we estimated the long-term
variability indices $V$ of the studied objects, their relative
variability amplitudes $V_{\chi}$, and the $\chi^2$ probabilities
$p$. Out of about two hundred considered sources, 41~proved to
have positive long-term variability indices, suggesting that these
sources may be variable. Fifteen objects can be considered to be
reliably variable according to the  $\chi^2$ criterion \mbox{$p >
0.98$}, three of these sources have $\chi^2$ probabilities  $ p
\ge 0.999$. The corresponding probabilities for six sources lie in
the $ 0.95  < p < 0.98$ interval, and those of the remaining 20
objects in  the  $0.73 \le p < 0.95$ interval. Twenty-four of 41
objects are variable or possibly variable in the optical range,
and five objects are known variable radio sources. We construct
the light curves and spectra for the sources with positive
long-term variability indices.
\end{abstract}

\maketitle

\section{INTRODUCTION}

Many radio sources exhibit flux density variations when observed at different epochs. This variability is due to
both external (scintillations) and internal factors, which are associated with the radiation generation
processes in the source itself.

Variable radio sources are associated with different classes of
objects, and their variability has different characteristic time
scales. Variable radio emission is observed in active galactic
nuclei (AGN), microquasars, pulsars, and stars. Some AGNs exhibit
intra-day variability (IDV) due to the scintillation of their very
compact components, caused by inhomogeneities of the intervening
medium between the observer and the object. Flux density variation
amplitudes may vary from several percent to several tens of
percent.

On time scales of one to several months, the variations of
synchrotron emission often correlate with those of optical and/or
x-ray radiation because of the nonuniform accretion rate and
interaction between the jet and the ambient medium in the
immediate vicinity of the nucleus. Variability on time scales of
several years may be due to more substantial changes in the
accretion rate, heating of the matter, and energy processing in
the accretion disk. Long-term variability of very bright AGNs with
fluxes above 1~Jy is a subject of systematic studies; however, few
such studies have been carried out for the population of fainter
AGNs because of the lack of observational data. At the same time,
the study of radio variability of the sources with weak flux
densities may prove to be a unique tool for investigating the
evolution of AGNs and the nature of this phenomenon. Note that
almost all the sources brighter than several mJy observed in
1.4~GHz surveys are  AGNs.

Variability studies performed by monitoring objects from the lists
of bright sources are carried out in many observatories worldwide,
including the Special Astrophysical Observatory of the Russian
Academy of Sciences, where such programs are performed on the
RATAN-600 radio telescope. Starting from 1998, long-term sets
(from one to three months) of multifrequency observations are
carried out at the Northern sector of RATAN-600 to study variable
objects. Bright discrete radio sources
with flat spectra are being investigated mainly. Such sources show variations on time
scales ranging from several tens of minutes to several decades.
The results of these long-term studies are reported, e.g.,

At the end of the 20-th century, an approach was developed that
uses the data of radio surveys to search for variable radio
sources. As an example of the application of this approach, we can
mention the study by De Vries et al.~\citet{Vr:Majorova_n}, who
used FIRST data (for the period from 1995 to 2002)  in the
vicinity of the South Galactic Pole to extract a sample of 123
radio sources with flux densities from  2 to 1000~mJy, found to
exhibit substantial variations at 1.4~GHz over a seven-year long
time interval, and a recent study by Thyagarajan et
al.~\cite{Th:Majorova_n}, who searched FIRST survey data to find
1627 variable and transient objects with flux densities up to
1~mJy, characteristic variability time scales ranging from several
minutes to several years, and flux variations ranging from 20\% to
a factor of 25.

De Vries et al.~\cite{Vr:Majorova_n} found that (1) the sample of
variable sources contains a substantially greater fraction of
quasars than the control sample with nonvariable sources, (2)
variable sources are almost twice more often identified with SDSS
than nonvariable sources, (3) the number of quasars is almost five
times greater than that of galaxies, and (4) the two samples do
not differ significantly in color.

In this paper we continue the research that we initiated
in~\cite{maj:Majorova_n}, i.e., a search for variable radio
sources based on the data of the ``Cold''
surveys~\cite{h:Majorova_n} carried out at 7.6~cm on the RATAN-600
radio telescope. We test the sources for variability, using the
criteria, including statistical ones, described by Majorova and
Zhelenkova~\cite{maj:Majorova_n}.

In our previous paper~\cite{maj:Majorova_n} we used the objects
from the RCR (RATAN Cold Refined)~\mbox{\cite{so2:Majorova_n}}
catalog selected using certain criteria to derive the calibration
curves and perform a detailed analysis of the errors of the
measured source flux densities in each of the considered surveys.
To assess variability, we compute the long-term variability index,
the probability of variability by $\chi^2$, and a number
of other parameters characterizing variability. In this paper we
search for variable objects, using a larger source sample compared
to the one used in our previous study~\cite{maj:Majorova_n}.

\section{USE OF SURVEYS TO SEARCH FOR VARIABLE SOURCES}

The periodicity, sensitivity, and observing frequencies effect the types of variable and transient sources detectable in the conducted surveys.

The radio source detection threshold of the ``Cold'' experiment
conducted in the 1980's at RATAN-600 at a frequency of
3.94~GHz~\mbox {\cite{h:Majorova_n,pa2:Majorova_n}} was equal to
about $10$~mJy~\cite{p3:Majorova_n,p4:Majorova_n}. The RC (RATAN
Cold) catalog based on this survey was later refined by conducting
a series of surveys in 1987--1999. These surveys were performed at
the same declination\footnote{The surveys were made at the
declination of  SS\,433 (${\rm Dec_{1980}}=4^{\circ}57'$).} and
frequency as the ``Cold-80'' survey, and had a sensitivity limit
of  about 10--15~mJy. The results of the reduction of these
surveys were reported in ~\mbox
{\cite{kbu:Majorova_n,bu:Majorova_n,so1:Majorova_n,so2:Majorova_n}}.
Soboleva, Majorova, and Zhelenkova~\mbox {\cite{so2:Majorova_n}}
published the \mbox{RCR-cata}\-log of objects observed within the
framework of the \mbox {1987--1999} surveys in the band  \mbox
{$7^{\rm h} \le$ RA $ < 17^{\rm h}$} as well as the results  of a
new reduction of the data of the ``\mbox{Cold-80}''
experiment.\!\footnote{Spectra of  RCR sources can be found at
{\tt http://www.sao.ru/hq/len/RCR/}}

It was found, after the identification of  RCR catalog objects~\mbox
{\cite{zhe1:Majorova_n,zhe2:Majorova_n}}, that practically all RCR sources are sufficiently bright for
a search to be made for long-term variability of AGNs with flux densities \mbox {$F
> 10$--$15$~mJy} as well as for possible transients.

To this end, in this study we use the data of 7.6-cm surveys conducted in  1980, 1988, 1993, and 1994.

A certain advantage of using surveys to study the variability of radio sources
is that during the survey the antenna of RATAN-600 radio telescope is focused on a certain elevation
and its configuration remains practically unchanged in the process of observations.
This reduces the errors due to the realignment of the antenna, and this is especially important
for the determination of flux densities of sufficiently faint sources.

Furthermore, because of the ``fan-shaped'' configuration of the
power-beam pattern (PBP) of\linebreak \mbox{RATAN-600} when
operated in the mode of single-sector
observations~\cite{e1:Majorova_n,e2:Majorova_n,e3:Majorova_n,m1:Majorova_n,m2:Majorova_n},
a large number of sources are found simultaneously within its
field of view. Thus at $7.6$~cm more than 30000 NVSS
objects~\cite{co1:Majorova_n} pass within the field of view in a
single crossing of the sky. Increasing the integration time via
repeated crossing of the same sky strip allows increasingly
fainter radio sources to be reliably studied.

Note that the data of the 1980--1994 surveys can be used to study the long-term variability of
radio sources on time scales of several years. Such variations are known
to be due to nonstationary processes in active galactic nuclei.

\section{SAMPLE OF RCR CATALOG SOURCES USED TO SEARCH FOR VARIABLE OBJECTS}

In this paper we search for variable sources, using the sample of
\mbox {RCR-catalog} radio sources with flux density data available
at three or more frequencies. We already studied 80 such sources
in our earlier paper~\cite{maj:Majorova_n}. Recall that out of 550
RCR-catalog objects, 245 have flux density data available only for
two frequencies (1.4~GHz~\cite{co1:Majorova_n} and
3.94~GHz~\cite{so2:Majorova_n}). These are mostly faint radio
sources with flux densities below 30~mJy.

Whereas in our previous study~\cite{maj:Majorova_n} we derived the
calibration curves based on bright objects with steep spectra with
minimum scatter of data points in order to minimize the
contamination by variable sources,\!\footnote{ Steep spectrum
radio sources rarely exhibit variations at frequencies above 1~GHz
except for the objects found to host a compact component, which is
responsible for flux
variations~\cite{Sp:Majorova_n,Al:Majorova_n}.} in this paper we
do not use criteria imposing restrictions on the spectral index,
compactness, morphological structure, etc.

As the initial data for the analysis of the variability of  \mbox
{RCR-catalog} sources, we use  several-day average observational
records that have undergone primary
reduction~\cite{kbu:Majorova_n}. After background subtraction we
identified the sources on the averaged scans via  \mbox {Gaussian
analysis}. We used standard software for the reduction of radio
astronomical observations~\cite{vo:Majorova_n}. A detailed
description of the technique used to reduce survey data has been
published in~\mbox{\cite{so2:Majorova_n,maj:Majorova_n}}, and the
technique of searching for variable sources using the data of
RATAN-600 surveys can be found in~\cite{maj:Majorova_n}.

When analyzing the sources in order to select the variable ones,
we studied objects with 3.94-GHz flux densities   \mbox {$F \ge
15$}~mJy, which are easily identifiable in the records and are
unblended with other sources.

From the entire list of RCR sources, we selected about 200 objects
that were observed in two or more sets of the ``Cold'' surveys and
meet the above requirements.

\begin{figure}[]
\onelinecaptionsfalse
\includegraphics[angle=0,width=0.3\textwidth,bb=0 15 304 257,clip]{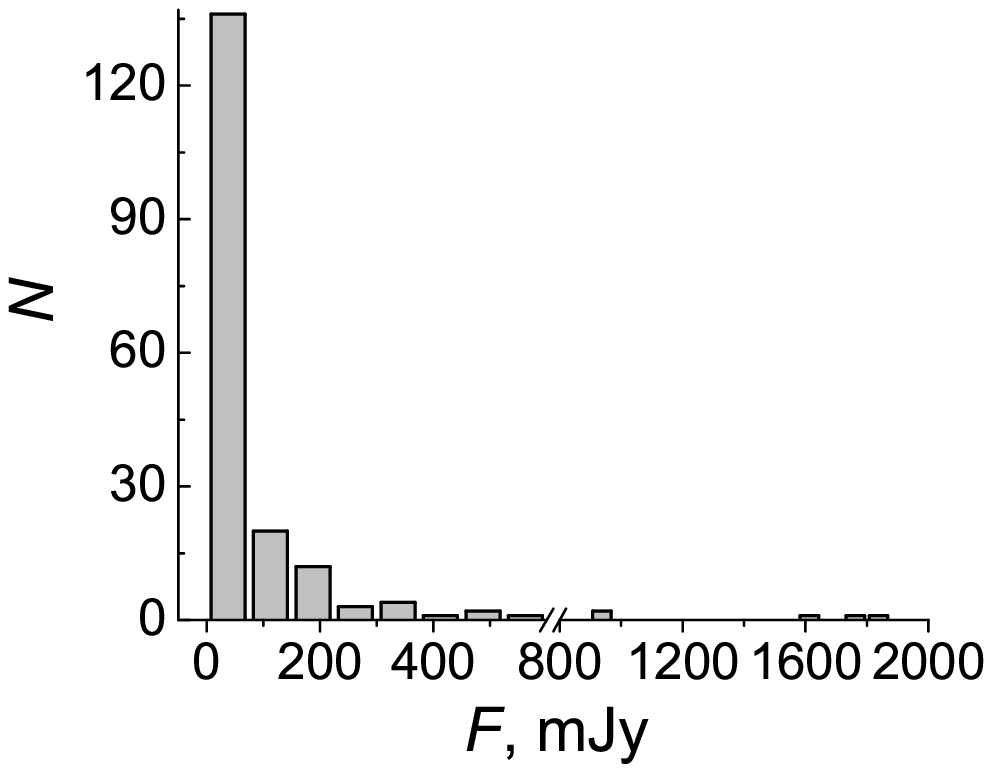}
\vbox{
\includegraphics[angle=0,width=0.3\textwidth,bb=0 15 304 257,clip]{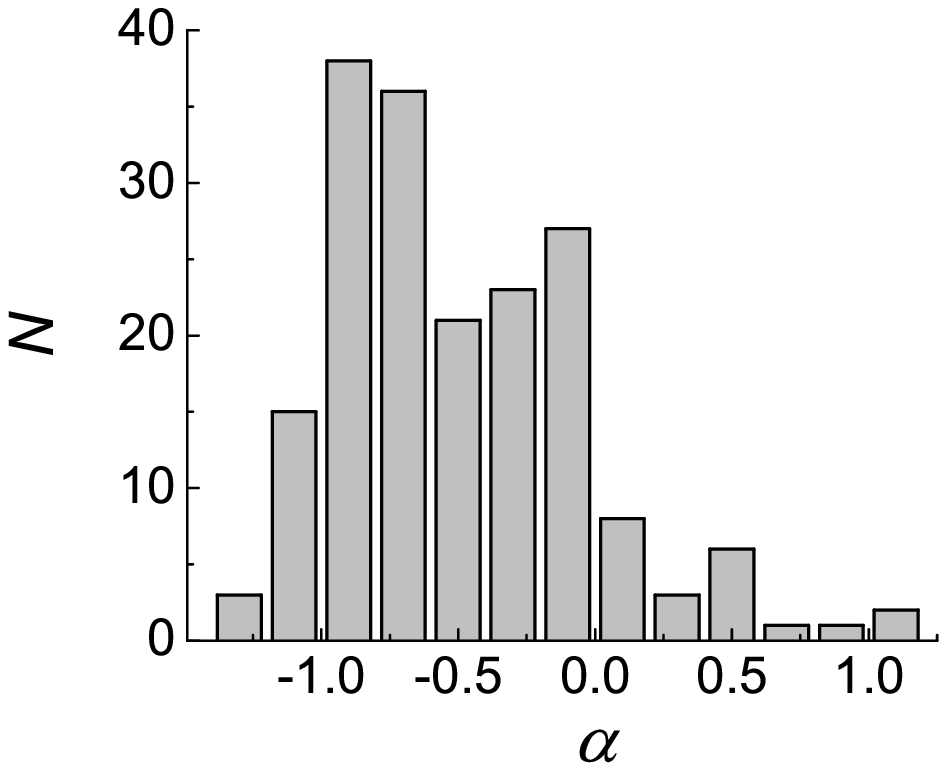}}
   \setcaptionmargin{5mm} \captionstyle{normal}
 \caption{
Histograms of the distribution of flux densities $F$ (the top
panel) and spectral indices $\alpha$ at 3.94~GHz (the bottom
panel) for the considered sample of RCR objects. }
\label{fig1:Majorova_n}
\end{figure}

Figure~\ref{fig1:Majorova_n} shows the histograms of the
distribution of flux densities $F$ (the top panel) and spectral
indices  $\alpha$ (the bottom panel) for this source
sample.\!\footnote{Here $\alpha$ is the spectral index at
3.94~GHz.} About half of the objects in this sample have flux
densities below 50~mJy, and  51 sources have flux densities above
70~mJy. The median spectral index $\alpha$ of the objects
considered is equal to $-0.61$.

\section{REVEALING VARIABLE OBJECTS AMONG THE SOURCES }

We use several criteria to select possibly variable sources. In particular, we
estimate the coefficients $V_{R}$~\cite{Vr:Majorova_n} and $V_{F}$~\cite{VF:Majorova_n},
and the long-term variability index $V$~\cite{V:Majorova_n} for each object of our sample.

We compute the coefficients by the following formulas:
\begin{align}
        V_{R} &= F_{i}/F_{j},
\label{2:Majorova_n}\\[+10pt]
    V_{F} &= \frac{F_{i}-F{j}}{\sqrt{(\sigma_{i}^2+\sigma_{j}^2)}},
\label{3:Majorova_n}\\[+10pt]
    V_{\phantom{F}} &= \frac{(F_{i}-\sigma_{i})-(F_{j}+\sigma_{j})}{(F_{i}-\sigma_{i})+(F_{j}+\sigma_{j})},
\label{4:Majorova_n}
\end{align}
where    $F_{i}$ and $F_{j}$ are the flux densities of the given
source, measured in the surveys of the $i$-th and \mbox{$j$-th}
cycles respectively, and  $\sigma_{i}$, $\sigma_{j}$ are the
absolute root-mean-square errors of the inferred flux densities
($i, j = 80$, $88$, $93$, and $94$).

The latter two criteria take into account the flux density
measurement errors and therefore  can be considered to be more
reliable for testing the sources for variability. One of the
source variability indicators is the positive estimate of its
long-term variability index for at least two surveys.

We computed the flux densities $F_{i}$ of radio sources, using
their antenna temperatures ${T_{a}}^{i}$, determined from averaged
records of the corresponding survey, and the calibration curves.
For a detailed description of the derivation of calibration curves
and selection of calibrating sources see~\cite{maj:Majorova_n}.

We compute the absolute $\sigma_{i}$ and relative ${\rm RMS}_{i}$
root mean square errors of the\pagebreak source flux density in
the $i$-th survey by the following formulas:
\begin{align}
   {\rm RMS}_{i}&=\sqrt{({{\rm RMS}_{i}}^{Ta})^2+({\rm RMS}^{k})^2},
\label{5:Majorova_n}\\[+10pt]
  \sigma_{i} &=F_{i}\times {\rm RMS}_{i},
\label{6:Majorova_n}
\end{align}
where ${{\rm RMS}_{i}}^{Ta}$ is the relative root mean square
error of the estimated antenna temperature of the source, and
${\rm RMS}^{k}$ is the relative root mean square error of the
derived calibration curve.
 $$
 {{\rm RMS}_{i}}^{Ta}=\sigma_{s}/{T_{a}}^{i},
 $$
where $\sigma_{s}$ is the dispersion of noise in the sky strip runs
of the survey considered.

The averaged  ${\rm RMS}^{k}$ values based on the sample of
calibrating sources for the  1980, 1988, 1993, and 1994 surveys
are listed in Table~1 in~\cite{maj:Majorova_n}.\!\footnote{ The
relative errors  \mbox {${\rm RMS}^{k}$} were averaged over the
\mbox {$ -30^{'} < dH < 30^{'}$}, \mbox {$ -15^{'} < dH <
15^{'}$}, \mbox {$-10^{'} < dH < 10^{'}$}, and \mbox {$-5^{'} < dH
< 5^{'}$} intervals as well as over the sample of calibrating
sources whose antenna temperatures in the records exceed
10$\sigma_{s}$.}

We considered the radio sources with flux density differences
between different surveys exceeding the sum of the root mean
square errors of the corresponding surveys to be possibly
variable. The long-term variability indices of such sources are
$V>0$.

We found 41 sources in the considered sample (consisting of about 200 objects of the RCR catalog)
to have positive long-term variability indices $V$. Columns~1, 2, 3, and 4 of Table~1 list the coordinates of these
objects (${\rm RA}_{2000}$, ${\rm Dec}_{2000}$) and their coefficients $V$ , $V_{R}$, and $V_{F}$,
respectively.

Table~1 also lists the following data based on all the surveys:
mean flux densities $\overline{F}$ of these objects (column~5),
absolute  $\sigma^{\rm set}$ and relative ${\rm RMS}^{\rm set}$
root mean square deviations  from the mean values $\overline{F}$
(columns~6 and 7 respectively), the angles
 $dH$ averaged
over all surveys (column~8),\!\footnote{ $dH= \Delta {\rm Dec} =
{\rm Dec}^{\rm ist}-{\rm Dec}^{0}$, where ${\rm Dec}^{0}$ and
${\rm Dec}^{\rm ist}$ are the declination of the central section
of the survey and that of the source, respectively. The latter
varies from survey to survey because of precession.} and spectral
indices $\alpha$ at 3.94~GHz (column~9). The asterisks (*) in
column~10 mark the radio sources that are identified in the
records with the highest confidence.

\section{ANALYSIS OF THE STATISTICAL PROPERTIES OF THE SOURCES}

To confirm the variability of the objects with positive long-term
variability indices, we use statistical techniques similar to
those employed in~\mbox
{\cite{maj:Majorova_n,g1:Majorova_n,kest:Majorova_n,fan:Majorova_n,sei:Majorova_n}}.
We  compute the absolute $\Delta F$ and relative  $V_{\chi}$
variability amplitudes as well as the weighted average flux
$\langle F \rangle$ of the source, weighted average root mean
square error $\langle \sigma \rangle$, and the $\chi^2$ criterion
for the number of degrees of freedom \mbox {$df=n-1$}, where $n$
is the number of surveys. We perform our computations by the
following formulas~\cite{sei:Majorova_n}:
\begin{align}
\langle F \rangle  &= \sum_{i}^n (F_{i}/\sigma_{i}^2) /
\sum_{i}^n\sigma_{i}^{-2},  \label{11:Majorova_n}\\[+10pt]
\langle \sigma \rangle    &= \left.\left( \sum_{i}^n
(1/\sigma_{i}^2)\right.\right)^{-0.5}  ,  \label{12:Majorova_n}\\[+10pt]
\chi^2 &= \sum_{i}^n (F_{i} - \langle F \rangle)^2 / \sigma_{i}^2,
\label{13:Majorova_n}
\end{align}
\begin{equation}
\Delta F = \left.\left( (n-1)[\chi^2 - (n-1)]/ \sum_{i}^n
(F_{i}/\sigma_{i}^2)\right.\right)^{0.5}, \notag
\end{equation}
\begin{equation}
V_{\chi} = \Delta F / \langle F \rangle . \label{14:Majorova_n}
\end{equation}

We summarize the results of these computations in Table~2. Columns~2, 3, and 4 list
the $\chi^2$ probabilities of source variability. The parameter $p_{df}$ provides a quantitative estimate
of the probability that a source whose flux density obeys a  $\chi^2$ distribution with
\mbox {$df=n-1$} degrees of freedom can be considered variable \mbox {($p = 1- \chi^2(n-1)$)}.
We computed the $df^{'}=df-1$ degrees of freedom $\chi^2$ probabilities for the sources
whose flux densities measured in the 1993 and 1994 surveys agree within the measurement errors.
In our computations we treated the averaged flux densities measured in 1993 and 1994 as a single data point, thereby
reducing the number of degrees of freedom by one. We list the $p_{df-1}$ values in column~3 of Table~2.
Column~4 lists the average probabilities \mbox {$\overline{p}=(p_{df}+p_{df-1})/2$}.

Column~5 lists the weighted average flux densities $\langle F
\rangle$, column~6 lists their absolute variability amplitudes
$\Delta F$, column~7 gives the parameter $V_{\chi}$, columns~8 and
9 give the absolute $\langle \sigma \rangle$ and relative $\langle
{\sigma \rangle}^{\rm otn} = \langle \sigma \rangle / \langle F
\rangle$ weighted average root mean square errors, column~10 lists
the $\chi^2$ values, and column~11 gives the number of degrees of
freedom $df=n-1$.

\begin{figure*}[]
\onelinecaptionsfalse \centerline{ \vbox{\vspace{13mm} \hbox{
\includegraphics[angle=0,width=0.31\textwidth,bb=0 25 316 274,clip]{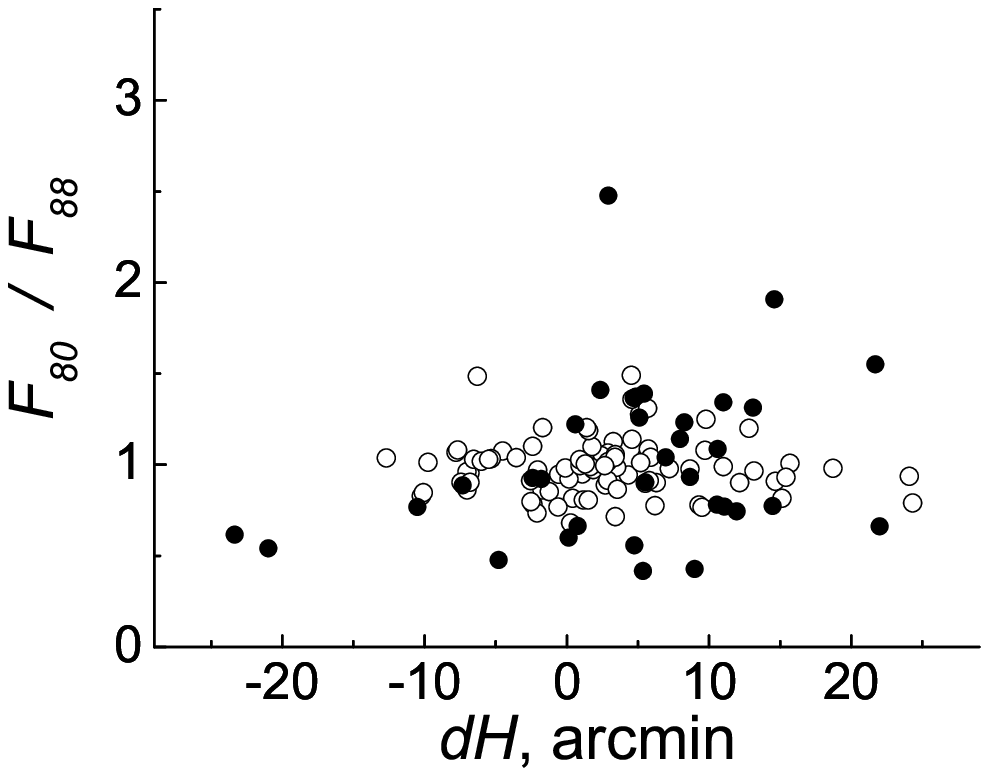}
\includegraphics[angle=0,width=0.31\textwidth,bb=0 25 316 274,clip]{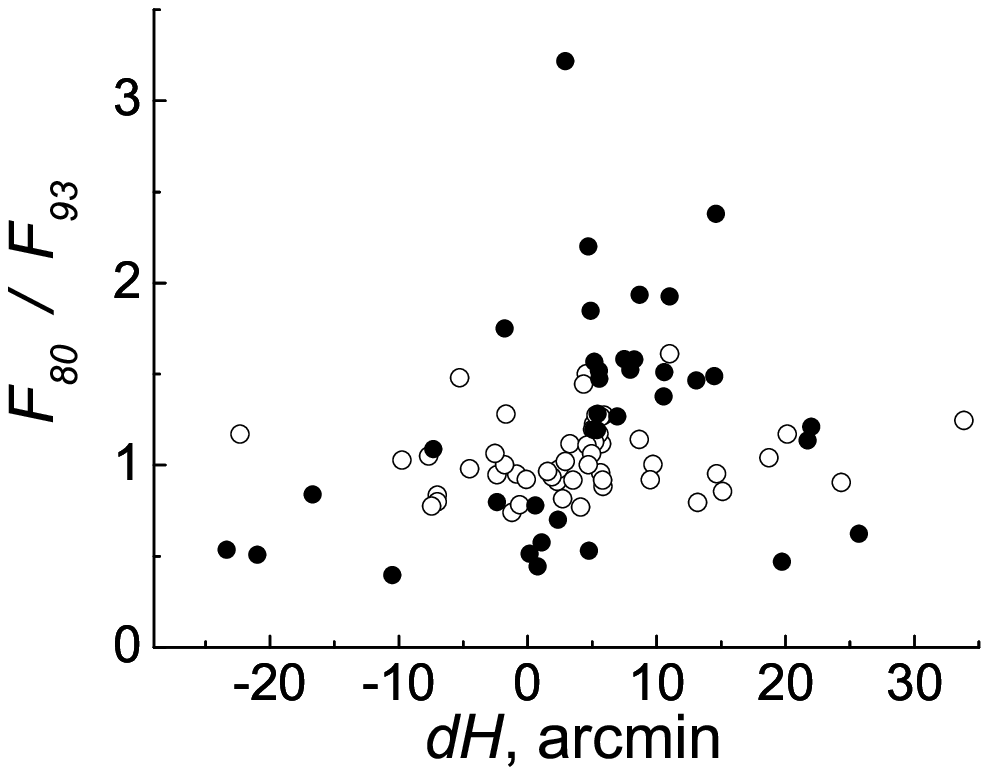}
\includegraphics[angle=0,width=0.31\textwidth,bb=0 25 316 274,clip]{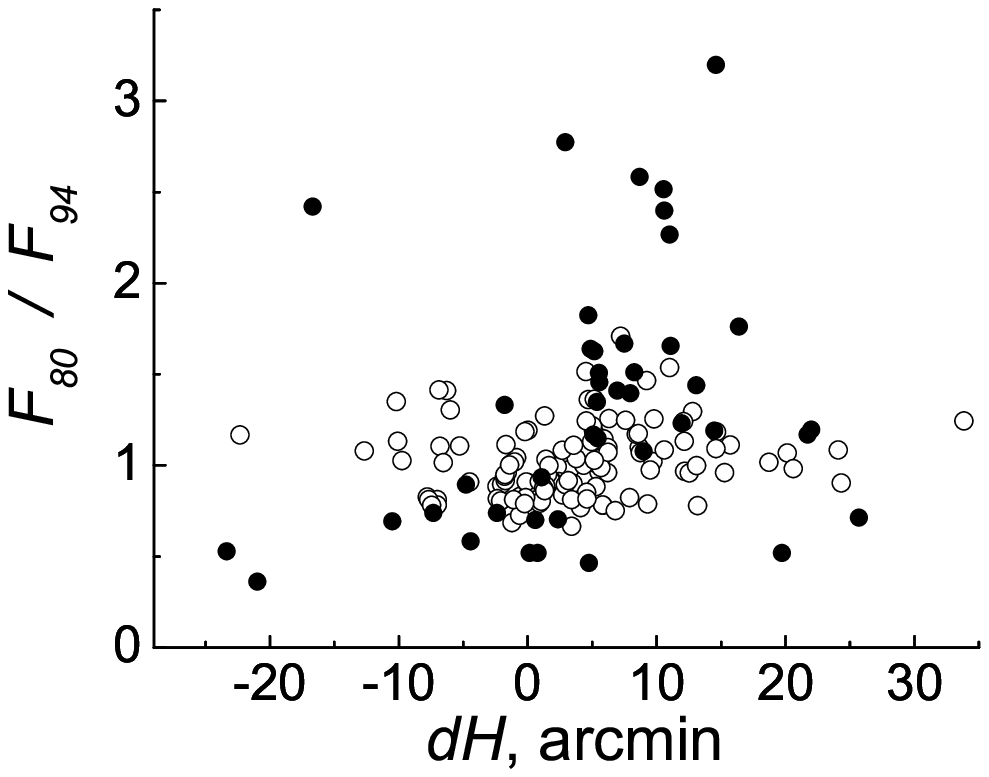}
 \vspace{5mm}
} \hbox{
\includegraphics[angle=0,width=0.31\textwidth,bb=0 25 316 274,clip]{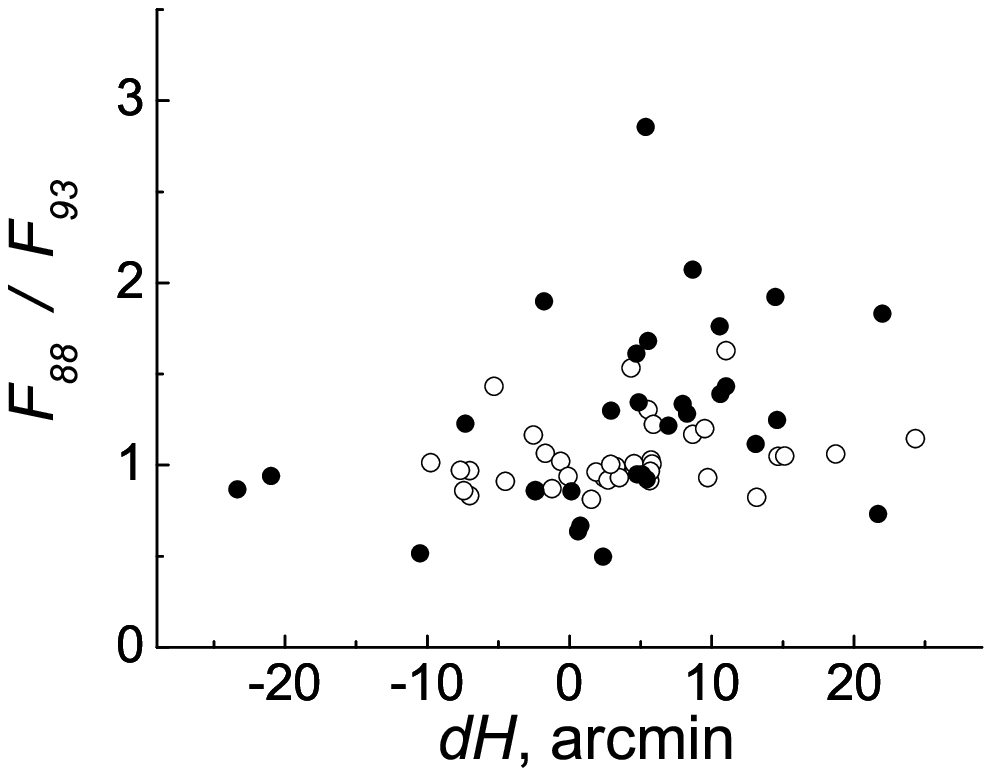}
\includegraphics[angle=0,width=0.31\textwidth,bb=0 25 316 274,clip]{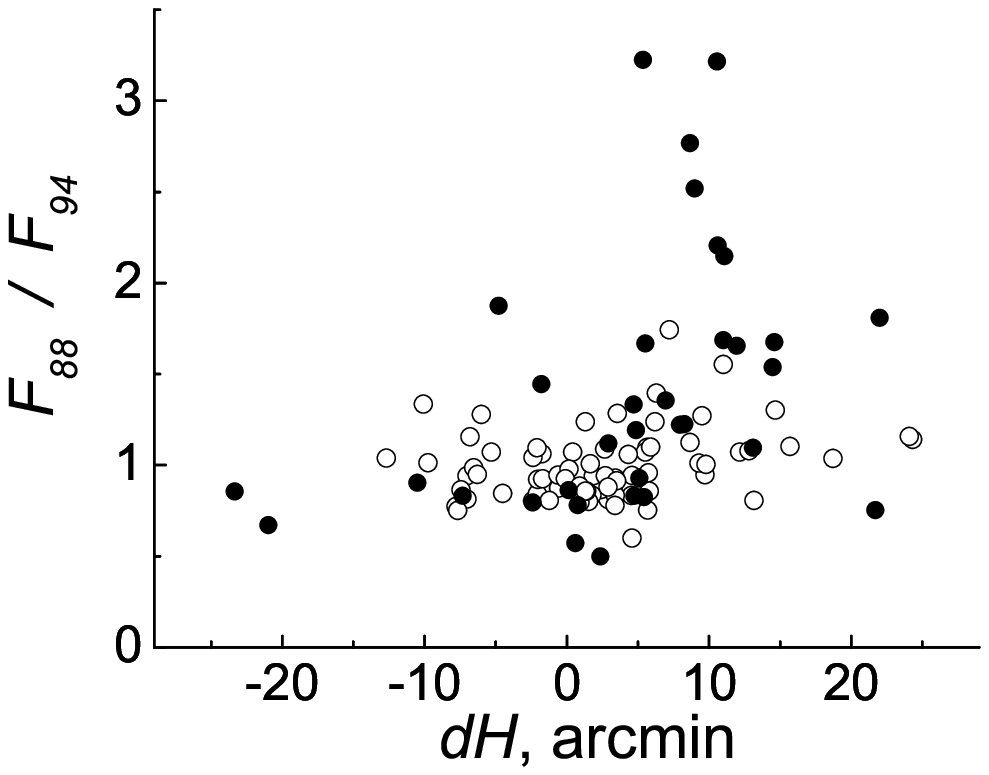}
\includegraphics[angle=0,width=0.31\textwidth,bb=0 25 316 274,clip]{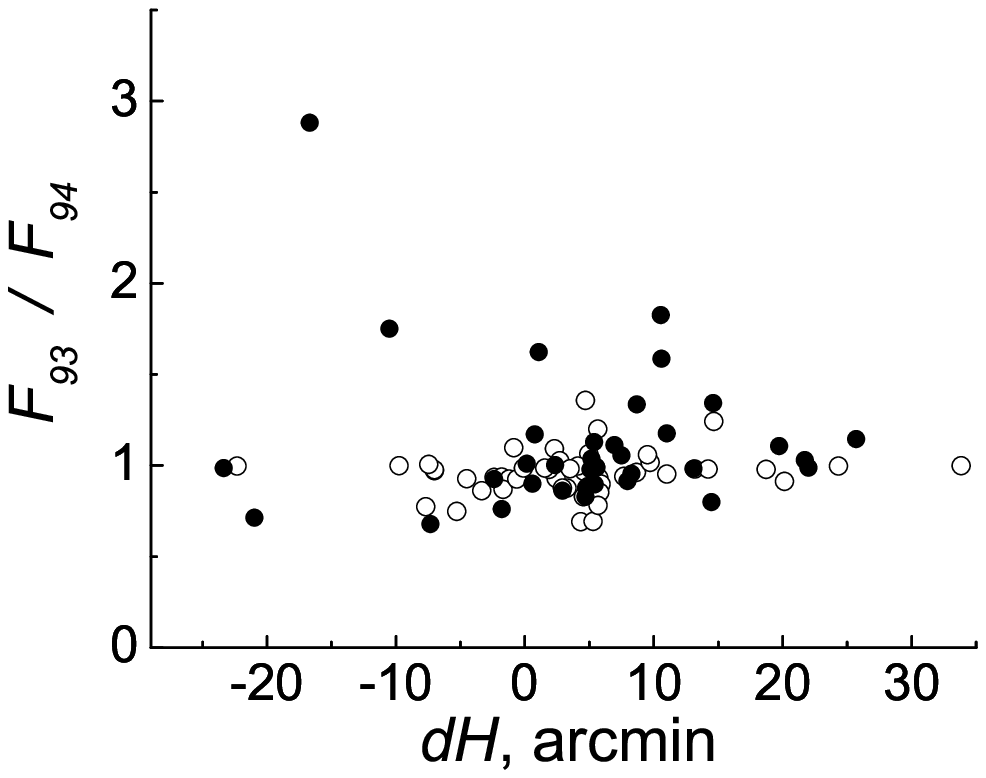}
} } } \setcaptionmargin{5mm} \captionstyle{normal}
 \vspace{5mm}
 \caption{
The $F_{i}/F_{j}$ flux density ratios for the studied sample of
RCR sources, obtained in different surveys ($i,j=1980$, $1988$,
$1993$, and $1994$). The filled circles show the  $F_{i}/F_{j}$
ratios for the objects in Table~1 (sources with $V>0$), and the
open circles---for objects with $V \le 0$. }
\label{fig2:Majorova_n}
 \vspace{13mm}
\end{figure*}

Figure~\ref{fig2:Majorova_n} shows the  $F_{i}/F_{j}$ flux density
ratios for the considered sample of \mbox {RCR sources} obtained
in different surveys ($i,j=1980$, $1988$, $1993$, and $1994$). The
filled circles show the  $F_{i}/F_{j}$ ratios for the objects in
Table~1 (sources with $V>0$), and the open circles---for objects
with $V \le 0$.

The smallest scatter of data points is between the 1993~ and~
1994~ surveys.~ For~ the~ sample~ of~ sources\pagebreak

\onecolumngrid

\setcaptionwidth{\linewidth}%
\setcaptionmargin{0mm} %
\onelinecaptionstrue \captionstyle{normal}
\medskip
\begin{longtable}{c|c|c|c|r|r|c|r|r|c}
\caption{ Parameters of RCR objects with positive long-term variability indices $V$}
\label{data:Majorova_n}\\
\hline
RCR & $V$       & $V_{R}$   & $V_{F}$  & \multicolumn{1}{c|}{$\overline{F}$,}  & \multicolumn{1}{c|}{$\sigma^{\rm set}$,}  & ${\rm RMS}^{\rm set}$  &\multicolumn{1}{c|}{$\overline{dH}$,}    &\multicolumn{1}{c|}{$\alpha$} &  Notes  \\
~~~${\rm RA}_{2000}$~~~~~~${\rm Dec}_{2000}$       &             &             &             & \multicolumn{1}{c|}{mJy}              &\multicolumn{1}{c|}{mJy}                  &                        &\multicolumn{1}{c|}{arcmin}               &             &    \\
\hline
(1)&(2)&(3)&(4)&\multicolumn{1}{c|}{(5)}&\multicolumn{1}{c|}{(6)}&(7)&\multicolumn{1}{c|}{(8)}&\multicolumn{1}{c|}{(9)}&(10)\\

\endfirsthead
\caption{(Contd.) }\\
\hline
RCR & $V$       & $V_{R}$   & $V_{F}$  & \multicolumn{1}{c|}{$\overline{F}$,}  & \multicolumn{1}{c|}{$\sigma^{\rm set}$,}  & ${\rm RMS}^{\rm set}$  &\multicolumn{1}{c|}{$\overline{dH}$,}    &\multicolumn{1}{c|}{$\alpha$} &  Notes  \\
~~~${\rm RA}_{2000}$~~~~~~${\rm Dec}_{2000}$       &             &             &             & \multicolumn{1}{c|}{mJy}              &\multicolumn{1}{c|}{mJy}                  &                        &\multicolumn{1}{c|}{arcmin}               &             &    \\
\hline
(1)&(2)&(3)&(4)&\multicolumn{1}{c|}{(5)}&\multicolumn{1}{c|}{(6)}&(7)&\multicolumn{1}{c|}{(8)}&\multicolumn{1}{c|}{(9)}&(10)\\
\hline

\endhead
\hline
\endfoot
\endlastfoot
\hline

J\,072919.57+044948.7        ~&~ $0.189$   ~&~ $2.09$      ~&~$2.61$    ~&~$ 51$   ~&~ $22 $    ~&~ $0.428$    ~&~$5.5  $           ~&~$-0.67$ ~&~ \\
J\,073357.46+045614.1        ~&~ $0.199$   ~&~ $1.90$      ~&~$3.99$    ~&~$265$   ~&~ $73 $    ~&~ $0.276$    ~&~$0.6  $           ~&~$ 0.11$ ~&~*\\
J\,075314.02+045129.4        ~&~ $0.025$   ~&~ $1.35$      ~&~$1.64$    ~&~$183$   ~&~ $25 $    ~&~ $0.138$    ~&~$4.2  $           ~&~$-0.34$ ~&~*\\
J\,080757.60+043234.6        ~&~ $0.275$   ~&~ $2.68$      ~&~$3.12$    ~&~$374$   ~&~ $142$    ~&~ $0.380$    ~&~$-23.0$           ~&~$-0.30$ ~&~ \\
J\,081218.14+050755.5        ~&~ $0.097$   ~&~ $1.92$      ~&~$1.91$    ~&~$118$   ~&~ $32 $    ~&~ $0.266$    ~&~$11.5 $           ~&~$-0.75$ ~&~ \\
J\,081626.62+045852.8        ~&~ $0.106$   ~&~ $1.67$      ~&~$2.42$    ~&~$ 53$   ~&~ $14 $    ~&~ $0.272$    ~&~$3.5  $           ~&~$-0.88$ ~&~ \\
J\,083148.89+042938.5        ~&~ $0.130$   ~&~ $1.89$      ~&~$2.32$    ~&~$949$   ~&~ $247$    ~&~ $0.261$    ~&~$-25.5$           ~&~$ 0.04$ ~&~*\\
J\,091636.22+044132.0        ~&~ $0.263$   ~&~ $2.52$      ~&~$2.90$    ~&~$171$   ~&~ $72 $    ~&~ $0.424$    ~&~$-13.0$           ~&~$-0.84$ ~&~ \\
J\,092355.77+045645.8        ~&~ $0.087$   ~&~ $1.85$      ~&~$1.88$    ~&~$20 $   ~&~~$6  $    ~&~ $0.305$    ~&~$3.4  $           ~&~$ 0.16$ ~&~ \\
J\,095218.73+050559.3        ~&~ $0.187$   ~&~ $2.41$      ~&~$2.15$    ~&~$60 $   ~&~ $34 $    ~&~ $0.565$    ~&~$9.9  $           ~&~$-0.43$ ~&~ \\
J\,100534.80+045119.8        ~&~ $0.104$   ~&~ $1.94$      ~&~$1.79$    ~&~$ 36$   ~&~ $10 $    ~&~ $0.270$    ~&~$-1.3 $           ~&~$-0.56$ ~&~ \\
J\,101603.12+051303.6        ~&~ $0.147$   ~&~ $1.81$      ~&~$2.74$    ~&~$574$   ~&~ $177$    ~&~ $0.309$    ~&~$17.7 $           ~&~$ 0.04$ ~&~*\\
J\,103846.84+051229.6        ~&~ $0.050$   ~&~ $1.55$      ~&~$1.81$    ~&~$455$   ~&~ $79 $    ~&~ $0.175$    ~&~$20.2 $           ~&~$ 0.24$ ~&~*\\
J\,104117.65+045306.4        ~&~ $0.157$   ~&~ $2.08$      ~&~$2.61$    ~&~$ 42$   ~&~ $13 $    ~&~ $0.307$    ~&~$-2.1 $           ~&~$-0.82$ ~&~ \\
J\,104527.19+045118.7        ~&~ $0.043$   ~&~ $1.75$      ~&~$1.70$    ~&~$ 32$   ~&~ $8  $    ~&~ $0.246$    ~&~$-1.9 $           ~&~$-0.85$ ~&~ \\
J\,105253.05+045735.3        ~&~ $0.039$   ~&~ $1.41$      ~&~$1.82$    ~&~$102$   ~&~ $16 $    ~&~ $0.160$    ~&~$3.9  $           ~&~$-0.15$ ~&~*\\
J\,105719.26+045545.4        ~&~ $0.200$   ~&~ $2.20$      ~&~$2.90$    ~&~$ 30$   ~&~ $11 $    ~&~ $0.350$    ~&~$2.0  $           ~&~$-0.02$ ~&~ \\
J\,113156.47+045549.3        ~&~ $0.050$   ~&~ $1.39$      ~&~$1.95$    ~&~$250$   ~&~ $36 $    ~&~ $0.145$    ~&~$3.8  $           ~&~$-0.84$ ~&~*\\
J\,114521.30+045526.7        ~&~ $0.018$   ~&~ $1.26$      ~&~$1.62$    ~&~$501$   ~&~ $51 $    ~&~ $0.102$    ~&~$3.5  $           ~&~$-0.33$ ~&~*\\
J\,114631.64+045818.2        ~&~ $0.056$   ~&~ $1.52$      ~&~$1.90$    ~&~$193$   ~&~ $37 $    ~&~ $0.192$    ~&~$4.9  $           ~&~$-0.21$ ~&~*\\
J\,115248.33+050057.2        ~&~ $0.171$   ~&~ $2.22$      ~&~$2.24$    ~&~$ 49$   ~&~ $16 $    ~&~ $0.332$    ~&~$7.5  $           ~&~$-0.89$ ~&~ \\
J\,115336.08+045505.2        ~&~ $0.175$   ~&~ $2.15$      ~&~$2.60$    ~&~$ 35$   ~&~ $10 $    ~&~ $0.300$    ~&~$1.6  $           ~&~$ 0.78$ ~&~ \\
J\,115851.23+045541.9        ~&~ $0.335$   ~&~ $3.09$      ~&~$3.91$    ~&~$ 21$   ~&~ $13 $    ~&~ $0.640$    ~&~$0.6  $           ~&~$-0.11$ ~&~ \\
J\,123507.25+045318.7        ~&~ $0.376$   ~&~ $3.22$      ~&~$4.63$    ~&~$ 38$   ~&~ $23 $    ~&~ $0.623$    ~&~$-0.1 $           ~&~$-0.05$ ~&~*\\
J\,123725.63+045741.6        ~&~ $0.032$   ~&~ $1.44$      ~&~$1.67$    ~&~$ 92$   ~&~ $18 $    ~&~ $0.198$    ~&~$4.1  $           ~&~$-1.15$ ~&~*\\
J\,123932.78+044305.3        ~&~ $0.027$   ~&~ $1.47$      ~&~$1.58$    ~&~$291$   ~&~ $55 $    ~&~ $0.191$    ~&~$-10.4$           ~&~$-0.13$ ~&~ \\
J\,124145.15+045924.5        ~&~ $0.227$   ~&~ $2.52$      ~&~$2.53$    ~&~$37 $   ~&~ $21 $    ~&~ $0.558$    ~&~$4.2  $           ~&~$-0.57$ ~&~ \\
J\,125755.32+045917.6        ~&~ $0.072$   ~&~ $1.58$      ~&~$2.02$    ~&~$153$   ~&~ $33 $    ~&~ $0.217$    ~&~$5.3  $           ~&~$-1.01$ ~&~*\\
J\,130631.65+050231.3        ~&~ $0.056$   ~&~ $1.67$      ~&~$1.67$    ~&~$ 61$   ~&~ $16 $    ~&~ $0.257$    ~&~$8.8  $           ~&~$-0.25$ ~&~ \\
J\,133541.21+050124.9        ~&~ $0.216$   ~&~ $2.77$      ~&~$2.11$    ~&~$34 $   ~&~ $14 $    ~&~ $0.406$    ~&~$4.1  $           ~&~$-0.96$ ~&~ \\
J\,133920.76+050159.3        ~&~ $0.116$   ~&~ $2.09$      ~&~$1.79$    ~&~$47 $   ~&~ $16 $    ~&~ $0.348$    ~&~$4.6  $           ~&~$-0.30$ ~&~ \\
J\,134201.37+050157.5        ~&~ $0.108$   ~&~ $1.77$      ~&~$2.08$    ~&~$68 $   ~&~ $21 $    ~&~ $0.309$    ~&~$6.1  $           ~&~$ 0.85$ ~&~ \\
J\,135050.06+045148.9        ~&~ $0.173$   ~&~ $2.11$      ~&~$2.41$    ~&~$30 $   ~&~ $10 $    ~&~ $0.336$    ~&~$-2.0 $           ~&~$ 0.44$ ~&~ \\
J\,141920.56+051110.6        ~&~ $0.104$   ~&~ $2.22$      ~&~$1.91$    ~&~$137$   ~&~ $48 $    ~&~ $0.351$    ~&~$17.0 $           ~&~$-0.27$ ~&~ \\
J\,142409.47+043451.7        ~&~ $0.231$   ~&~ $3.01$      ~&~$2.46$    ~&~$317$   ~&~ $148$    ~&~ $0.467$    ~&~$-20.0$           ~&~$-0.20$ ~&~ \\
J\,145032.99+050824.6        ~&~ $0.082$   ~&~ $1.76$      ~&~$1.90$    ~&~$121$   ~&~ $46 $    ~&~ $0.377$    ~&~$11.9 $           ~&~$-0.32$ ~&~*\\
J\,151141.19+051809.4        ~&~ $0.019$   ~&~ $1.40$      ~&~$1.44$    ~&~$395$   ~&~ $91 $    ~&~ $0.230$    ~&~$23.3 $           ~&~$ 1.11$ ~&~ \\
J\,161015.24+044923.5        ~&~ $0.080$   ~&~ $1.80$      ~&~$1.90$    ~&~$ 33$   ~&~ $13 $    ~&~ $0.403$    ~&~$-4.3 $           ~&~$-0.20$ ~&~ \\
J\,161637.49+045932.8        ~&~ $0.067$   ~&~ $1.48$      ~&~$2.19$    ~&~$932$   ~&~ $217$    ~&~ $0.233$    ~&~$3.8  $           ~&~$ 0.22$ ~&~*\\
J\,163106.83+050119.2        ~&~ $0.110$   ~&~ $1.67$      ~&~$2.21$    ~&~$ 79$   ~&~ $23 $    ~&~ $0.299$    ~&~$4.2  $           ~&~$-0.90$ ~&~*\\
J\,165643.92+050014.0        ~&~ $0.096$   ~&~ $1.63$      ~&~$2.33$    ~&~$ 53$   ~&~ $15 $    ~&~ $0.288$    ~&~$3.8  $           ~&~$-0.73$ ~&~*\\
\hline
\end{longtable}
\twocolumngrid

 \onecolumngrid
\setcaptionwidth{\linewidth}%
\setcaptionmargin{0mm} %
\onelinecaptionstrue \captionstyle{normal}
\medskip
\begin{longtable}{c|c|c|c|r|r|c|r|c|r|c}
\caption{ Statistical properties of  RCR objects with positive long-term variability
indices $V$}
\label{data:Majorova_n}\\
 \hline
RCR & $p_{df}$  &$p_{df-1}$ &$\overline{p}$, &\multicolumn{1}{c|}{$\langle F \rangle$,} &\multicolumn{1}{c|}{$\Delta F$,} &$V_\chi$   &\multicolumn{1}{c|}{$\langle \sigma \rangle$,} &${\langle \sigma \rangle}^{\rm otn}$  &\multicolumn{1}{c|}{$\chi^2$}  &$df$    \\
~~~${\rm RA}_{2000}$~~~~~~${\rm Dec}_{2000}$ &  & & &\multicolumn{1}{c|}{mJy} &\multicolumn{1}{c|}{mJy} &  &\multicolumn{1}{c|}{mJy}      & & &   \\
\hline
(1)&(2)&(3)&(4)&\multicolumn{1}{c|}{(5)}&\multicolumn{1}{c|}{(6)}&(7)&\multicolumn{1}{c|}{(8)}&(9)&\multicolumn{1}{c|}{(10)}&(11)\\
\endfirsthead
\caption{(Contd.) }\\
 \hline
RCR & $p_{df}$  &$p_{df-1}$ &$\overline{p}$, & \multicolumn{1}{c|}{$\langle F \rangle$,} &\multicolumn{1}{c|}{$\Delta F$,} &$V_\chi$   &\multicolumn{1}{c|}{$\langle \sigma \rangle$,} &${\langle \sigma \rangle}^{\rm otn}$  &\multicolumn{1}{c|}{$\chi^2$}  &$df$    \\
~~~${\rm RA}_{2000}$~~~~~~${\rm Dec}_{2000}$ &  & & &\multicolumn{1}{c|}{mJy} &\multicolumn{1}{c|}{mJy} &  &\multicolumn{1}{c|}{mJy}      & & &   \\
\hline
(1)&(2)&(3)&(4)&\multicolumn{1}{c|}{(5)}&\multicolumn{1}{c|}{(6)}&(7)&\multicolumn{1}{c|}{(8)}&(9)&\multicolumn{1}{c|}{(10)}&(11)\\
\hline
\endhead
\hline
\endfoot
\endlastfoot
\hline

J\,072919.57+044948.7      ~  &~ $0.966$ ~ &~ $     $ ~ &~ $0.966$  ~    &~$ 42$ ~ &~$ 16$  ~  &~ $0.385$  ~ &~$ 5$ ~  &~ $0.124$  ~ &~$ 6.82$  ~ &~$2$   \\
J\,073357.46+045614.1      ~  &~ $1    $ ~ &~ $0.999$ ~ &~ $0.999$  ~    &~$266$ ~ &~$103$  ~  &~ $0.388$  ~ &~$15$ ~  &~ $0.056$  ~ &~$18.80$  ~ &~$3$   \\
J\,075314.02+045129.4      ~  &~ $0.653$ ~ &~ $0.796$ ~ &~ $0.725$  ~    &~$173$ ~ &~$ 10$  ~  &~ $0.057$  ~ &~$10$ ~  &~ $0.059$  ~ &~$ 3.31$  ~ &~$3$   \\
J\,080757.60+043234.6      ~  &~ $0.991$ ~ &~ $0.995$ ~ &~ $0.993$  ~    &~$302$ ~ &~$179$  ~  &~ $0.594$  ~ &~$37$ ~  &~ $0.124$  ~ &~$10.67$  ~ &~$3$   \\
J\,081218.14+050755.5      ~  &~ $0.738$ ~ &~ $0.906$ ~ &~ $0.823$  ~    &~$110$ ~ &~$ 21$  ~  &~ $0.194$  ~ &~$12$ ~  &~ $0.112$  ~ &~$ 4.01$  ~ &~$3$   \\
J\,081626.62+045852.8      ~  &~ $0.974$ ~ &~ $0.990$ ~ &~ $0.983$  ~    &~$ 51$ ~ &~$ 17$  ~  &~ $0.337$  ~ &~$ 4$ ~  &~ $0.078$  ~ &~$ 9.25$  ~ &~$3$   \\
J\,083148.89+042938.5      ~  &~ $0.960$ ~ &~ $0.984$ ~ &~ $0.972$  ~    &~$859$ ~ &~$316$  ~  &~ $0.367$  ~ &~$79$ ~  &~ $0.092$  ~ &~$ 8.31$  ~ &~$3$   \\
J\,091636.22+044132.0      ~  &~ $0.965$ ~ &~ $     $ ~ &~ $0.965$  ~    &~$136$ ~ &~$ 62$  ~  &~ $0.453$  ~ &~$15$ ~  &~ $0.111$  ~ &~$ 8.55$  ~ &~$3$   \\
J\,092355.77+045645.8      ~  &~ $0.810$ ~ &~ $0.890$ ~ &~ $0.850$  ~    &~$ 18$ ~ &~$~5 $  ~  &~ $0.270$  ~ &~$ 2$ ~  &~ $0.117$  ~ &~$ 3.62$  ~ &~$3$   \\
J\,095218.73+050559.3      ~  &~ $0.922$ ~ &~ $0.976$ ~ &~ $0.949$  ~    &~$ 45$ ~ &~$~18$  ~  &~ $0.403$  ~ &~$ 7$ ~  &~ $0.161$  ~ &~$ 5.13$  ~ &~$3$   \\
J\,100534.80+045119.8      ~  &~ $0.955$ ~ &~ $0.981$ ~ &~ $0.968$  ~    &~$ 29$ ~ &~$ 11$  ~  &~ $0.377$  ~ &~$ 3$ ~  &~ $0.096$  ~ &~$ 8.12$  ~ &~$3$   \\
J\,101603.12+051303.6      ~  &~ $0.966$ ~ &~ $0.987$ ~ &~ $0.976$  ~    &~$536$ ~ &~$179$  ~  &~ $0.334$  ~ &~$43$ ~  &~ $0.081$  ~ &~$ 8.67$  ~ &~$3$   \\
J\,103846.84+051229.6      ~  &~ $0.703$ ~ &~ $0.839$ ~ &~ $0.771$  ~    &~$442$ ~ &~$ 54$  ~  &~ $0.121$  ~ &~$37$ ~  &~ $0.084$  ~ &~$ 3.70$  ~ &~$3$   \\
J\,104117.65+045306.4      ~  &~ $0.982$ ~ &~ $0.994$ ~ &~ $0.989$  ~    &~$ 35$ ~ &~$ 14$  ~  &~ $0.414$  ~ &~$ 3$ ~  &~ $0.087$  ~ &~$10.56$  ~ &~$3$   \\
J\,104527.19+045118.7      ~  &~ $0.673$ ~ &~ $0.808$ ~ &~ $0.741$  ~    &~$ 28$ ~ &~$  6$  ~  &~ $0.202$  ~ &~$ 3$ ~  &~ $0.105$  ~ &~$ 4.37$  ~ &~$3$   \\
J\,105253.05+045735.3      ~  &~ $0.808$ ~ &~ $0.895$ ~ &~ $0.852$  ~    &~$101$ ~ &~$ 15$  ~  &~ $0.149$  ~ &~$ 7$ ~  &~ $0.065$  ~ &~$ 4.74$  ~ &~$3$   \\
J\,105719.26+045545.4      ~  &~ $0.978$ ~ &~ $0.990$ ~ &~ $0.984$  ~    &~$ 28$ ~ &~$ 12$  ~  &~ $0.427$  ~ &~$ 3$ ~  &~ $0.108$  ~ &~$10.42$  ~ &~$3$   \\
J\,113156.47+045549.3      ~  &~ $0.738$ ~ &~ $0.842$ ~ &~ $0.790$  ~    &~$239$ ~ &~$ 26$  ~  &~ $0.110$  ~ &~$15$ ~  &~ $0.064$  ~ &~$ 4.01$  ~ &~$3$   \\
J\,114521.30+045526.7      ~  &~ $0.548$ ~ &~ $0.729$ ~ &~ $0.639$  ~    &~$487$ ~ &~$ 30$  ~  &~ $0.061$  ~ &~$27$ ~  &~ $0.055$  ~ &~$ 2.63$  ~ &~$3$   \\
J\,114631.64+045818.2      ~  &~ $0.854$ ~ &~ $0.924$ ~ &~ $0.889$  ~    &~$190$ ~ &~$ 31$  ~  &~ $0.165$  ~ &~$12$ ~  &~ $0.062$  ~ &~$ 5.36$  ~ &~$3$   \\
J\,115248.33+050057.2      ~  &~ $0.974$ ~ &~ $     $ ~ &~ $0.974$  ~    &~$ 39$ ~ &~$ 20$  ~  &~ $0.513$  ~ &~$ 5$ ~  &~ $0.119$  ~ &~$ 9.23$  ~ &~$3$   \\
J\,115336.08+045505.2      ~  &~ $0.974$ ~ &~ $0.989$ ~ &~ $0.981$  ~    &~$ 33$ ~ &~$ 13$  ~  &~ $0.393$  ~ &~$ 3$ ~  &~ $0.083$  ~ &~$ 9.59$  ~ &~$3$   \\
J\,115851.23+045541.9      ~  &~ $0.999$ ~ &~ $1    $ ~ &~ $0.999$  ~    &~$ 19$ ~ &~$ 17$  ~  &~ $0.882$  ~ &~$ 2$ ~  &~ $0.128$  ~ &~$18.87$  ~ &~$3$   \\
J\,123507.25+045318.7      ~  &~ $1    $ ~ &~ $1    $ ~ &~ $1    $  ~    &~$ 33$ ~ &~$ 25$  ~  &~ $0.753$  ~ &~$ 3$ ~  &~ $0.093$  ~ &~$25.06$  ~ &~$3$   \\
J\,123725.63+045741.6      ~  &~ $0.680$ ~ &~ $0.828$ ~ &~ $0.754$  ~    &~$ 88$ ~ &~$  8$  ~  &~ $0.089$  ~ &~$ 6$ ~  &~ $0.072$  ~ &~$ 3.51$  ~ &~$3$   \\
J\,123932.78+044305.3      ~  &~ $0.844$ ~ &~ $     $ ~ &~ $0.840$  ~    &~$281$ ~ &~$ 40$  ~  &~ $0.135$  ~ &~$19$ ~  &~ $0.069$  ~ &~$ 3.70$  ~ &~$3$   \\
J\,124145.15+045924.5      ~  &~ $0.961$ ~ &~ $     $ ~ &~ $0.961$  ~    &~$ 29$ ~ &~$ 14$  ~  &~ $0.491$  ~ &~$ 5$ ~  &~ $0.163$  ~ &~$ 6.54$  ~ &~$2$   \\
J\,125755.32+045917.6      ~  &~ $0.868$ ~ &~ $0.939$ ~ &~ $0.904$  ~    &~$148$ ~ &~$ 27$  ~  &~ $0.184$  ~ &~$10$ ~  &~ $0.066$  ~ &~$ 5.59$  ~ &~$3$   \\
J\,130631.65+050231.3      ~  &~ $0.764$ ~ &~ $     $ ~ &~ $0.764$  ~    &~$ 54$ ~ &~$  9$  ~  &~ $0.159$  ~ &~$ 6$ ~  &~ $0.119$  ~ &~$ 2.89$  ~ &~$2$   \\
J\,133541.21+050124.9      ~  &~ $0.902$ ~ &~ $0.933$ ~ &~ $0.764$  ~    &~$ 25$ ~ &~$ 13$  ~  &~ $0.519$  ~ &~$ 4$ ~  &~ $0.165$  ~ &~$ 6.30$  ~ &~$3$   \\
J\,133920.76+050159.3      ~  &~ $0.871$ ~ &~ $     $ ~ &~ $0.917$  ~    &~$ 36$ ~ &~$ 11$  ~  &~ $0.307$  ~ &~$ 5$ ~  &~ $0.149$  ~ &~$ 4.13$  ~ &~$2$   \\
J\,134201.37+050157.5      ~  &~ $0.857$ ~ &~ $0.931$ ~ &~ $0.871$  ~    &~$ 60$ ~ &~$ 17$  ~  &~ $0.279$  ~ &~$ 6$ ~  &~ $0.103$  ~ &~$ 5.45$  ~ &~$3$   \\
J\,135050.06+045148.9      ~  &~ $0.972$ ~ &~ $0.989$ ~ &~ $0.894$  ~    &~$ 23$ ~ &~$ 11$  ~  &~ $0.481$  ~ &~$ 3$ ~  &~ $0.113$  ~ &~$ 9.03$  ~ &~$3$   \\
J\,141920.56+051110.6      ~  &~ $0.845$ ~ &~ $0.943$ ~ &~ $0.980$  ~    &~$131$ ~ &~$ 27$  ~  &~ $0.203$  ~ &~$22$ ~  &~ $0.167$  ~ &~$ 3.72$  ~ &~$2$   \\
J\,142409.47+043451.7      ~  &~ $0.988$ ~ &~ $     $ ~ &~ $0.988$  ~    &~$296$ ~ &~$143$  ~  &~ $0.484$  ~ &~$39$ ~  &~ $0.131$  ~ &~$ 8.82$  ~ &~$2$   \\
J\,145032.99+050824.6      ~  &~ $0.925$ ~ &~ $     $ ~ &~ $0.925$  ~    &~$111$ ~ &~$ 25$  ~  &~ $0.235$  ~ &~$18$ ~  &~ $0.164$  ~ &~$ 3.61$  ~ &~$1$   \\
J\,151141.19+051809.4      ~  &~ $0.725$ ~ &~ $0.876$ ~ &~ $0.801$  ~    &~$366$ ~ &~$ 49$  ~  &~ $0.133$  ~ &~$44$ ~  &~ $0.122$  ~ &~$ 2.59$  ~ &~$2$   \\
J\,161015.24+044923.5      ~  &~ $0.919$ ~ &~ $     $ ~ &~ $0.919$  ~    &~$ 31$ ~ &~$  7$  ~  &~ $0.223$  ~ &~$ 5$ ~  &~ $0.159$  ~ &~$ 3.07$  ~ &~$1$   \\
J\,161637.49+045932.8      ~  &~ $0.948$ ~ &~ $0.984$ ~ &~ $0.967$  ~    &~$920$ ~ &~$198$  ~  &~ $0.214$  ~ &~$71$ ~  &~ $0.077$  ~ &~$ 5.92$  ~ &~$2$   \\
J\,163106.83+050119.2      ~  &~ $0.920$ ~ &~ $0.972$ ~ &~ $0.947$  ~    &~$ 72$ ~ &~$ 17$  ~  &~ $0.241$  ~ &~$ 7$ ~  &~ $0.097$  ~ &~$ 5.09$  ~ &~$2$   \\
J\,165643.92+050014.0      ~  &~ $0.954$ ~ &~ $0.988$ ~ &~ $0.971$  ~    &~$ 51$ ~ &~$ 14$  ~  &~ $0.269$  ~ &~$ 5$ ~  &~ $0.093$  ~ &~$ 6.18$  ~ &~$2$   \\
\hline
\end{longtable}
\twocolumngrid

\noindent with $V \le 0$ the standard deviations of flux density
ratios from the corresponding mean ratio are equal to 0.11, 0.15,
0.20, 0.19, 0.18, and 0.19, for $F_{94}/F_{93}$, $F_{80}/F_{88}$,
$F_{80}/F_{93}$, $F_{80}/F_{94}$, $F_{88}/F_{93}$, and
$F_{88}/F_{94}$, respectively. The standard deviations for the
sample of possibly variable objects  ($V>0$) are equal to 0.40,
0.44, 0.63, 0.70, 0.53, and 0.72, for $F_{94}/F_{93}$,
$F_{80}/F_{88}$, $F_{80}/F_{93}$, $F_{80}/F_{94}$,
$F_{88}/F_{93}$, and $F_{88}/F_{94}$, respectively.

The differences of flux density ratios in the 1993 and 1994 surveys for most of the possibly variable objects
from Table~1 do not exceed the standard error of flux determination or are close to it, suggesting that the sources
of our sample are variable mostly on time scales of  6--8 years. The exceptions are objects
J\,091636+044132, J\,115248+050057, J\,123932+044305, and J\,142409+043451.

It follows from a comparison of the data listed in Tables~1~and~2 that the weighted average
flux density ratios $\langle F \rangle$ computed by formula~(\ref{11:Majorova_n}) practically
coincide with $\overline{F}$ within the errors. Also close are the  $\sigma^{\rm set}$ (Table~1)
and $\Delta F$ (Table~2) values for individual sources, and  $V_{\chi}$ and \mbox{${\rm
RMS}^{\rm set}= \sigma^{\rm set}/\overline{F}$}.

\begin{figure}[b]
\setcaptionmargin{5mm}
 \onelinecaptionsfalse \centerline{ \vbox{
\hbox{
\includegraphics[angle=0,width=0.31\textwidth,bb=0 35 313 265,clip]{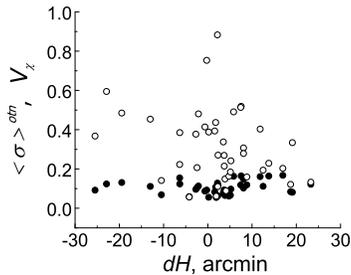}
} } } \setcaptionmargin{5mm} \captionstyle{normal}
 \caption{
Dependence of the relative variability amplitude $V_{\chi}$ and
relative weighted average standard error of the measured flux
densities ${\langle \sigma \rangle}^{\rm otn}$ on angle $dH$ for
sources with $V>0$. The open and filled circles show the
dependences for $V_{\chi}$ and  ${\langle \sigma \rangle}^{\rm
otn}$ respectively. } \label{fig3:Majorova_n}
\end{figure}

For the sources listed in Tables~1~and~2 the relative variability
amplitudes $V_{\chi}$ and relative standard deviations from the
mean, ${\rm RMS}^{\rm set}$, mostly exceed substantially the
corresponding relative weighted standard errors of the measured
flux densities ${\langle \sigma \rangle}^{\rm otn}$. This is
evident from Fig.~\ref{fig3:Majorova_n}, which shows how the above
quantities depend on angle $dH$: ${\langle \sigma \rangle}^{\rm
otn}$ (the filled circles) and $V_{\chi}$ (the open triangles).
The relative standard deviations $\overline{{\rm RMS}^{\rm set}}$
and  $\overline{V_{\chi}}$ averaged\linebreak over all sources
with $V>0$ are equal to\linebreak $0.315\pm0.125$ and
$0.317\pm0.182$ respectively, and\linebreak
\mbox{$\overline{{\langle \sigma \rangle}^{\rm otn} } =
0.105\pm0.0332$}.

\begin{figure}[]
\setcaptionmargin{5mm}
 \onelinecaptionsfalse \centerline{ \vbox{
\includegraphics[angle=0,width=0.3\textwidth,bb=0 15 304 257,clip]{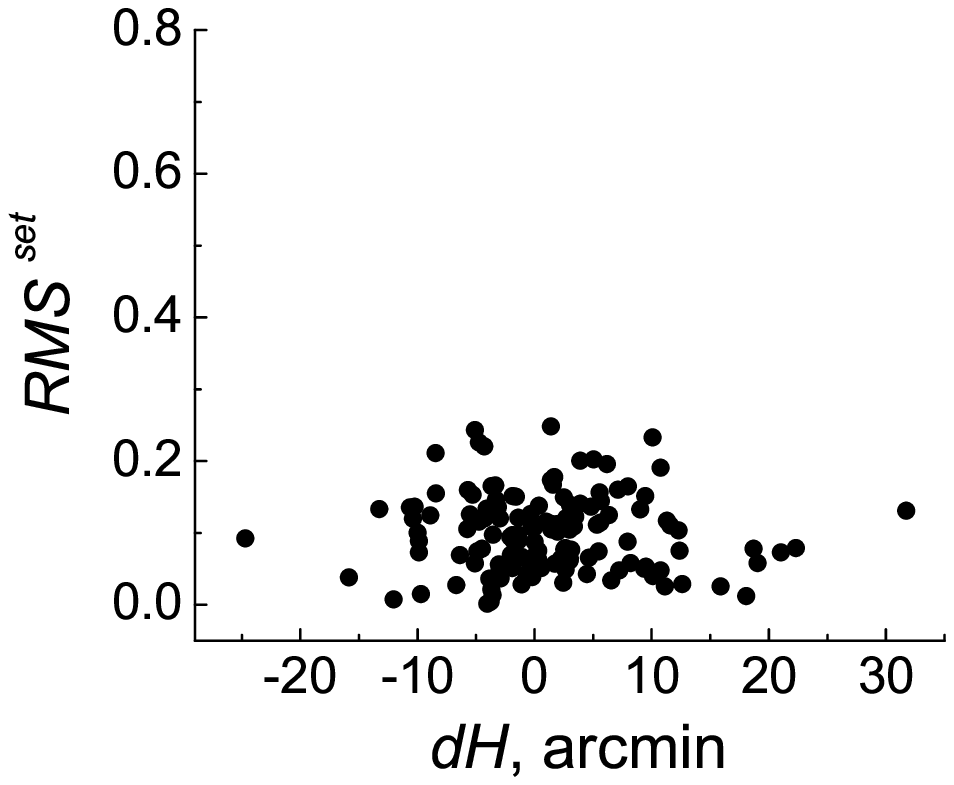}
\includegraphics[angle=0,width=0.3\textwidth,bb=0 15 304 257,clip]{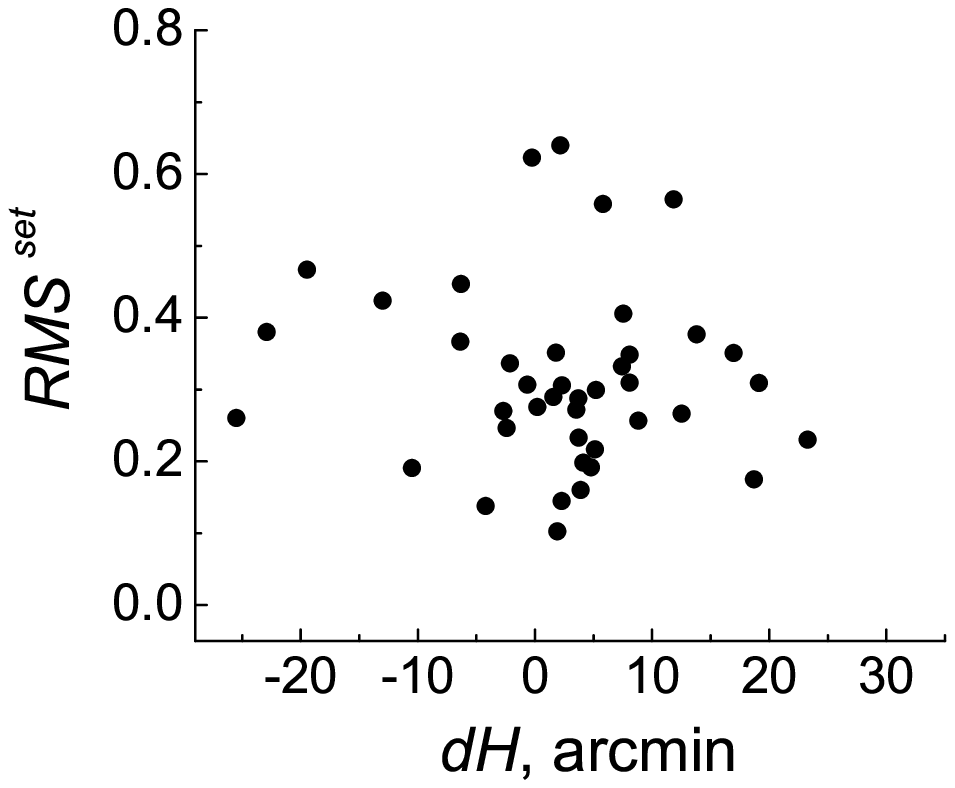}
} }  \setcaptionmargin{5mm} \captionstyle{normal}
 \caption{
Relative standard deviation ${\rm RMS}^{\rm set}$
from the mean value averaged over all surveys  as a function of angle $dH$ for sources with
$V>0$ (the upper panel) and $V \le 0$ (the lower panel). } \label{fig4:Majorova_n}
\end{figure}

Let us now compare the ${\rm RMS}^{\rm set}$  values for sources
with positive and negative long-term variability indices. The
upper and lower panels in Fig.~\ref{fig4:Majorova_n} show the
dependences of  ${\rm RMS}^{\rm set}$ on $dH$ for sources with
$V>0$ and $V \le 0$ respectively. The ${\rm RMS}^{\rm set}$ values
for possibly variable objects ($V>0$) exceed significantly the
${\rm RMS}^{\rm set}$ for objects with \mbox {$V \le 0$}. The same
is true for the relative variability amplitudes  $V_{\chi}$ of
variable and nonvariable sources. The mean $\overline{{\rm
RMS}^{\rm set}}$ value averaged over all nonvariable sources  ($V
\le 0$)  is $0.101\pm0.062$ which is comparable with the averaged
$\overline{{\langle \sigma \rangle}^{\rm otn} }$ value.

\begin{figure}[]
\onelinecaptionsfalse \centerline{ \vbox{
\includegraphics[angle=0,width=0.31\textwidth,bb=0 35 313 265,clip]{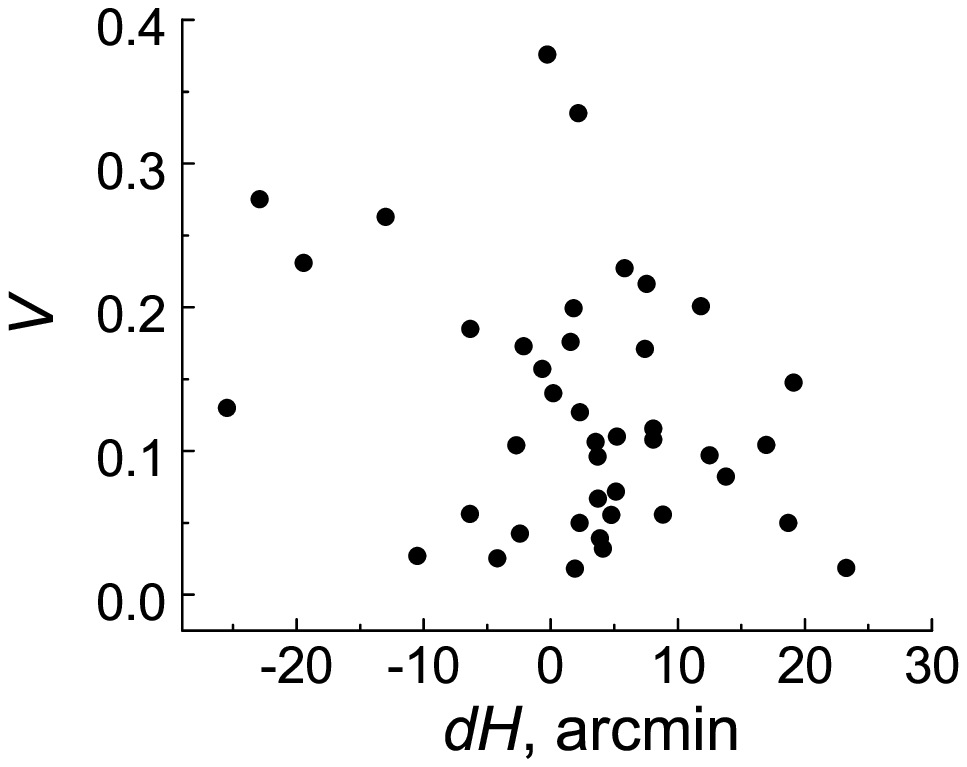}
\includegraphics[angle=0,width=0.31\textwidth,bb=0 35 313 265,clip]{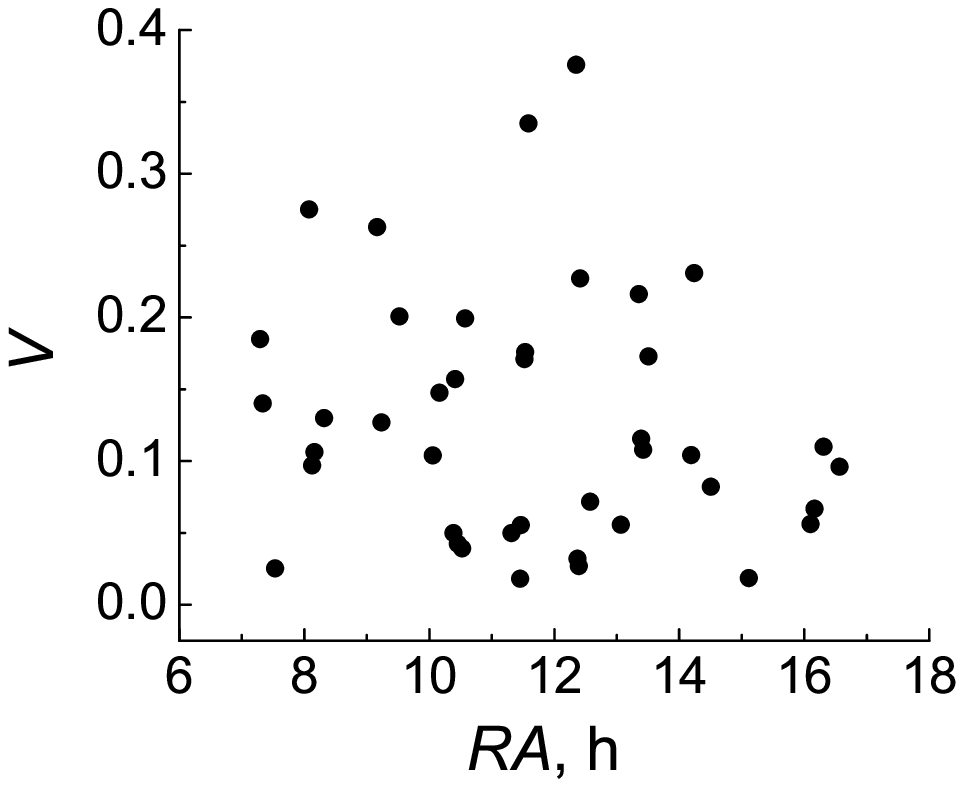}
} }  \setcaptionmargin{5mm} \captionstyle{normal}
 \caption{
The long-term variability indices $V$ for objects from  Table~1 as a function of angle $dH$
(the upper panel) and  $\rm RA$ (the lower panel). }
\label{fig5:Majorova_n}
\end{figure}

Let us now consider the distribution of right ascensions,
declinations (offsets relative to the central section of the
survey), flux densities, and spectral indices of the objects with
positive long-term variability indices.

The upper and lower panels in Fig.~\ref{fig5:Majorova_n} show the
parameter $V$ plotted as a function of angle  $dH$ and RA
respectively. Although the sources are distributed sufficiently
uniformly in the interval of angles considered, there is a trend
for the increase of the number of sources toward the central cross
section of the survey which can be explained by the fact that
fainter objects can be detected in the vicinity of the central
cross section. Variable sources are distributed uniformly in right
ascension.

\begin{figure*}[]
\onelinecaptionsfalse \centerline{ \vbox{\vspace{0mm} \hbox{
\includegraphics[angle=0,width=0.30\textwidth,clip]{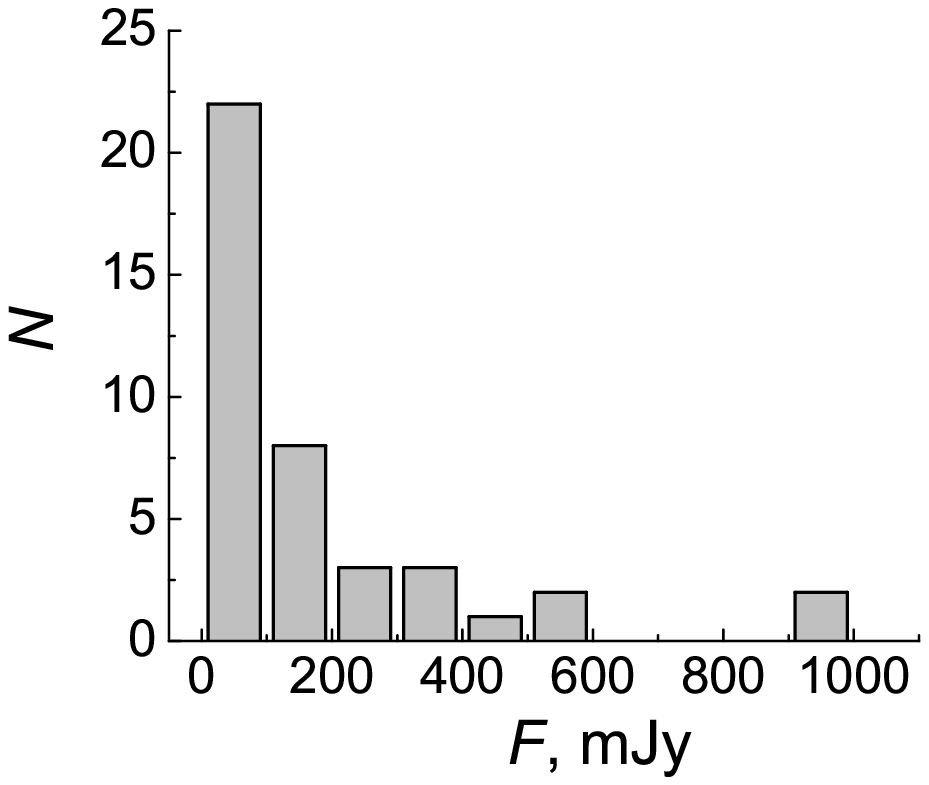}
\includegraphics[angle=0,width=0.31\textwidth,clip]{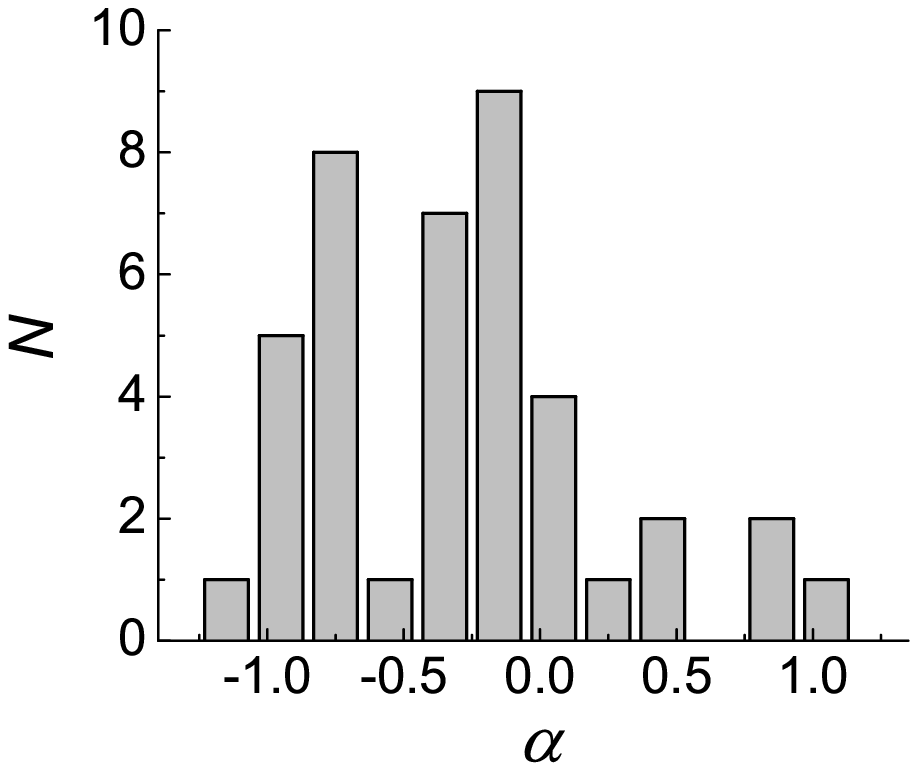}
\includegraphics[angle=0,width=0.30\textwidth,clip]{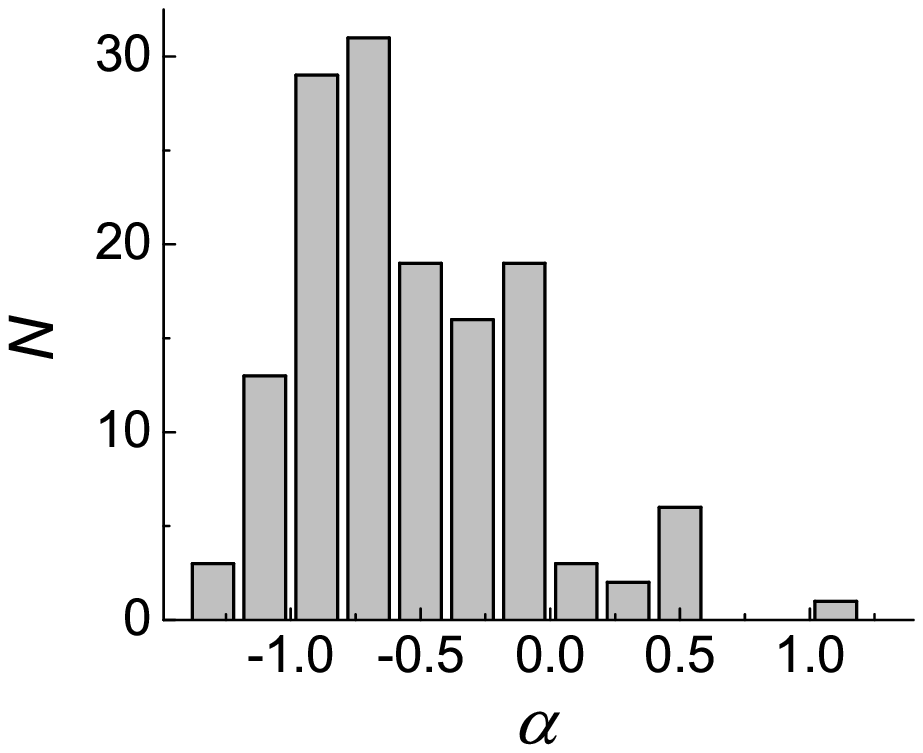}
} } } \setcaptionmargin{5mm} \captionstyle{normal}
 \caption{ Histograms of the flux densities $F$ (the left panel) and
spectral indices $\alpha$ (the central panel) of possibly variable
objects (with $V>0$) from Table~1. The right panel shows the
histogram of the spectral indices of nonvariable sources (with
$V<0$). } \label{fig6:Majorova_n}
\end{figure*}

Figure~\ref{fig6:Majorova_n} shows the histograms of the distribution of flux densities $F$ (the left panel)
and spectral indices  $\alpha$ (the central panel) for the sample of possibly variable objects. The right panel shows the histogram of the distribution
of spectral indices for nonvariable sources (with $V<0$). Almost half of the sources with
positive long-term variability indices are bright objects with flux densities above 100~mJy,
about one third of the sources have flux densities  $F \le 40$~mJy. The median spectral index
of this sample is equal to $-0.3$. The distribution exhibits two conspicuous maxima: at
$\alpha=-0.75$ and $\alpha=-0.15$.

The median spectral index of nonvariable sources is equal to $-0.65$.
This subsample is dominated by objects with negative spectral indices.

\begin{figure}[]
\onelinecaptionsfalse \centerline{ \vbox{\vspace{2mm}
\includegraphics[angle=0,width=0.31\textwidth,bb=0 30 314 272,clip]{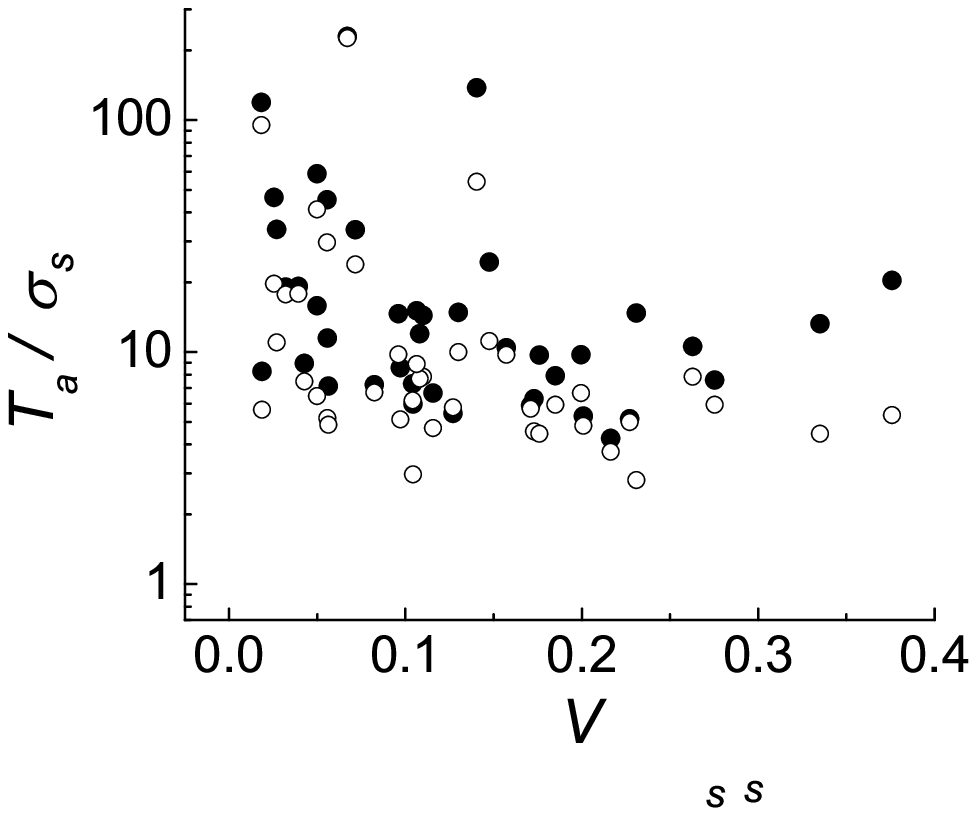}
\includegraphics[angle=0,width=0.31\textwidth,bb=0 30 314 272,clip]{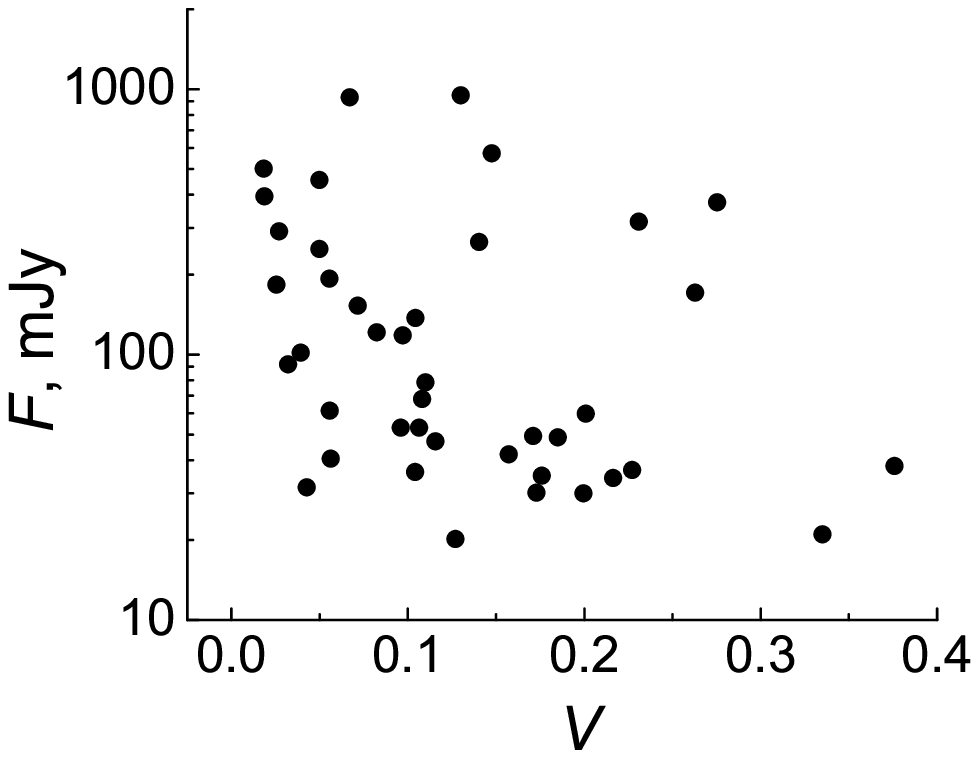}
} } \setcaptionmargin{5mm} \captionstyle{normal}
 \caption{ Dependence of the long-term variability indices  $V$ of
the sources from Table~1 on the signal-to-noise ratio
($T_{a}/\sigma_{s}$) (the upper panel) and flux density~$F$ (the
lower panel). } \label{fig7:Majorova_n}
\end{figure}

The upper and lower panels of Fig.~\ref{fig7:Majorova_n} show the
dependence of the long-term variability index on the
signal-to-noise ratio $T_{a}/\sigma_{s}$ and source flux
density~$F$ respectively. We computed the coefficients  $V$ using
the antenna temperatures determined in two surveys, and therefore
each $V$ value has two corresponding  $T_{a}/\sigma_{s}$ ratios,
which are shown in Fig.~\ref{fig7:Majorova_n} by the filled and
open circles respectively. Noteworthy is the fact that the sources
with long-term variability indices in the  $ 0 < V < 0.18$
interval have signal-to-noise ratios  $T_{a}/\sigma_{s}$ ranging
from 3 to 200, whereas the $T_{a}/\sigma_{s}$ ratios for objects
with $V > 0.18$ do not exceed 20. Note that only three sources
have antenna temperatures at 3$\sigma_{s}$ in one of the surveys,
whereas the remaining sources have $T_{a} \ge 5\sigma_{s}$.

\section{VARIABLE SOURCES OF THE  RCR CATALOG}

Let us now discuss the criteria that allow the objects with $V>0$,
listed in Tables~1 and 2, to be considered variable. Kesteven,
Bridle, and Brandie~\cite{kest:Majorova_n}, and Fanti et
al.~\cite{fan:Majorova_n} considered a source to be possibly and
reliably variable if its  $\chi^2$ probability satisfied the
condition $0.1\% \le 1-p \le 1\%$ ($0.99 \le  p < 0.999$) and $
1-p \le 0.1\%$ ($ p \ge 0.999$) respectively.

Seielstad, Pearson, and Readhead~~\cite{sei:Majorova_n}, and
Gorshkov and Konnikova~\cite{g1:Majorova_n} considered the objects
with $p \ge 0.985$ and \mbox {$p \ge 0.98$} to be reliably
variable, and those with \mbox {$0.95 \le p < 0.98$} possibly
variable.

Three sources (J\,073357+045614, J\,115851+\linebreak045541,
J\,123507+045318) out of 41 objects listed in Tables~1 and 2
satisfy the condition $ p \ge 0.999$.\!\footnote{The source
J\,073357+045614 satisfies this condition if the flux density
measured during the 1993 set is taken into account. Without the
1993 set the $\chi^2$ probability for this object is $p=0.966$.}
Twelve and six sources have $\chi^2$ probabilities in the  \mbox
{$ 0.98  <  p < 0.999$} and  \mbox {$ 0.95 <  p < 0.98$} intervals
respectively. So, 15~sources can be considered to be reliably
variable in accordance with the criteria proposed
in~\mbox{\cite{kest:Majorova_n,fan:Majorova_n,sei:Majorova_n,g1:Majorova_n}},
and six sources can be considered to be possibly variable. Another
six sources have \mbox {$ 0.90 \le p < 0.95$}.

These results are mostly based on the $p_{df-1}$ values computed
using the averaged flux densities measured in the  1993 and 1994
surveys. For the objects with the 1993 and 1994 survey flux
density differences exceeding the measurement errors of $F$, we
use the $p_{df}$ probabilities. Thus, the source variability found
in this study has a typical time scale of  6--8 years.

If we base our analysis on the average probabilities
$\overline{p}$ listed in Table~2, the same three sources satisfy
the condition  $ p \ge 0.999$, seven sources satisfy the condition
$ 0.98  <  p < 0.999$, and 11 sources satisfy the condition $
0.947  \le  p < 0.98$. That is, the total number of reliably
variable and possibly variable objects (21~sources) remains the
same, and only the proportion changes. All these radio sources,
except one, have long-term variability indices $V \gtrsim 0.1$.

Ting et al.~\cite{VF:Majorova_n} used the coefficient $V_{F}$ as
the variability criterion. They  considered a source to be variable if its
parameter $V_{F}$ was greater than or equal to 3. For such objects
the difference between the flux densities measured in different
surveys exceeds 3$\sigma$ (formula~(\ref{3:Majorova_n})). Out of
the 21 most likely variable sources \mbox{($ 0.95  <  p \le
0.999$)}, four meet this criterion, two more have $V_{F} \ge 2.9$,
and all the remaining sources, except one, have $2.1 <  V_{F} <
2.9$.

Consider now the use of the relative variability amplitude
$V_{\chi}$ as a criterion. An analysis of the data reported by
Seielstad, Pearson, and Readhead~\cite{sei:Majorova_n} showed that
for most of the sources with $p \ge 0.985$ this parameter is no
less than 0.2. Among the 21~sources claiming to be variable with
the probability $p > 0.95$, 18 objects have $V_{\chi} > 0.3$, and
three have relative variability amplitudes in the interval
\mbox{$0.2 < V_{\chi} < 0.3$}. The latter three sources pass
rather close to the central cross section of the survey, and their
long-term variability indices are equal to  0.067, 0.096, and
0.11.

And now a few words about the parameter $V_{R}$, which characterizes the maximum to minimum flux density ratio
for a particular source measured in different surveys. All the most likely variable objects
have  $V_{R} \gtrsim 1.5$.

\begin{figure}[]
\onelinecaptionsfalse \centerline{ \vbox{ \vspace{-2mm}
\includegraphics[angle=0,width=0.31\textwidth,bb=0 30 314 272,clip]{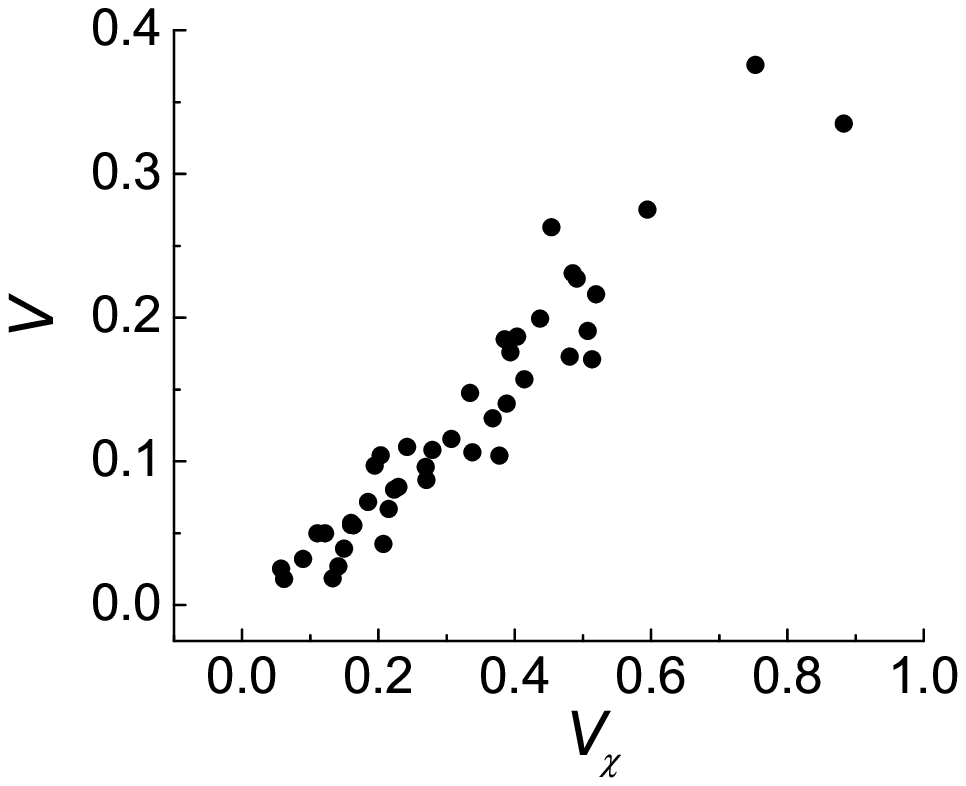}
\includegraphics[angle=0,width=0.31\textwidth,bb=0 30 314 272,clip]{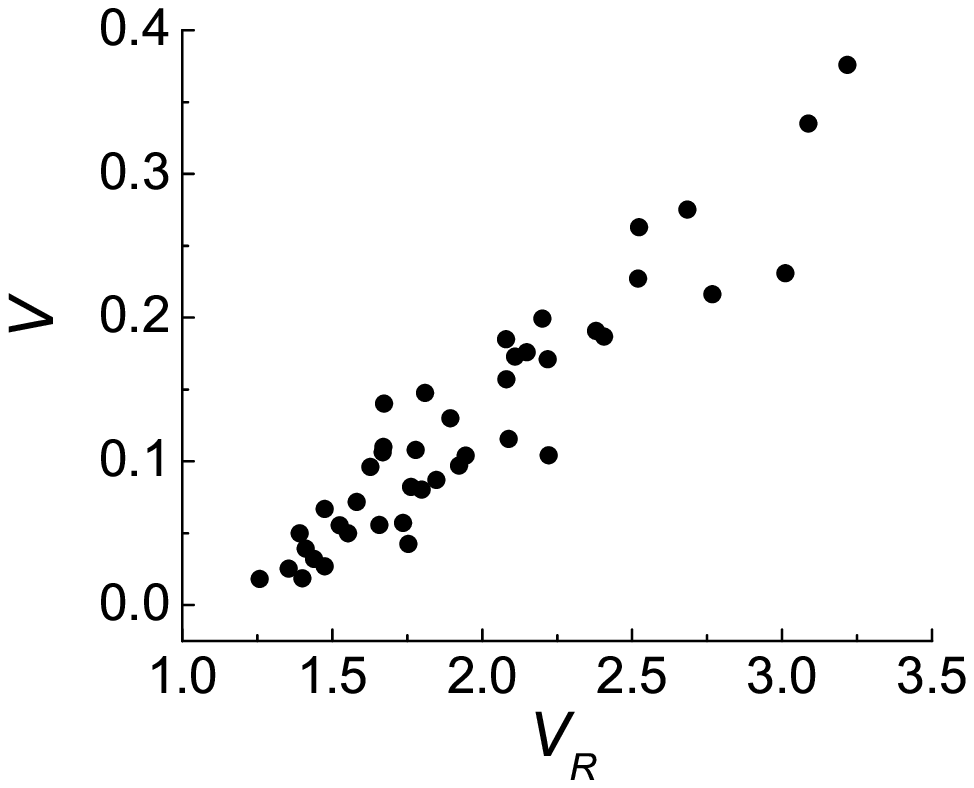}
} }  \setcaptionmargin{5mm} \captionstyle{normal}
 \caption{
Dependence of the long-term variability index  $V$ of the sources
from Table~1 on variability amplitude $V_{\chi}$ (the upper panel)
and parameter $V_{R}$ (the lower panel). } \label{fig8:Majorova_n}
\end{figure}

Note that the variability criteria  $V$, $V_{\chi}$, $V_{F}$, and
$V_{R}$, considered here, are interrelated.
Figure~\ref{fig8:Majorova_n} shows, as an example, the dependences
of the long-term variability indices  $V$ of the sources from
Table~1 on variability amplitude  $V_{\chi}$ (the upper panel) and
parameter $V_{R}$ (the lower panel).

An analysis of the obtained results suggests that all the sources
listed in Tables~1 and 2 (a total of 41~objects) can be suspected
of variability, because the differences of their flux densities
measured in different surveys exceed the total flux measurement
errors. However, the reliability of this conclusion (i.e., that
the object is variable) is different for different sources. Among
the 41 sources, 15 are reliably variable with the probabilities $p
> 0.98$, and three are variable with a probability of $ p \ge
0.999$. Six sources with probabilities in the  \mbox {$ 0.95  <  p
< 0.98$} interval are possibly variable in accordance with the
criteria proposed in~\mbox{\cite{sei:Majorova_n,g1:Majorova_n}}.
The variability probabilities for the remaining 20~objects from
Table~1 lie in the \mbox {$0.73 \le p < 0.95$} interval.

Among the  21 sources with $p > 0.95$, one third have flux
densities above 150~mJy, seven have flux densities in the  \mbox
{40 < $ \overline{F}$ < 100~mJy} interval, and seven objects have
flux densities in the \mbox {20 < $ \overline{F}$ < 40~mJy}
interval. Most of these sources are objects with nearly flat
spectra, and two have inverted spectra. One third of the sources
have spectral indices  $\alpha < -0.67$.

Figure~\ref{fig9:Majorova_n} shows the light curves (the left
panel) and spectra (the right panel) for the sources with $V>0$.
Figure~\ref{fig10:Majorova_n} shows, as an example, the light
curves and spectra of the variable flat-spectrum radio sources
J\,155035+052710 and J\,165833+051516, which have long-term
variability indices  $V<0$, according to the data of our surveys.
The parameter $V$ for  J\,165833+051516 becomes positive if we set
the relative standard error of flux density measurement equal to
the averaged value  $\overline{{\rm RMS}^{\rm set}}=0.10$ obtained
from the sample of objects with $V \le 0$.

\begin{figure*}
\setcaptionmargin{5mm} \onelinecaptionstrue \centerline{ \vbox{
\hbox{
\includegraphics[angle=0,width=0.31\textwidth,clip]{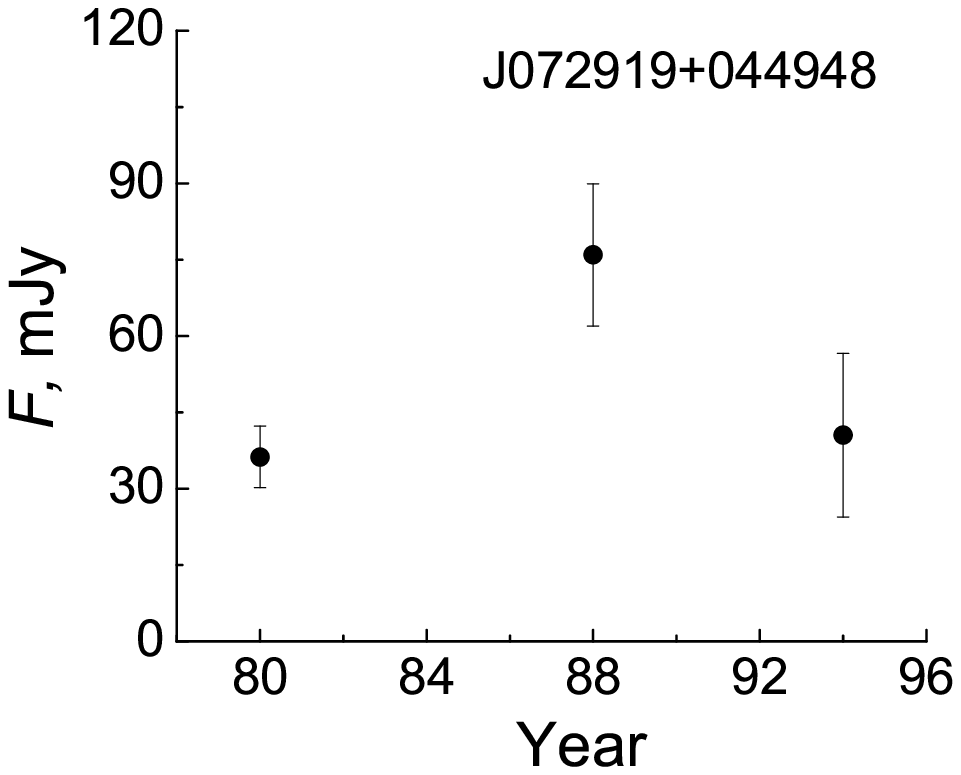}\hspace{20mm}
\includegraphics[angle=0,width=0.31\textwidth,clip]{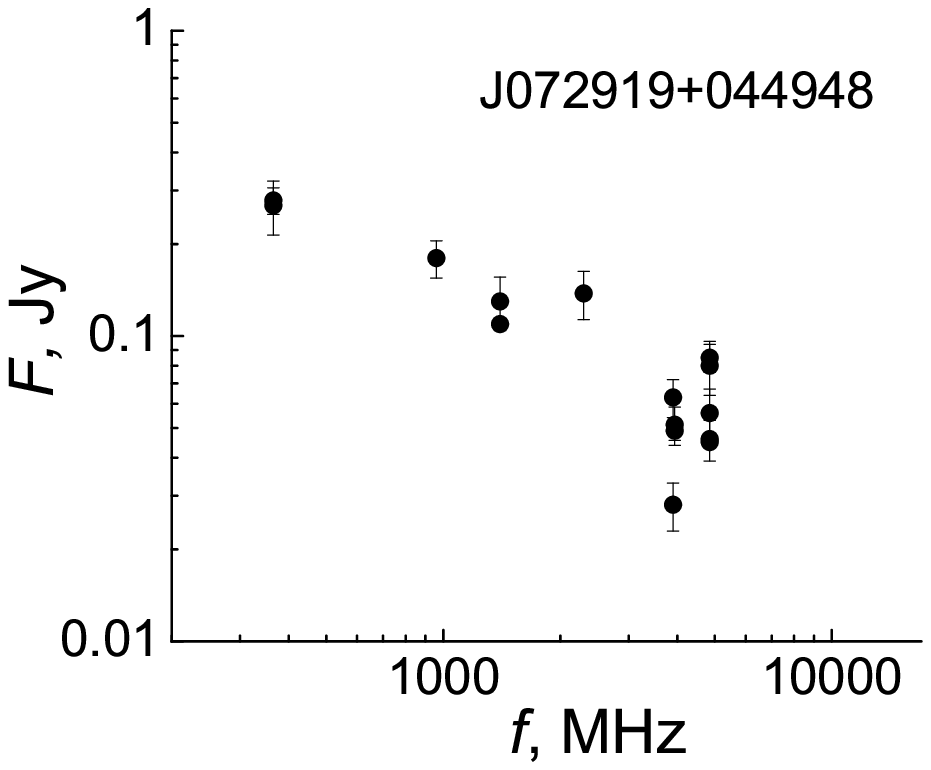}
}  \hbox{
\includegraphics[angle=0,width=0.31\textwidth,clip]{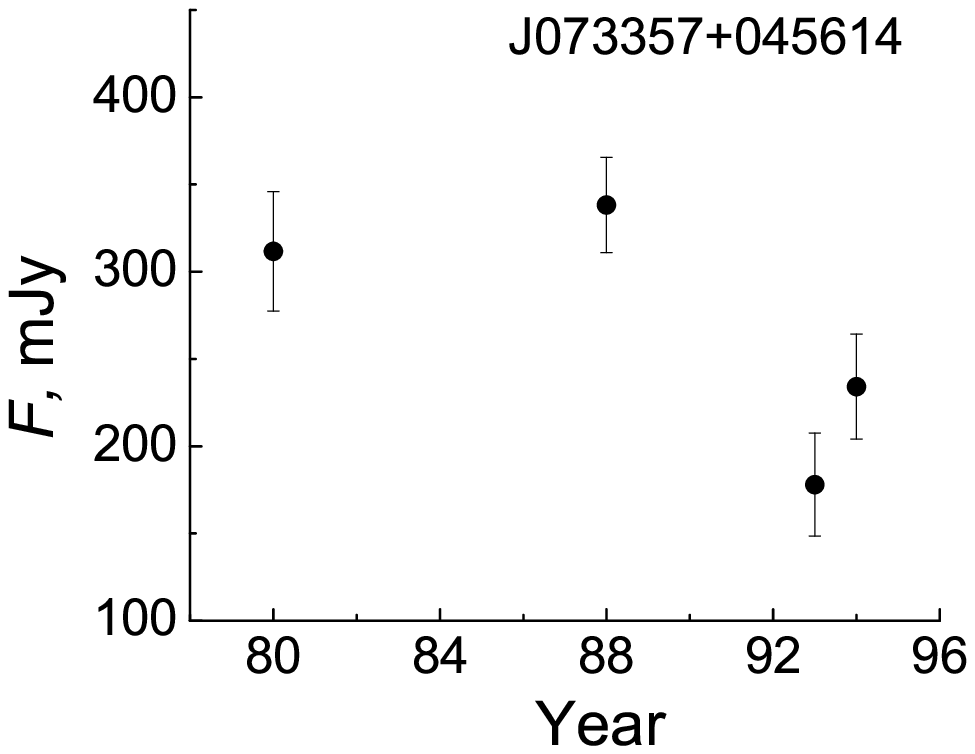}\hspace{21mm}
\includegraphics[angle=0,width=0.31\textwidth,clip]{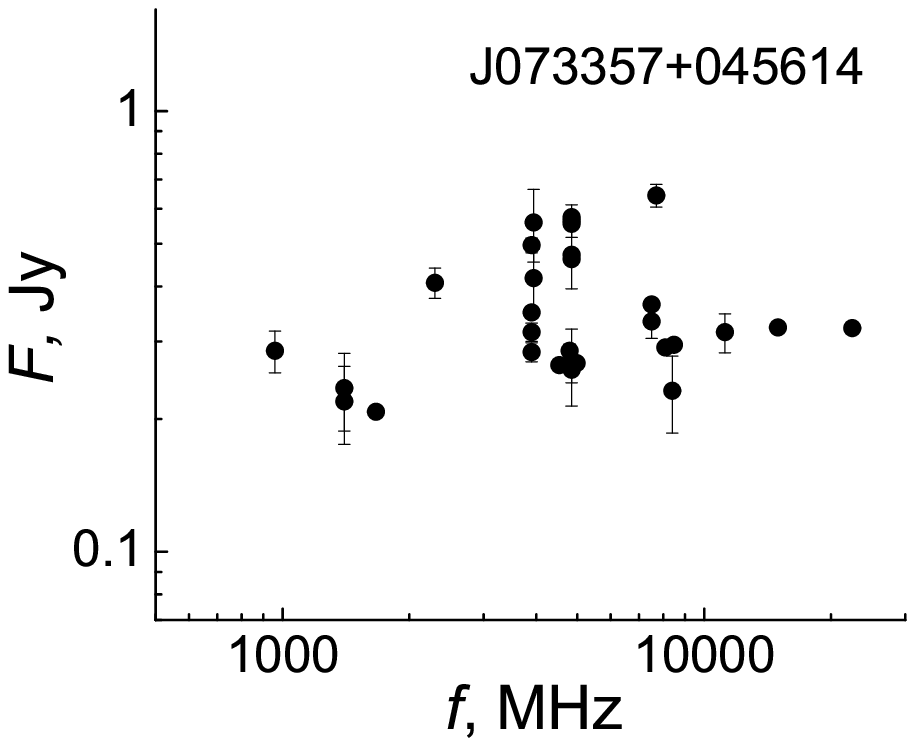}
}  \hbox{
\includegraphics[angle=0,width=0.31\textwidth,clip]{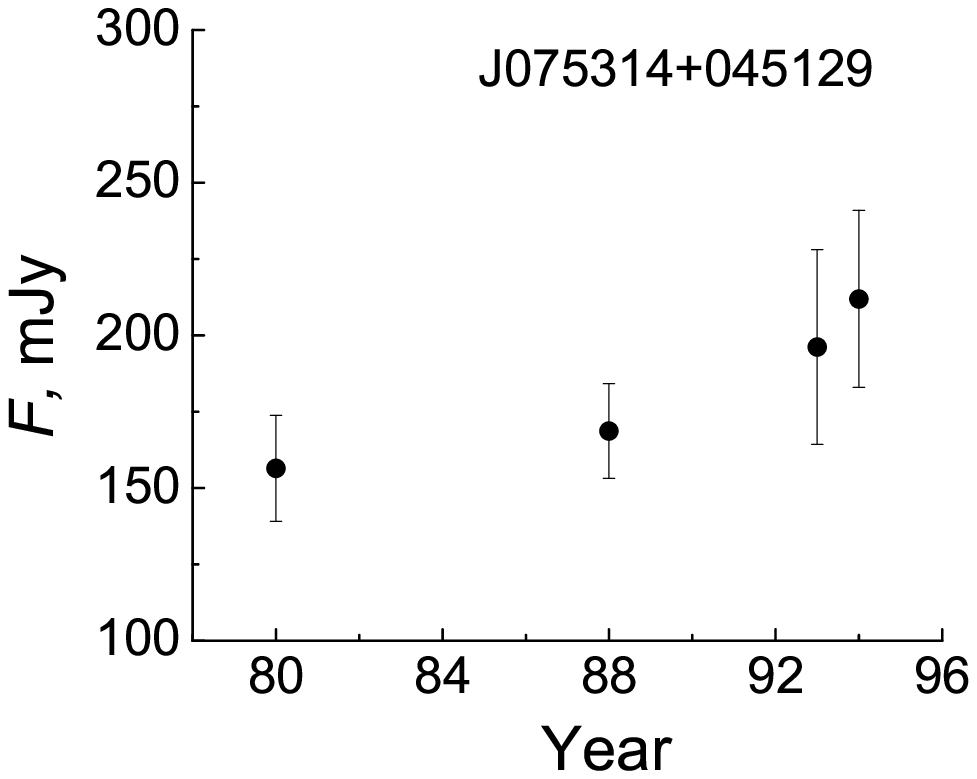}\hspace{20mm}
\includegraphics[angle=0,width=0.31\textwidth,clip]{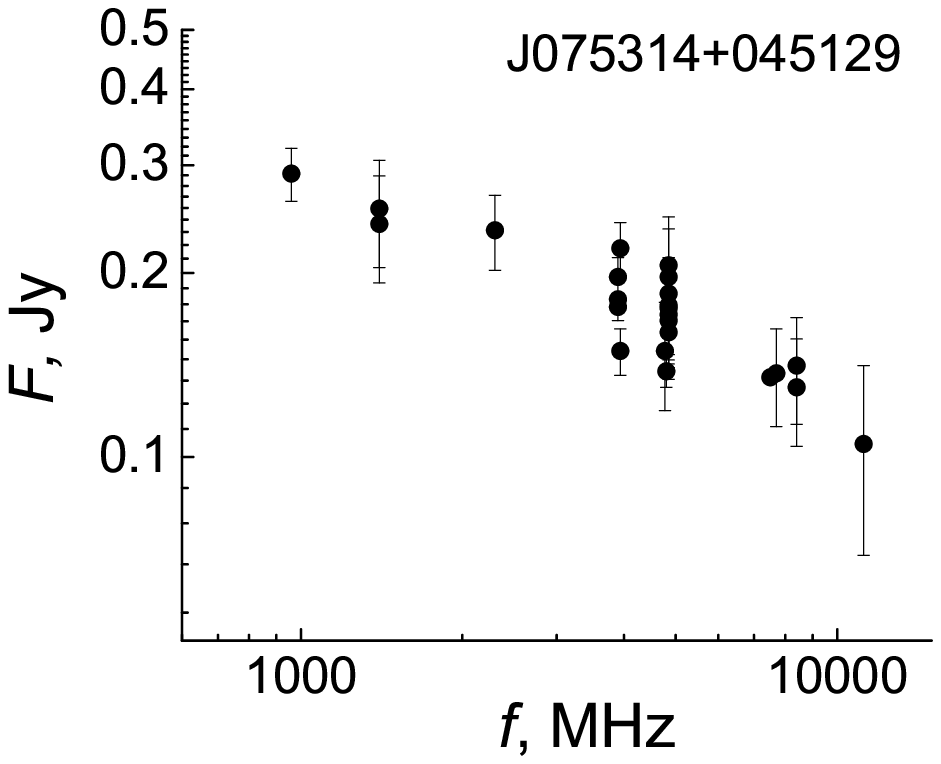}
}  \hbox{
\includegraphics[angle=0,width=0.31\textwidth,clip]{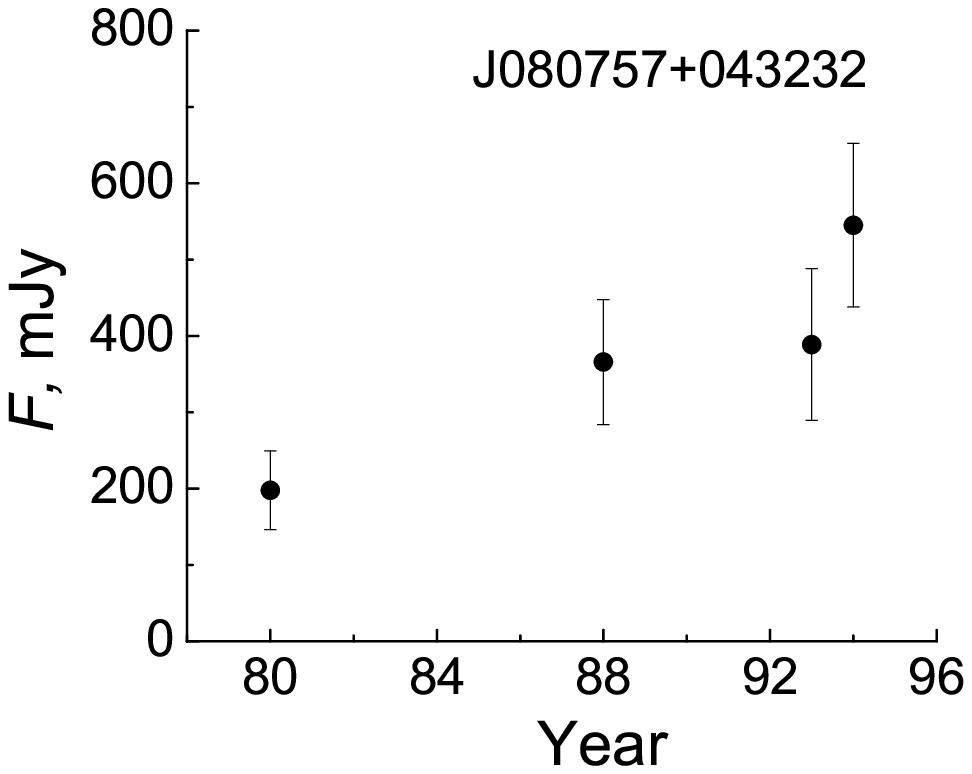}\hspace{21.5mm}
\includegraphics[angle=0,width=0.31\textwidth,clip]{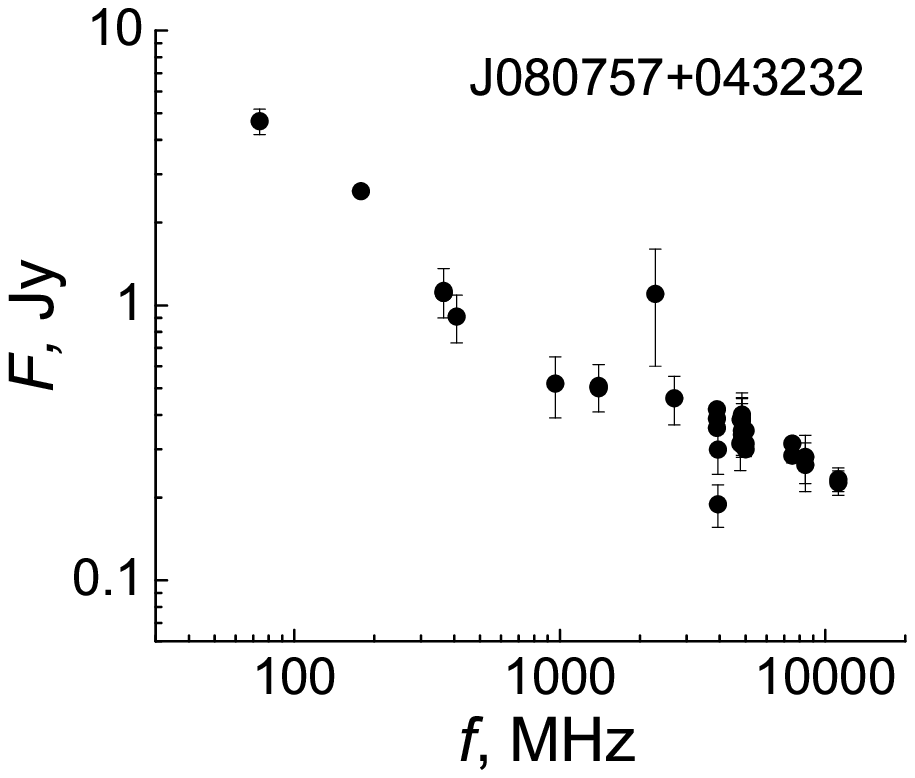}
}  \hbox{
\includegraphics[angle=0,width=0.31\textwidth,clip]{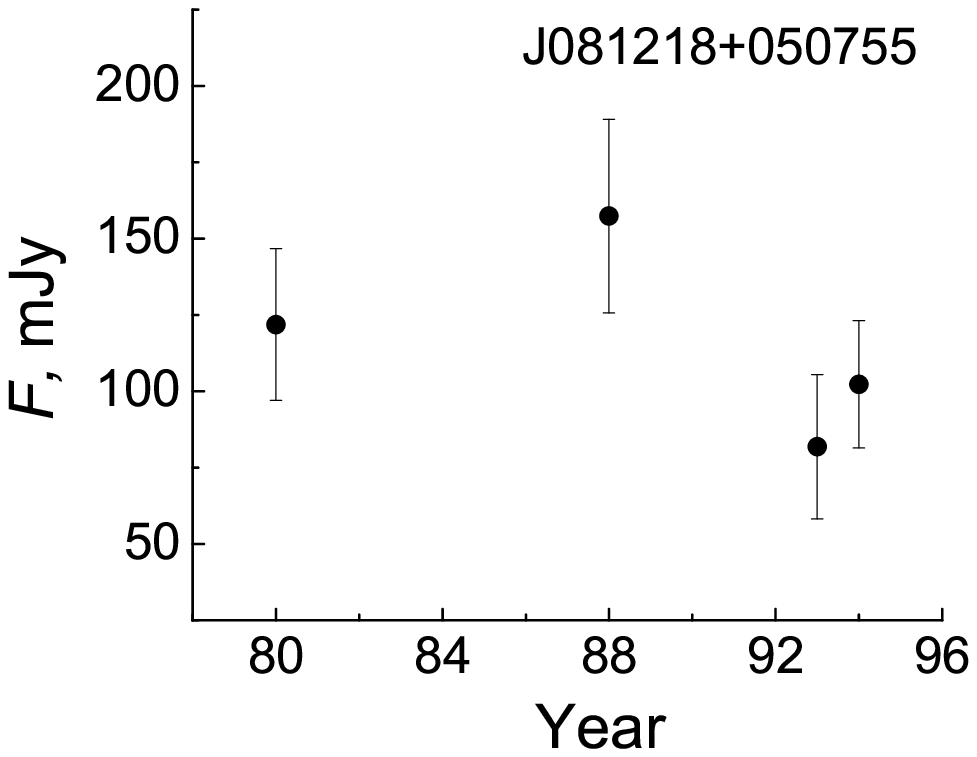}\hspace{21mm}
\includegraphics[angle=0,width=0.31\textwidth,clip]{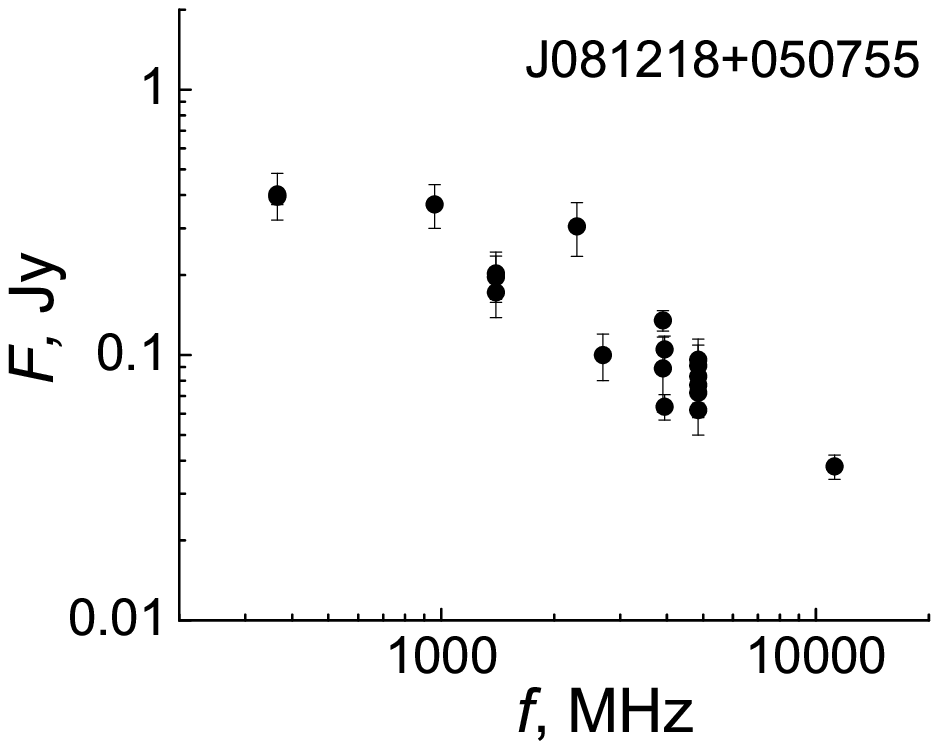}}}}
 \caption{ Light curves (the left panel) and spectra (the right
panel) of radio sources with $V>0$. } \label{fig9:Majorova_n}
\end{figure*}

\addtocounter{figure}{-1}
\begin{figure*}
\setcaptionmargin{5mm} \onelinecaptionstrue \centerline{ \vbox{
\hbox{
\includegraphics[angle=0,width=0.31\textwidth,clip]{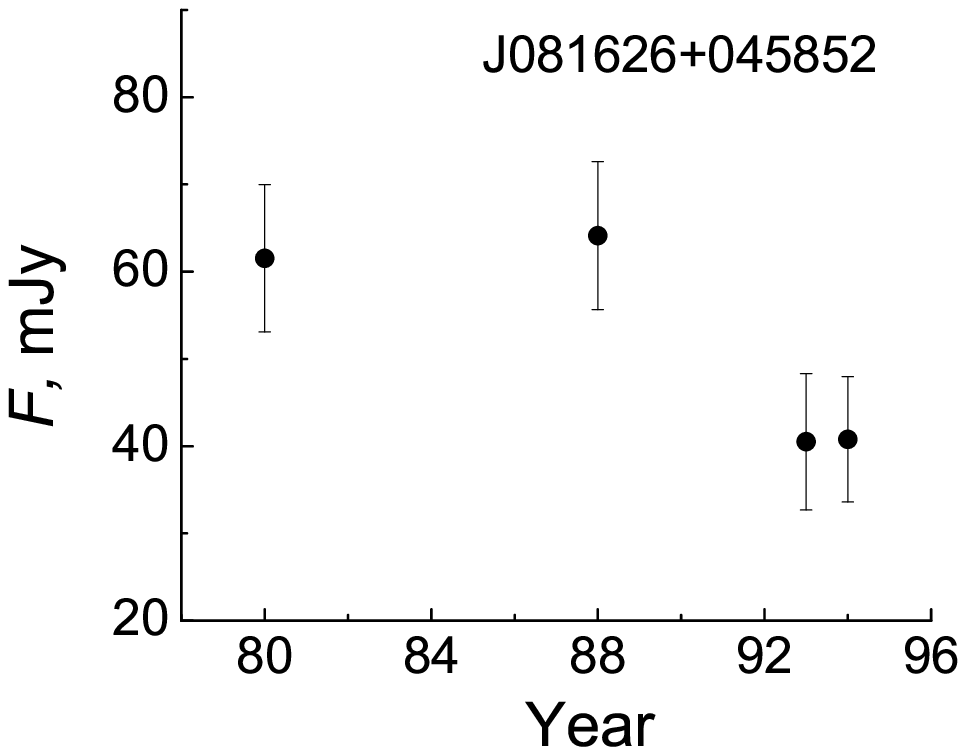}\hspace{20mm}
\includegraphics[angle=0,width=0.31\textwidth,clip]{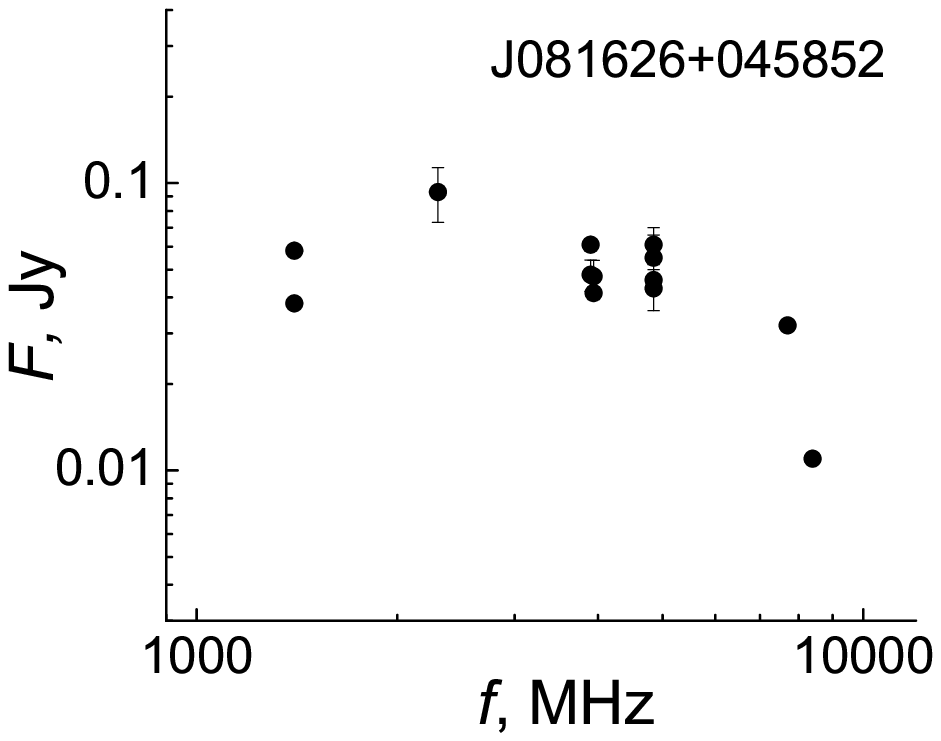}
}  \hbox{
\includegraphics[angle=0,width=0.31\textwidth,clip]{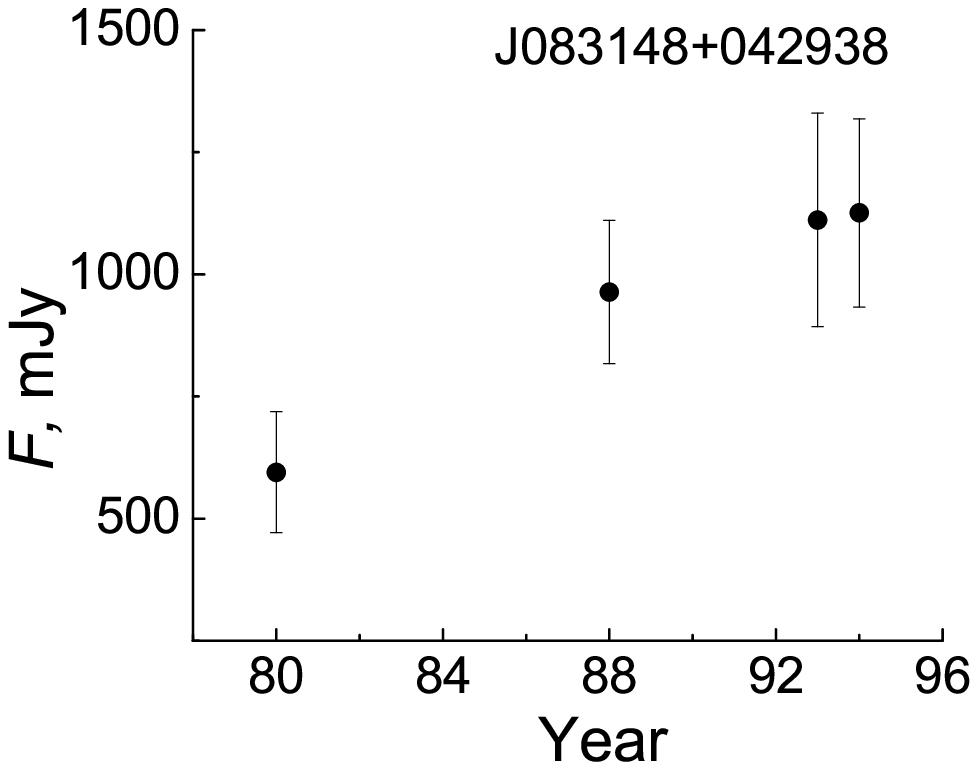}\hspace{20mm}
\includegraphics[angle=0,width=0.31\textwidth,clip]{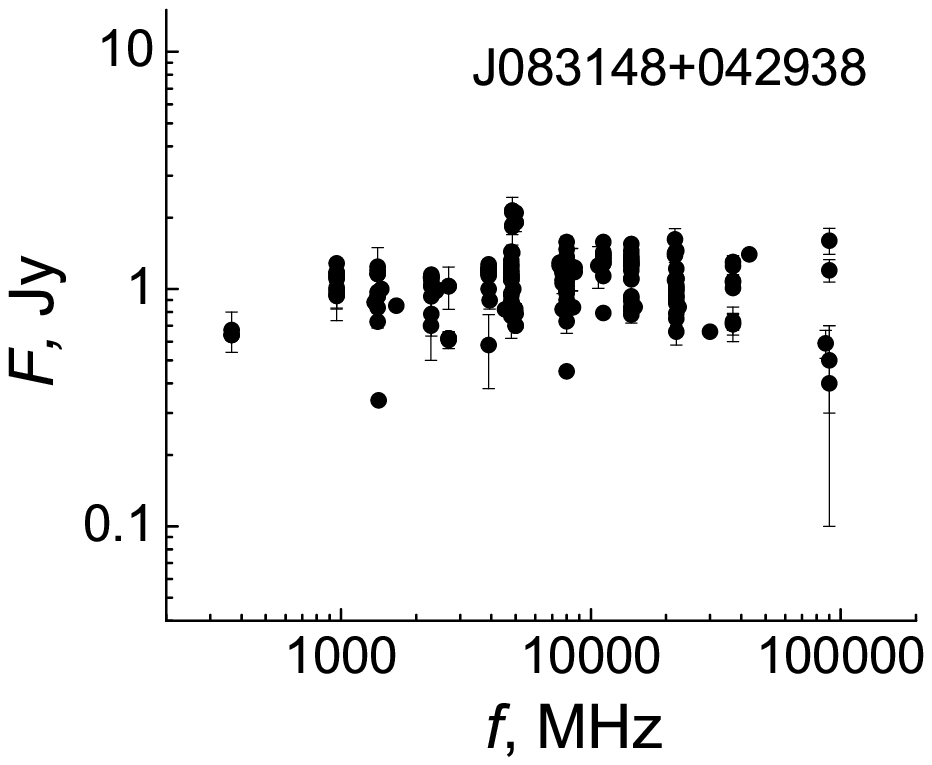}
}  \hbox{
\includegraphics[angle=0,width=0.31\textwidth,clip]{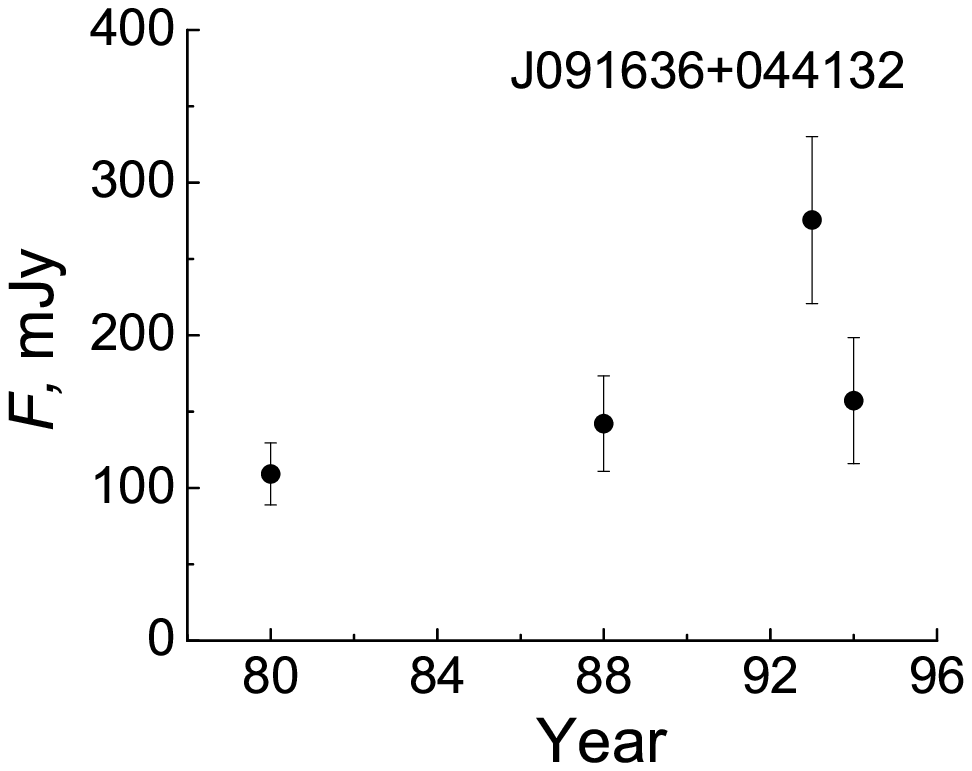}\hspace{20mm}
\includegraphics[angle=0,width=0.31\textwidth,clip]{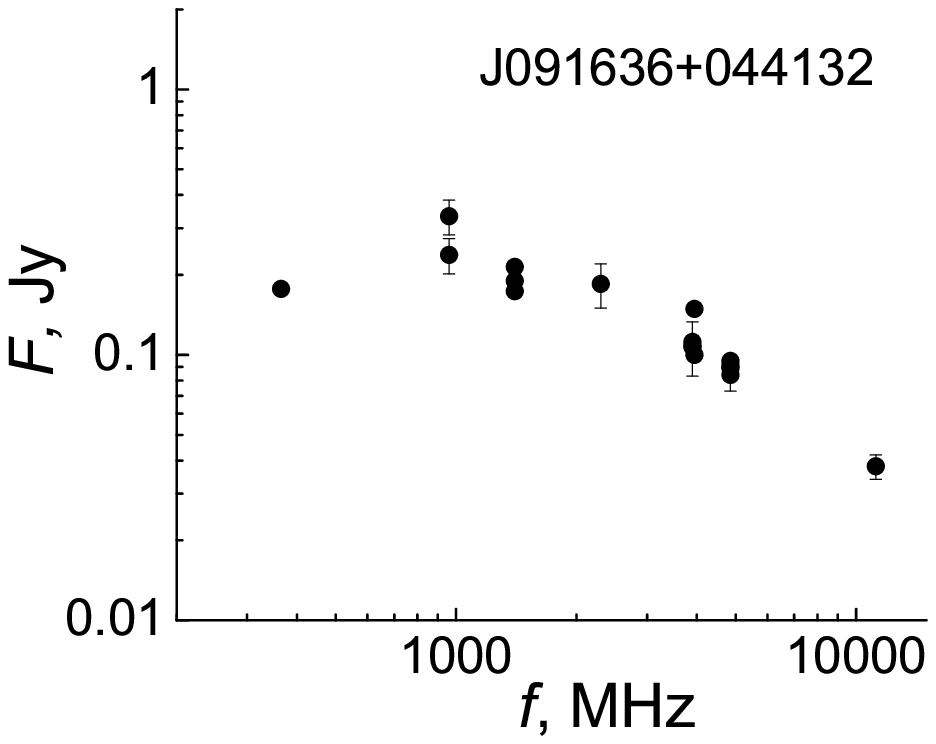}
}  \hbox{
\includegraphics[angle=0,width=0.31\textwidth,clip]{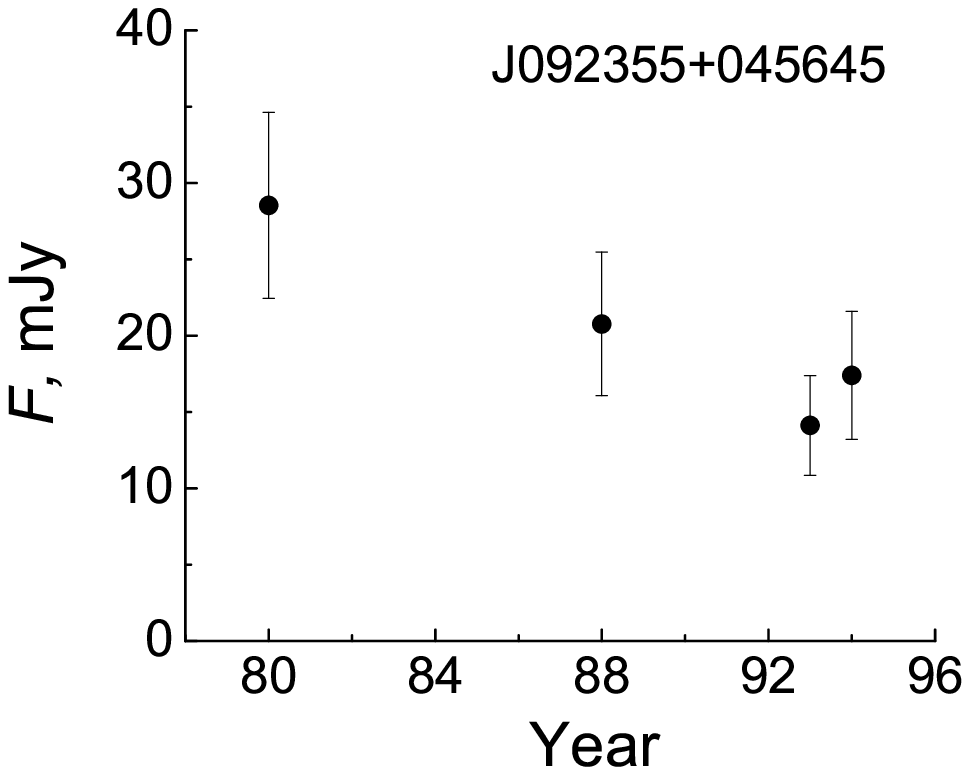}\hspace{20mm}
\includegraphics[angle=0,width=0.31\textwidth,clip]{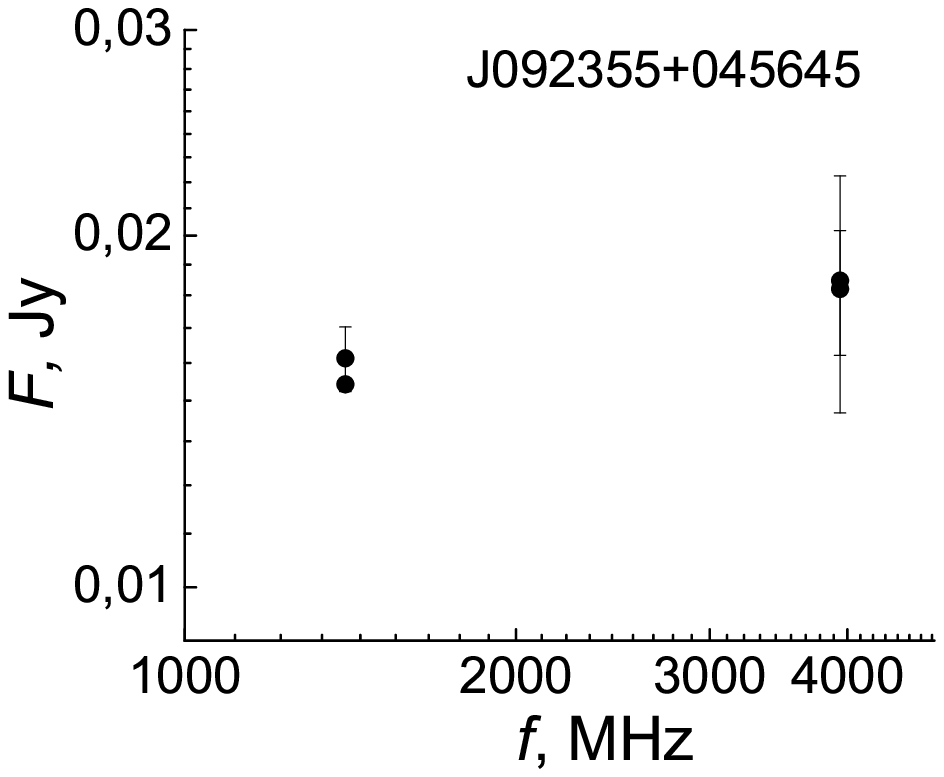}
} \hbox{
\includegraphics[angle=0,width=0.31\textwidth,clip]{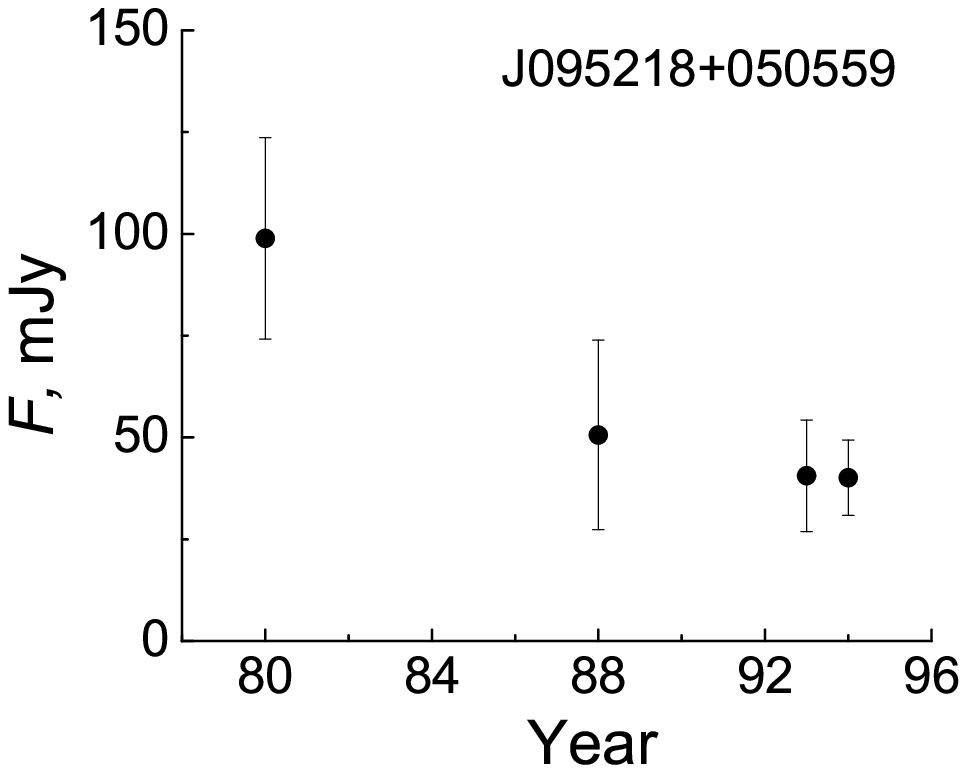}\hspace{21mm}
\includegraphics[angle=0,width=0.31\textwidth,clip]{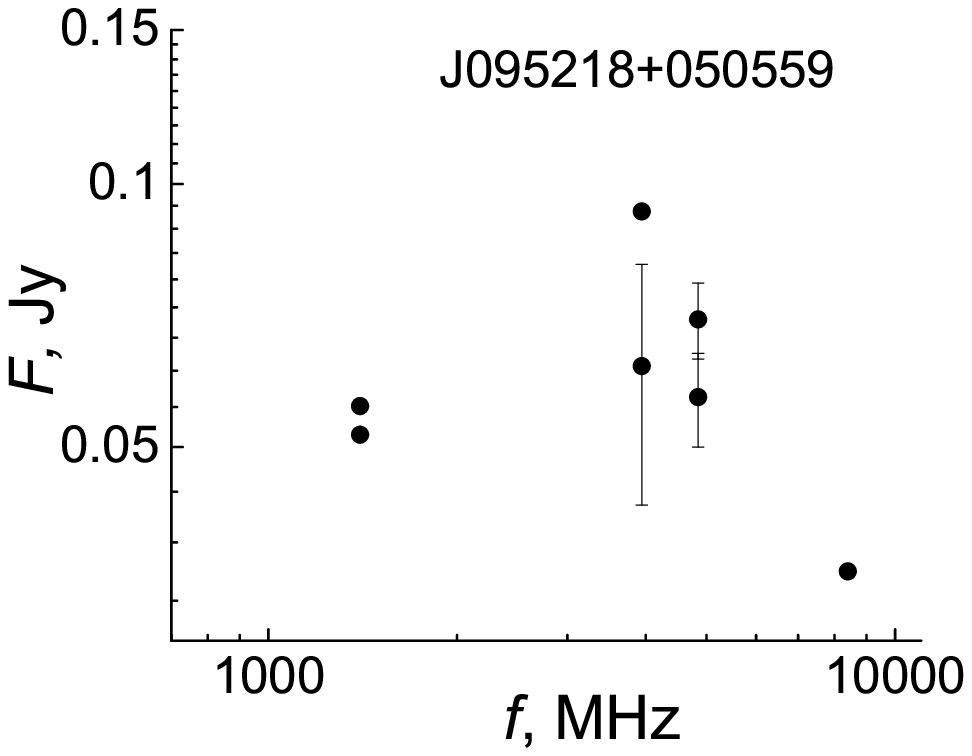}}}}
\caption{(Contd.)}
\end{figure*}

\addtocounter{figure}{-1}
\begin{figure*}
\setcaptionmargin{5mm} \onelinecaptionstrue \centerline{ \vbox{
\hbox{
\includegraphics[angle=0,width=0.31\textwidth,clip]{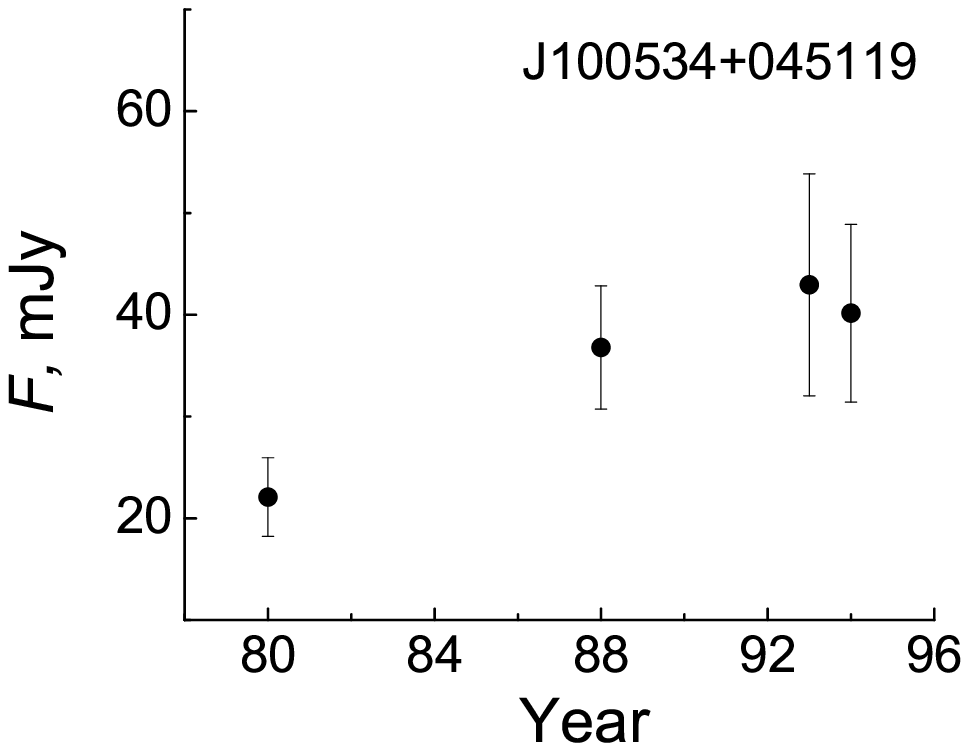}\hspace{20mm}
\includegraphics[angle=0,width=0.31\textwidth,clip]{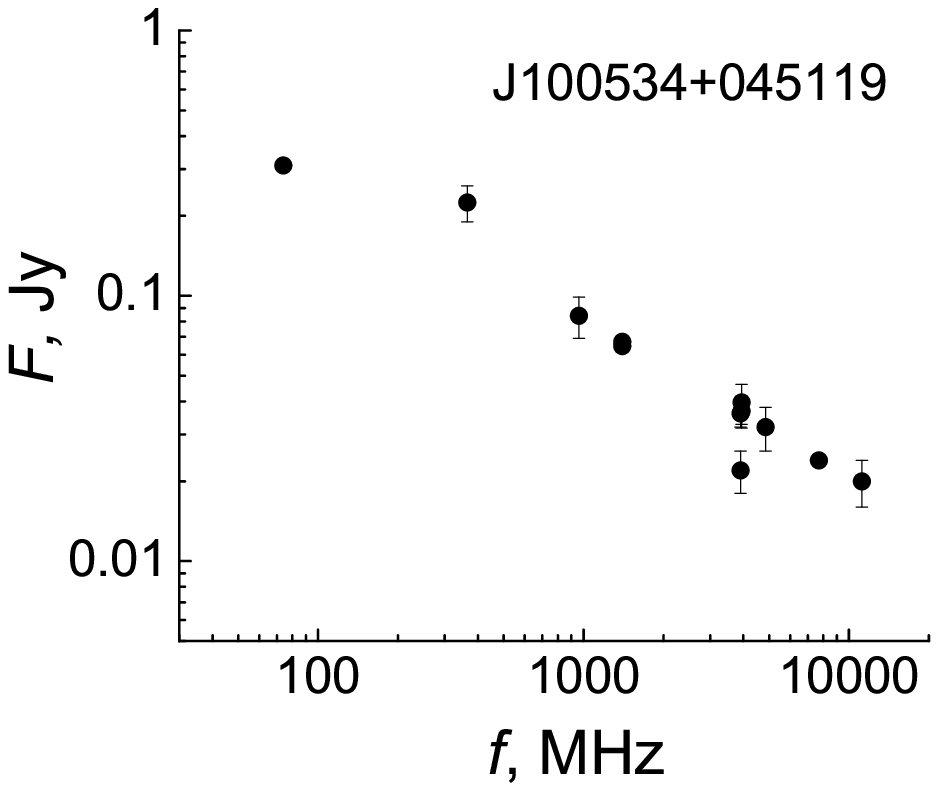}
} \hbox{
\includegraphics[angle=0,width=0.31\textwidth,clip]{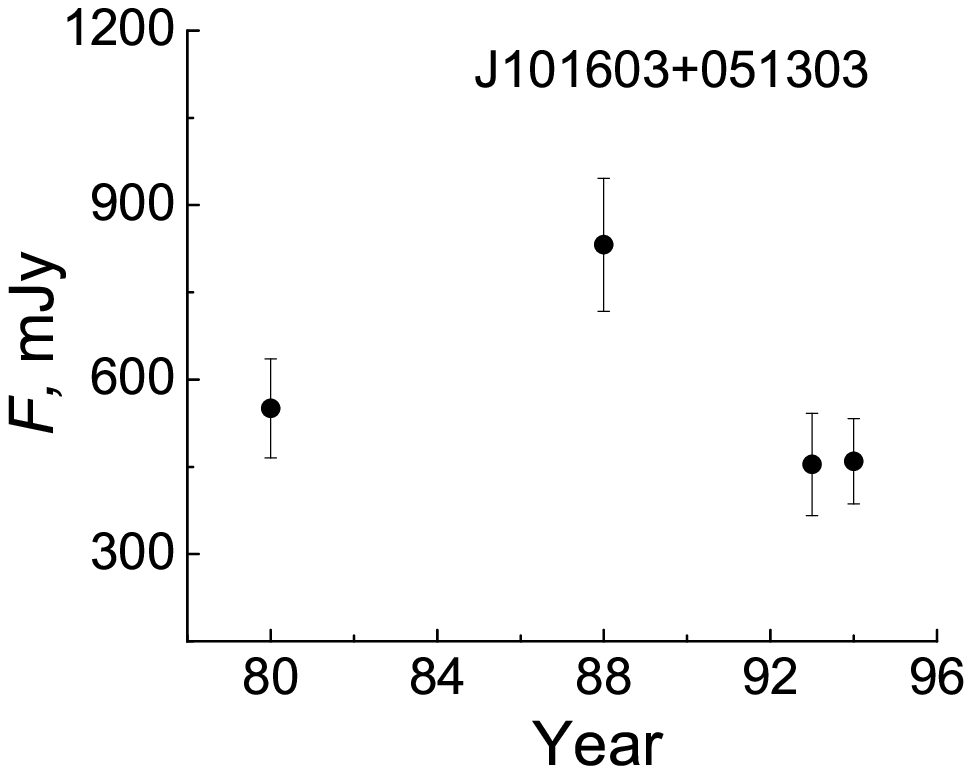}\hspace{21.5mm}
\includegraphics[angle=0,width=0.31\textwidth,clip]{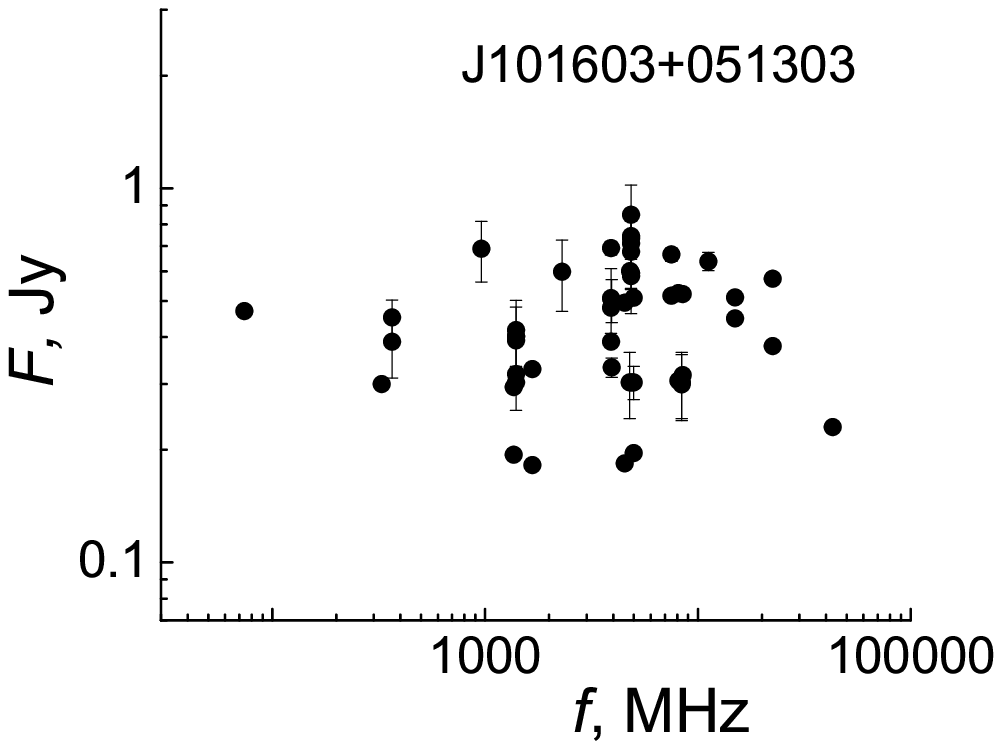}
} \hbox{
\includegraphics[angle=0,width=0.31\textwidth,clip]{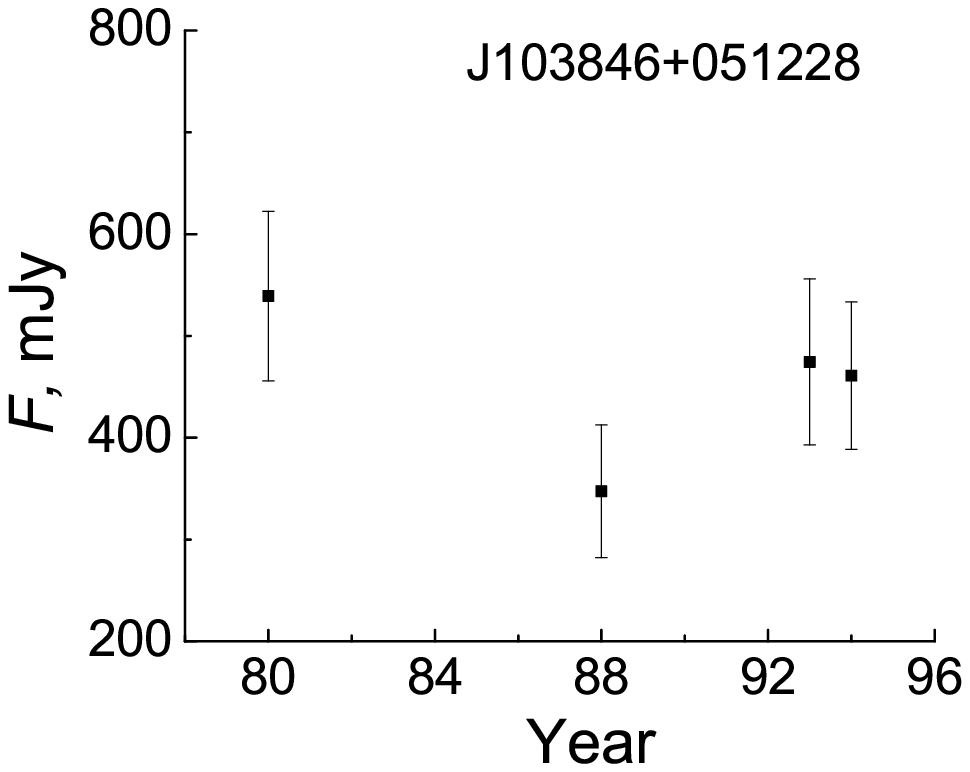}\hspace{21mm}
\includegraphics[angle=0,width=0.31\textwidth,clip]{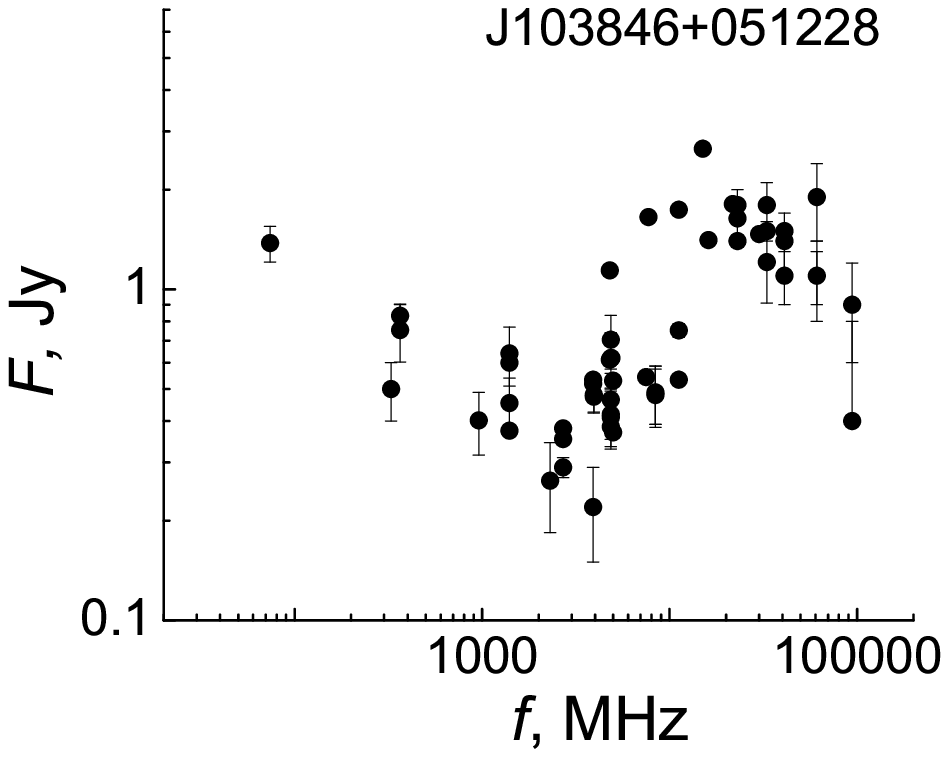}
} \hbox{
\includegraphics[angle=0,width=0.31\textwidth,clip]{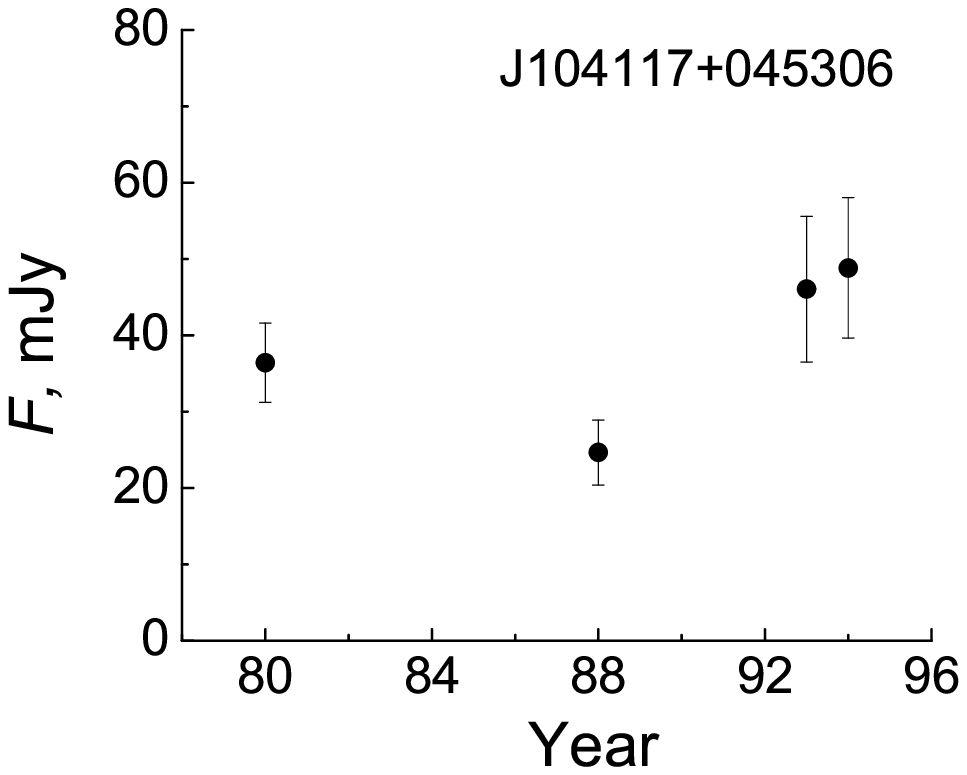}\hspace{21mm}
\includegraphics[angle=0,width=0.31\textwidth,clip]{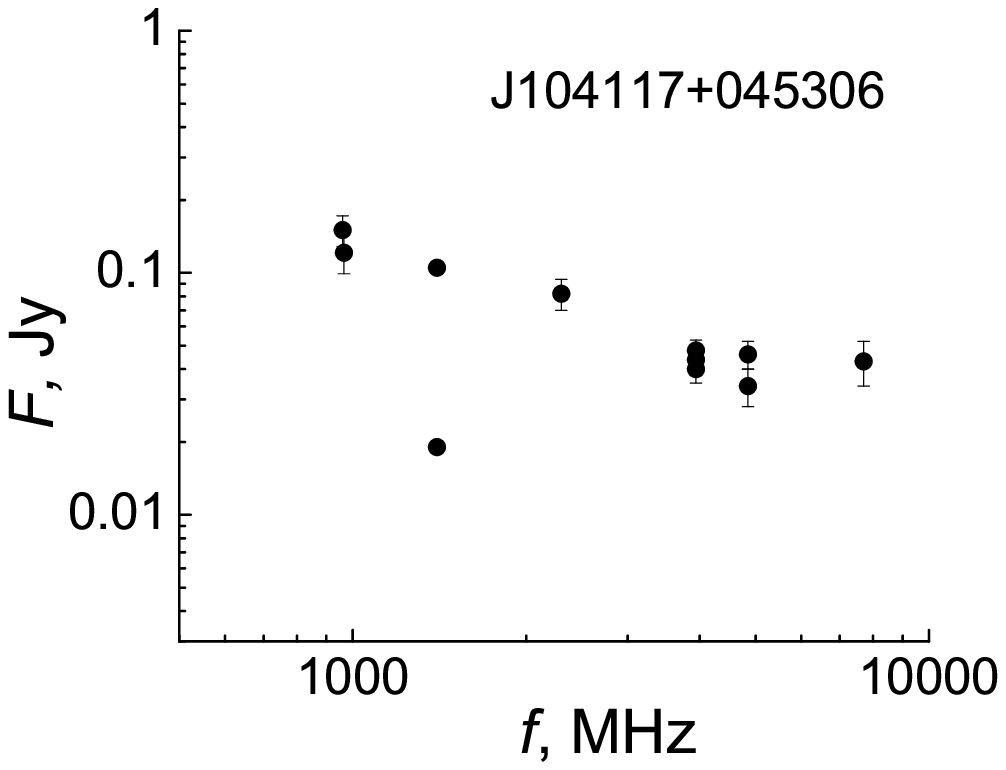}
} \hbox{
\includegraphics[angle=0,width=0.31\textwidth,clip]{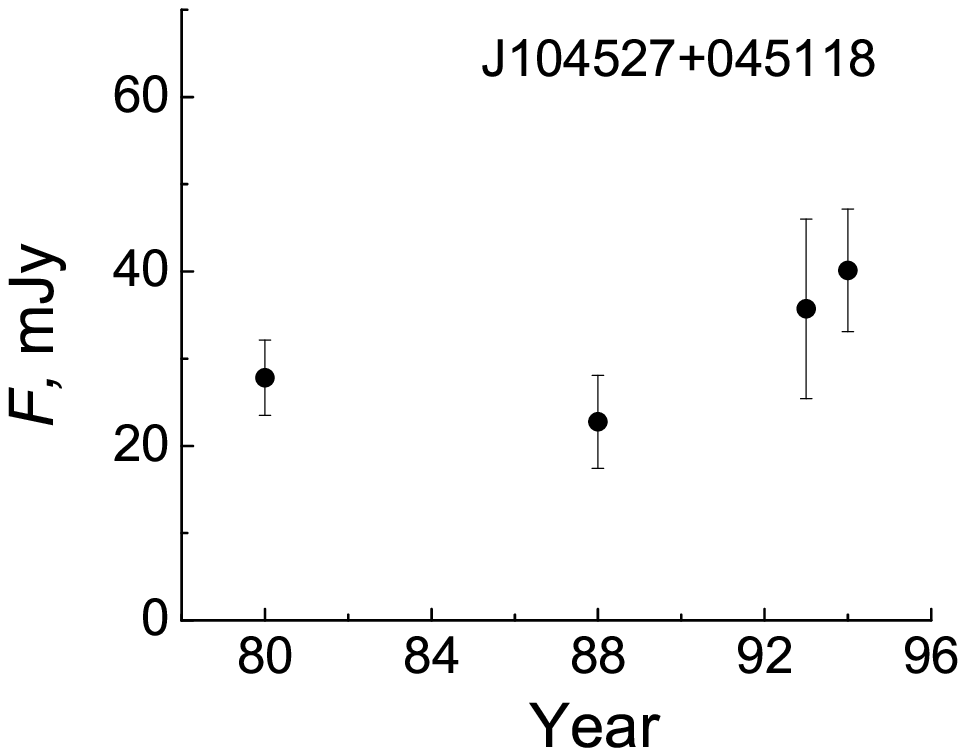}\hspace{20.5mm}
\includegraphics[angle=0,width=0.31\textwidth,clip]{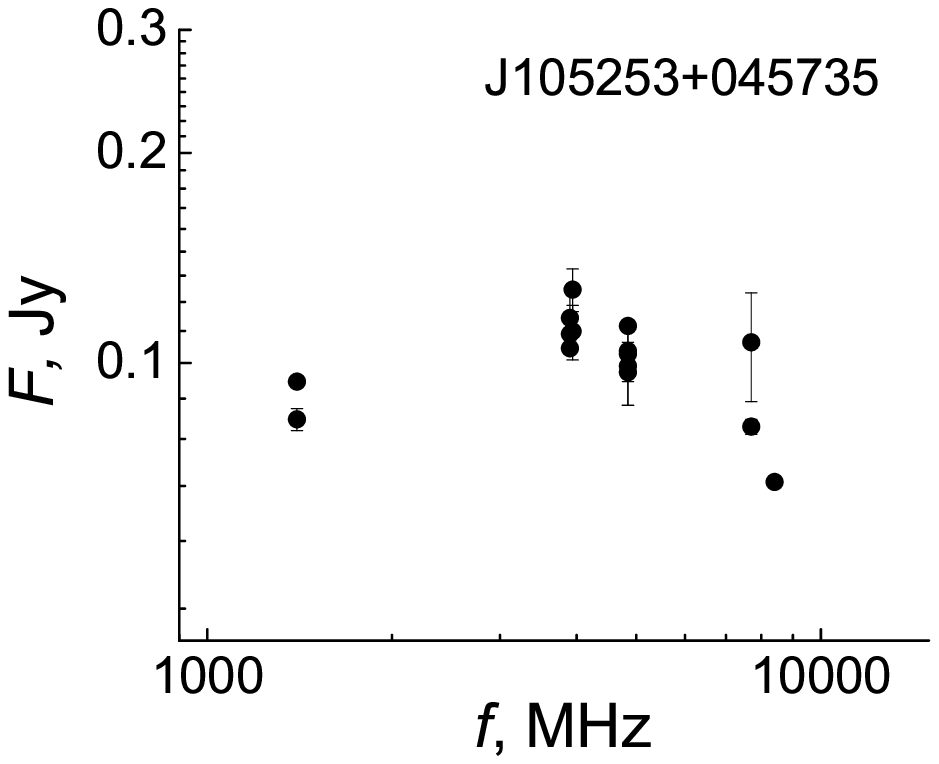}}}}
\caption{(Contd.)}
\end{figure*}

\addtocounter{figure}{-1}
\begin{figure*}
\setcaptionmargin{5mm} \onelinecaptionstrue \centerline{ \vbox{
\hbox{
\includegraphics[angle=0,width=0.31\textwidth,clip]{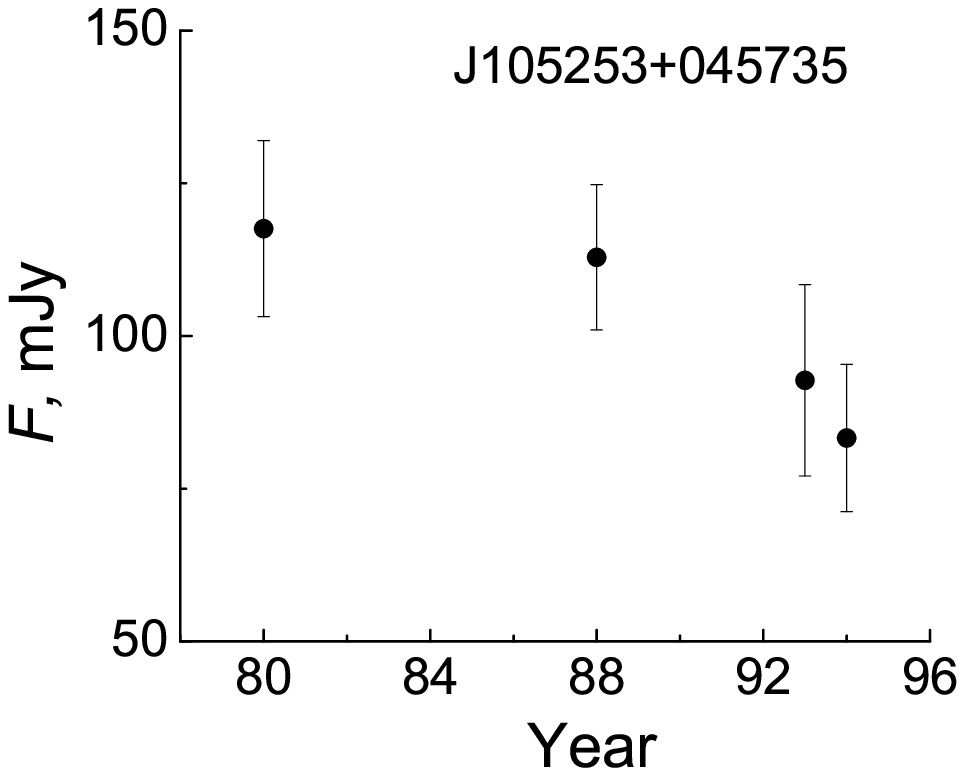}\hspace{20mm}
\includegraphics[angle=0,width=0.31\textwidth,clip]{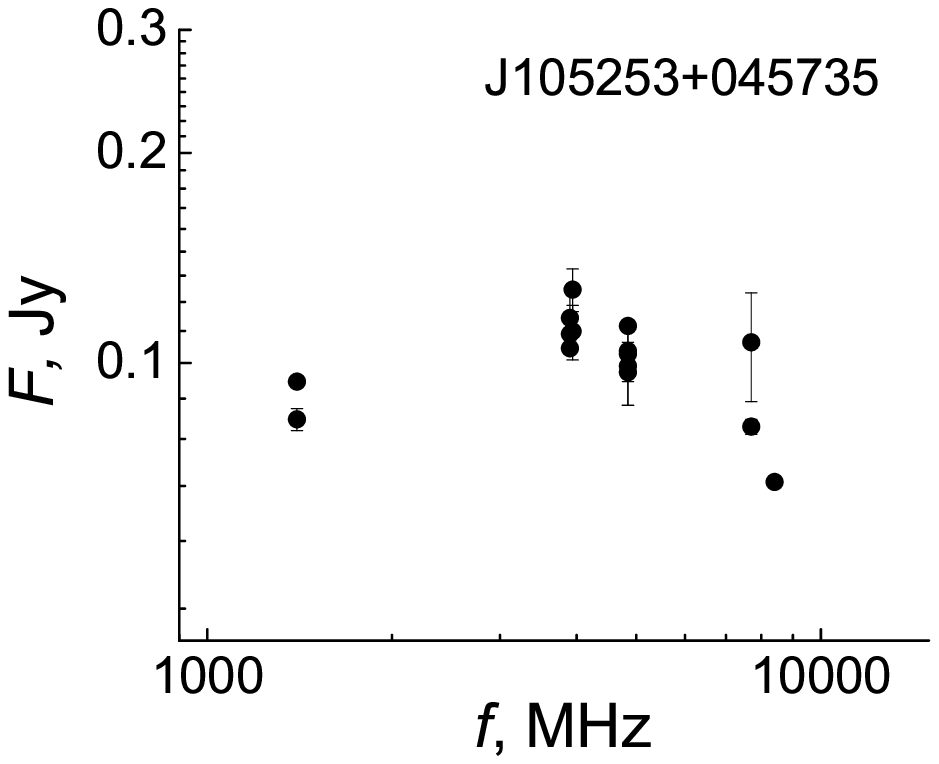}
} \hbox{
\includegraphics[angle=0,width=0.31\textwidth,clip]{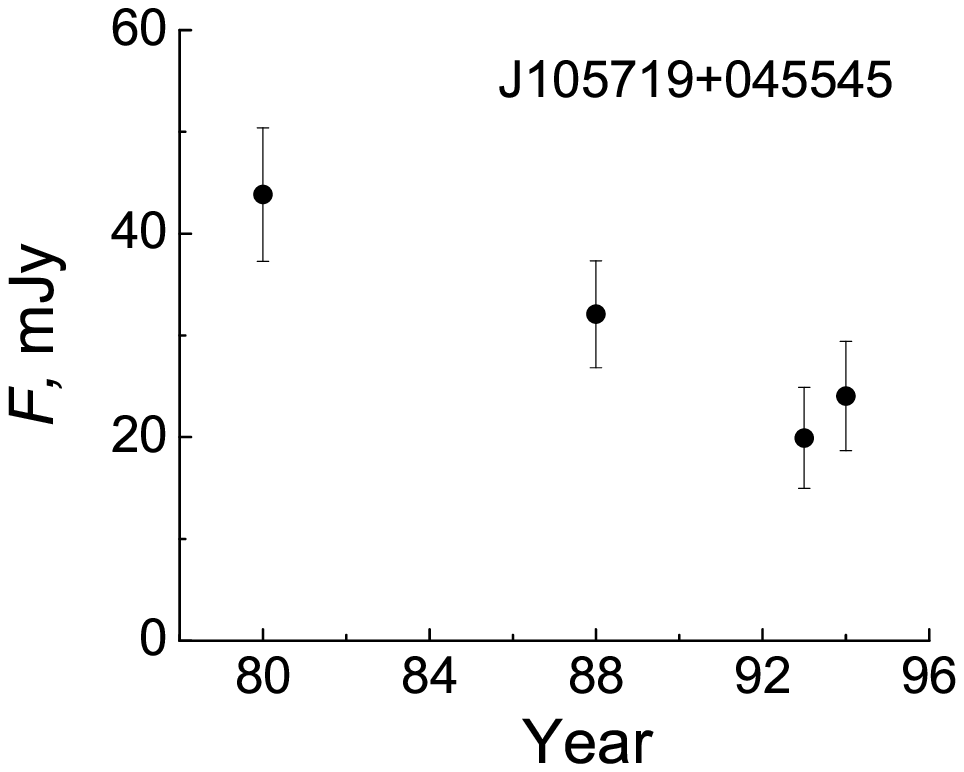}\hspace{20.5mm}
\includegraphics[angle=0,width=0.31\textwidth,clip]{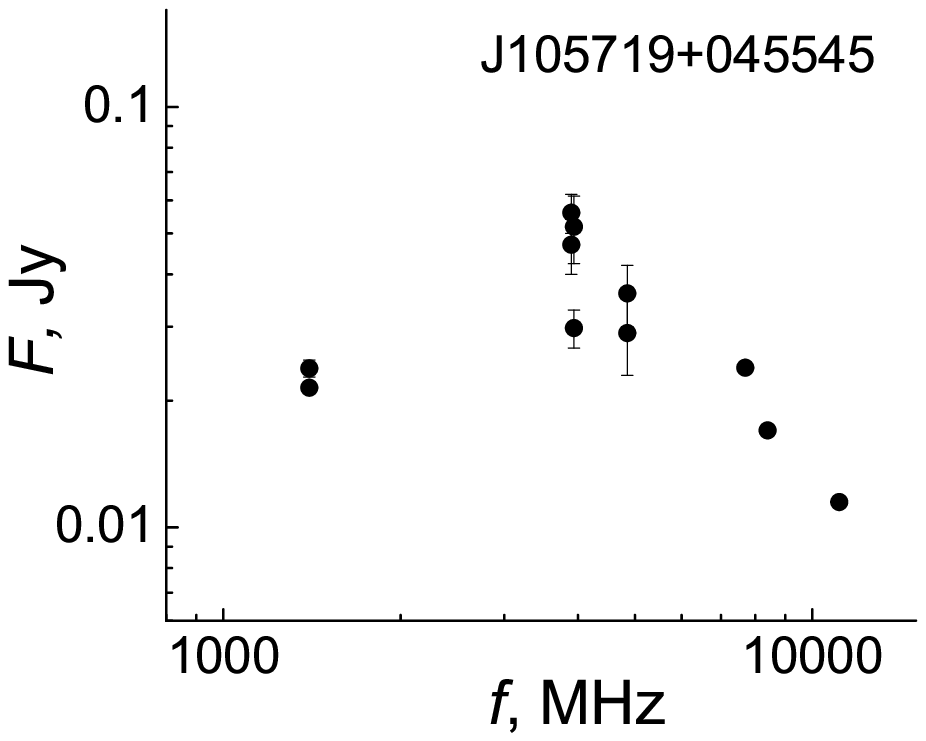}
} \hbox{
\includegraphics[angle=0,width=0.31\textwidth,clip]{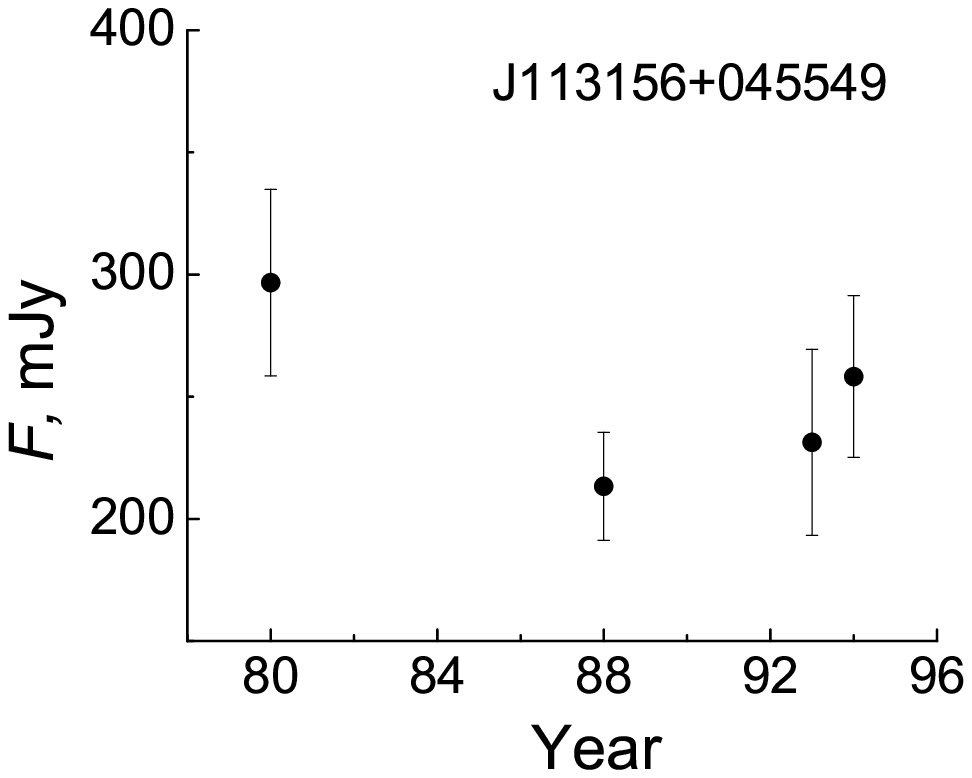}\hspace{20mm}
\includegraphics[angle=0,width=0.31\textwidth,clip]{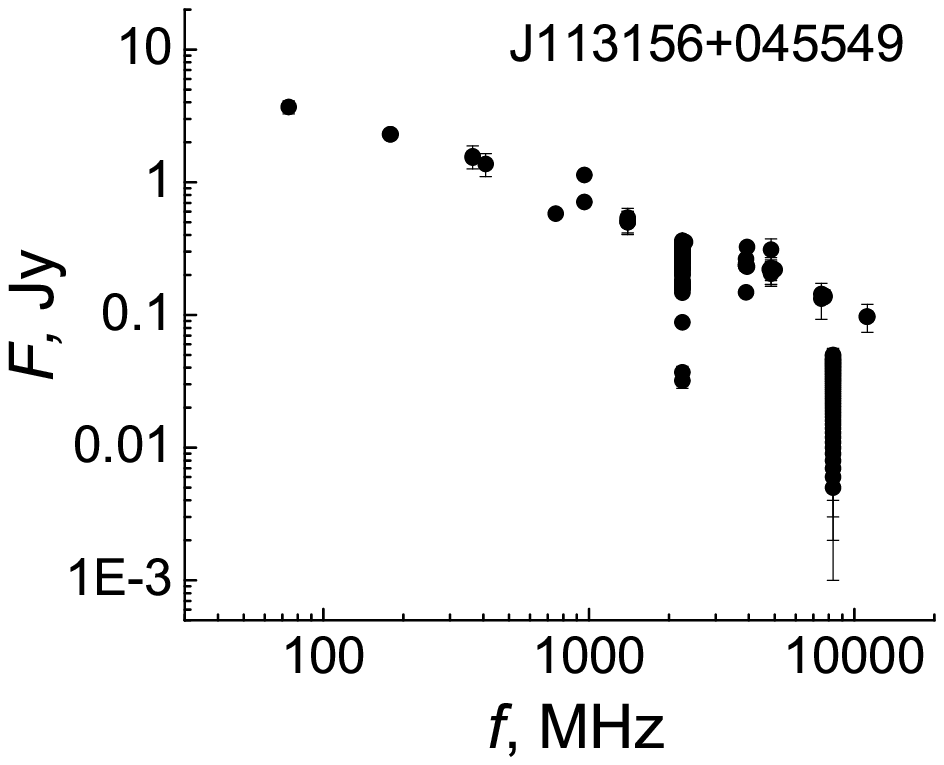}
} \hbox{
\includegraphics[angle=0,width=0.31\textwidth,clip]{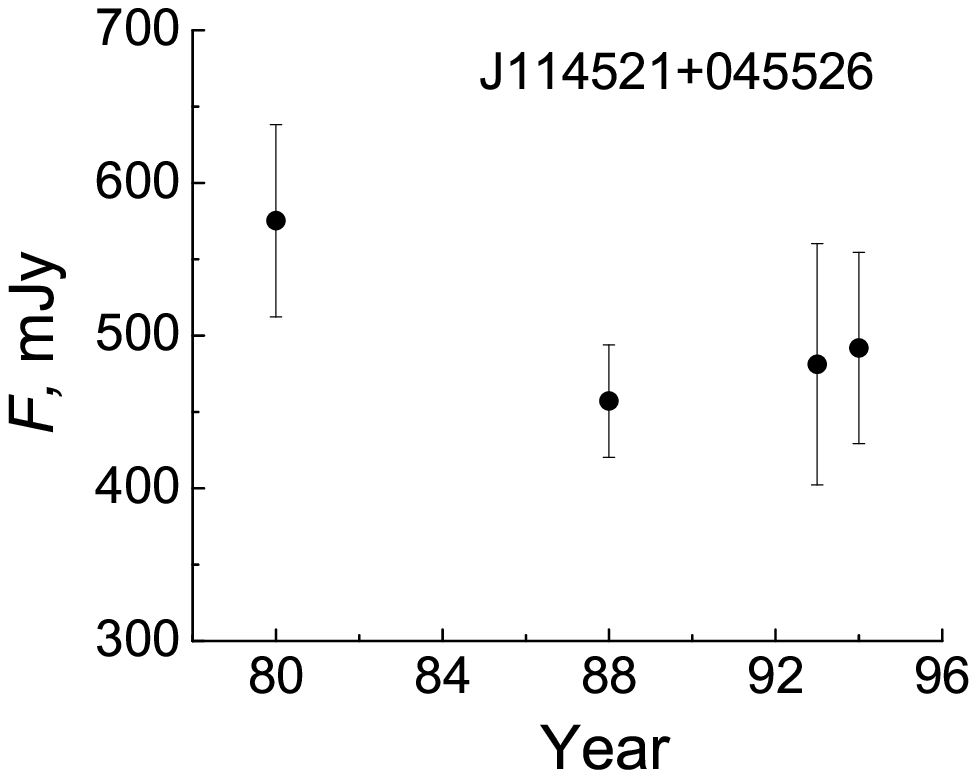}\hspace{21mm}
\includegraphics[angle=0,width=0.31\textwidth,clip]{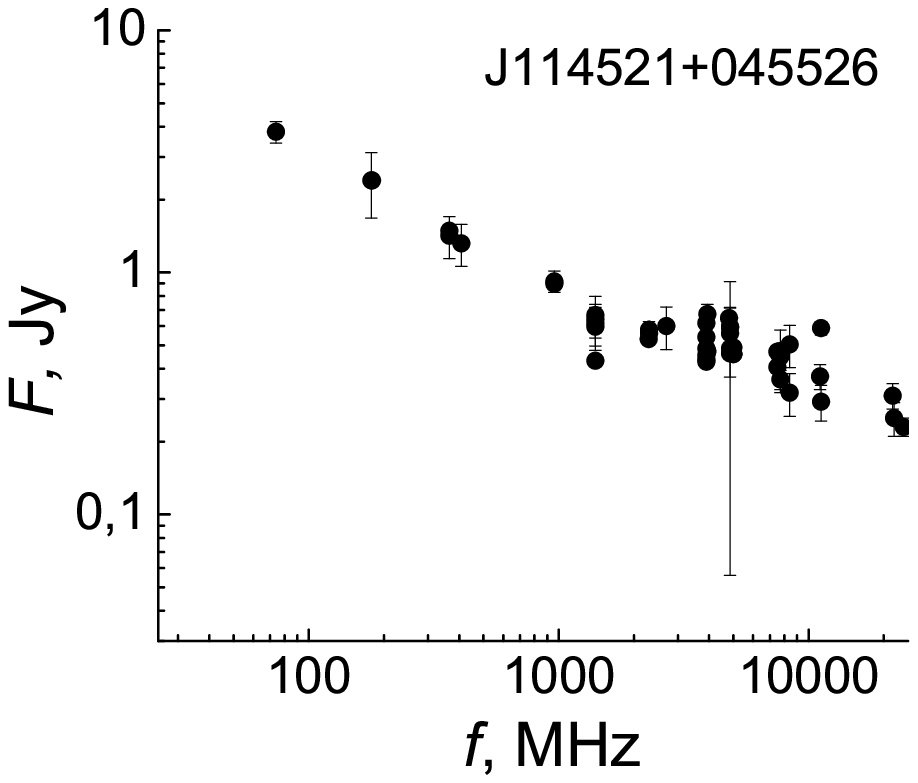}
} \hbox{
\includegraphics[angle=0,width=0.31\textwidth,clip]{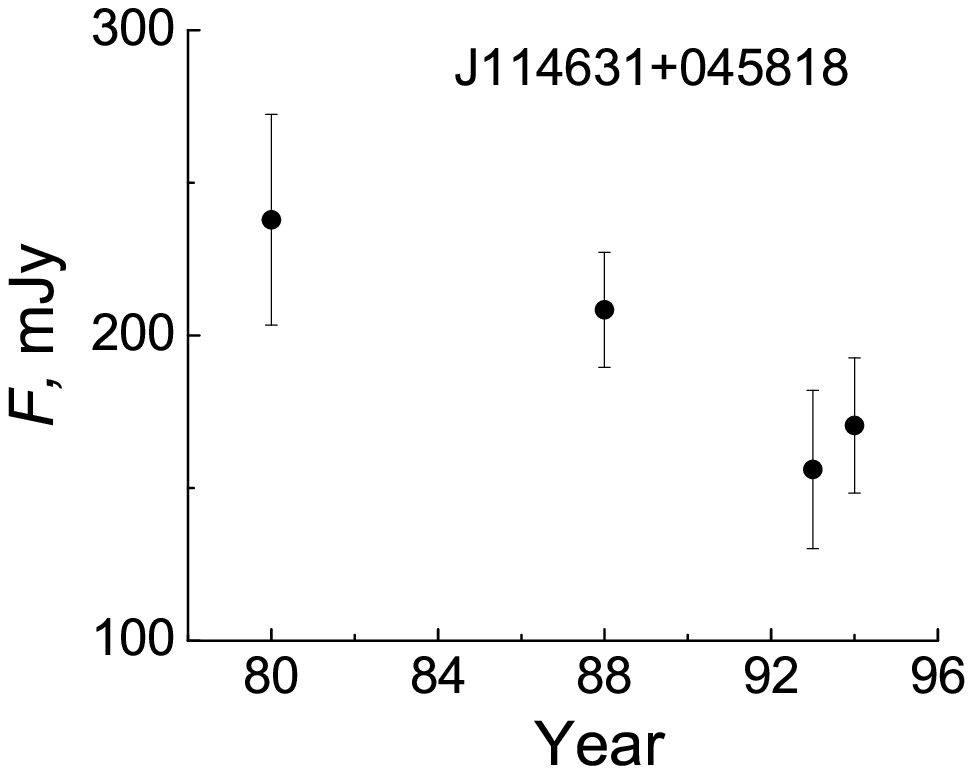}\hspace{21mm}
\includegraphics[angle=0,width=0.31\textwidth,clip]{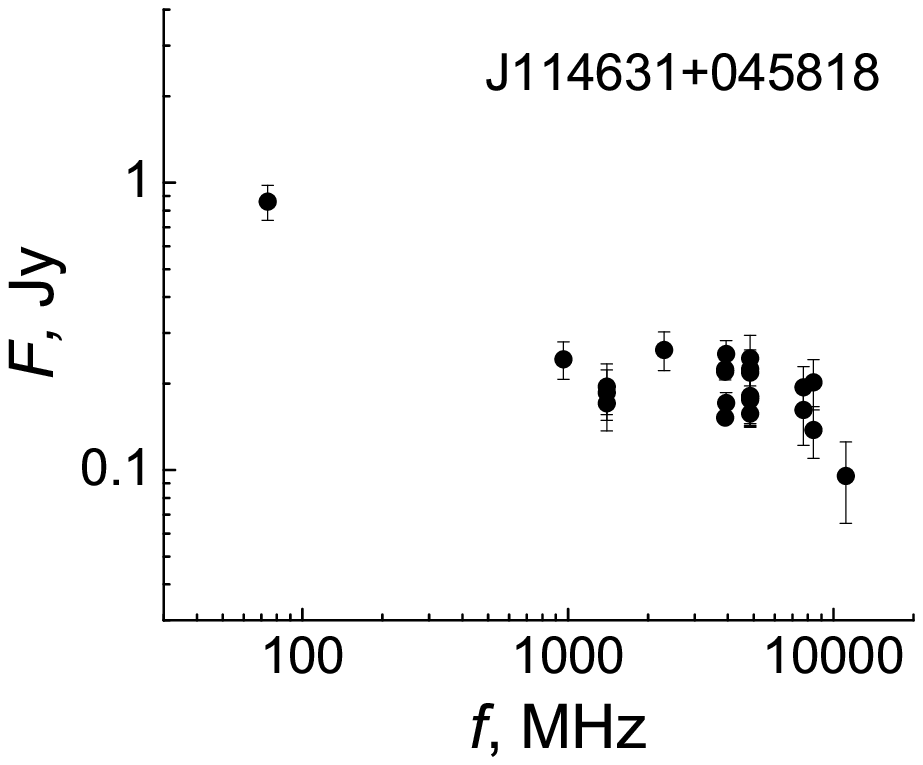}}}}
\caption{(Contd.)}
\end{figure*}

\addtocounter{figure}{-1}
\begin{figure*}
\setcaptionmargin{5mm} \onelinecaptionstrue \centerline{ \vbox{
\hbox{
\includegraphics[angle=0,width=0.31\textwidth,clip]{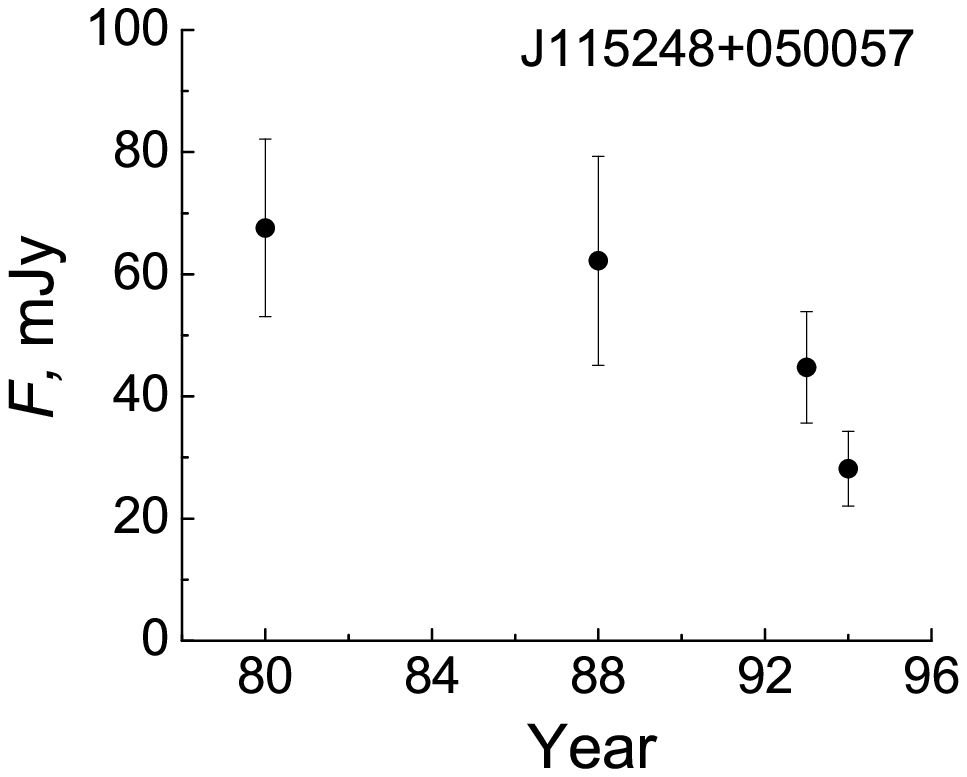}\hspace{20mm}
\includegraphics[angle=0,width=0.31\textwidth,clip]{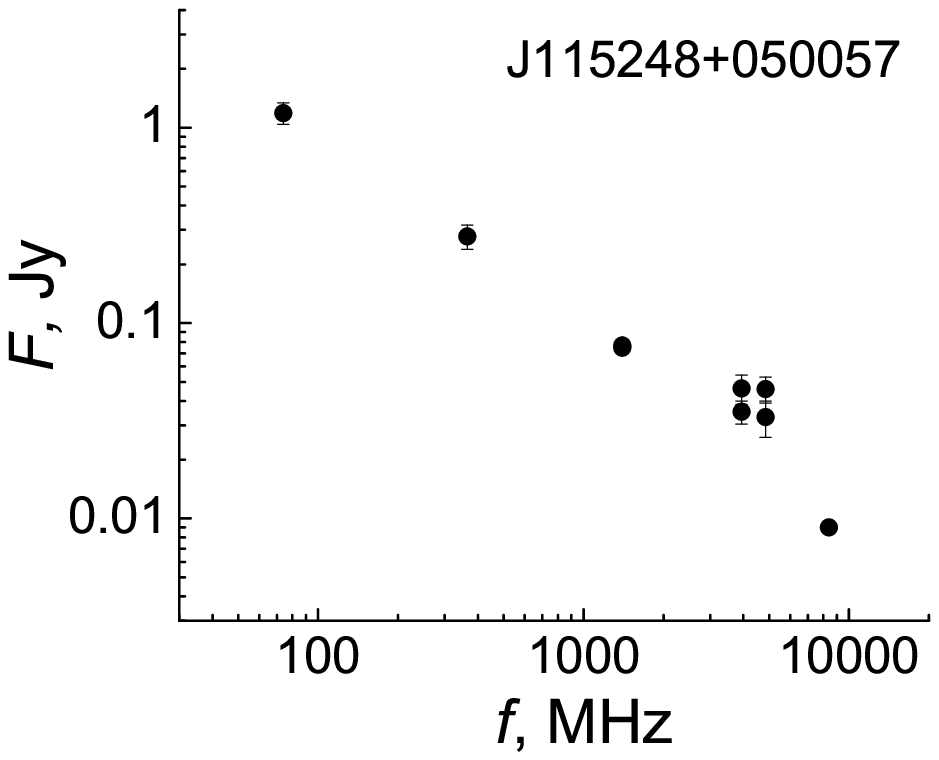}
} \hbox{
\includegraphics[angle=0,width=0.31\textwidth,clip]{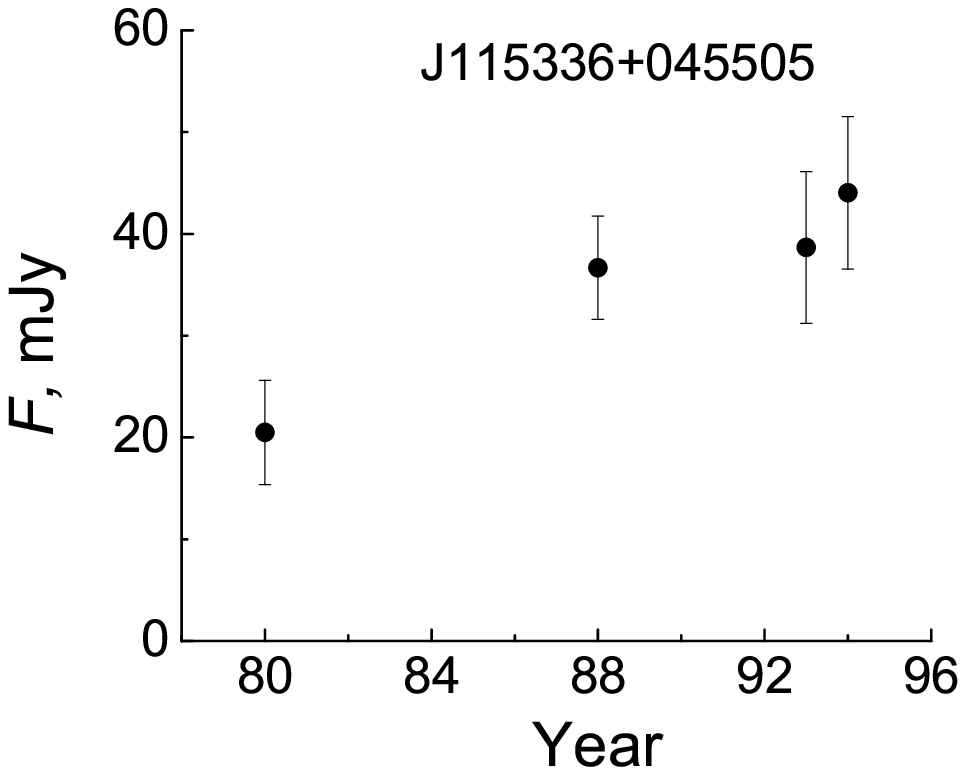}\hspace{20mm}
\includegraphics[angle=0,width=0.31\textwidth,clip]{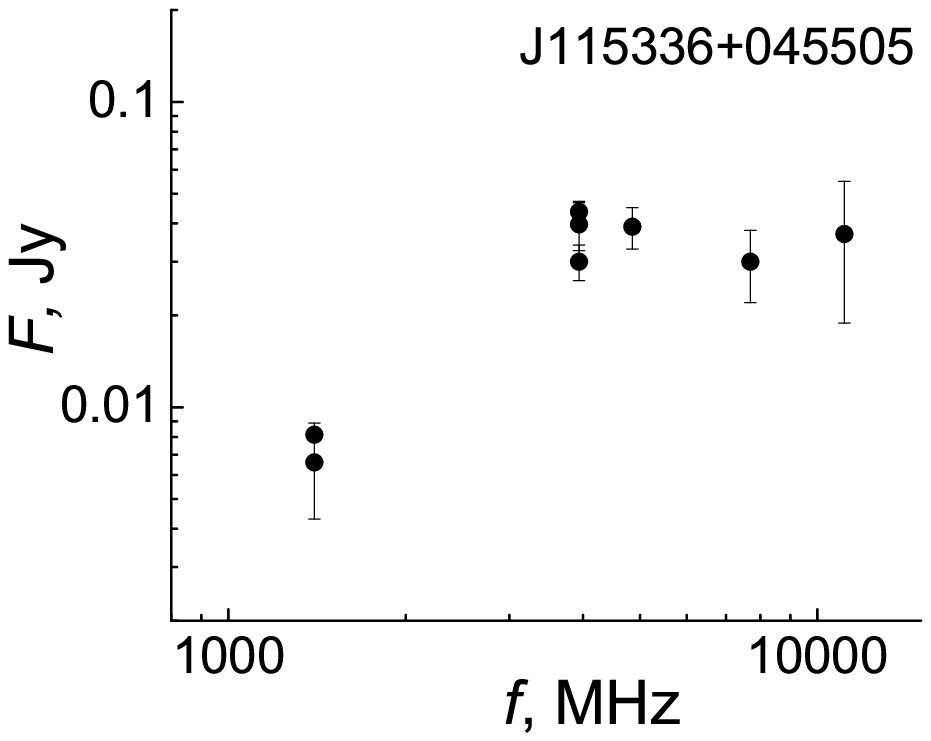}
} \hbox{
\includegraphics[angle=0,width=0.31\textwidth,clip]{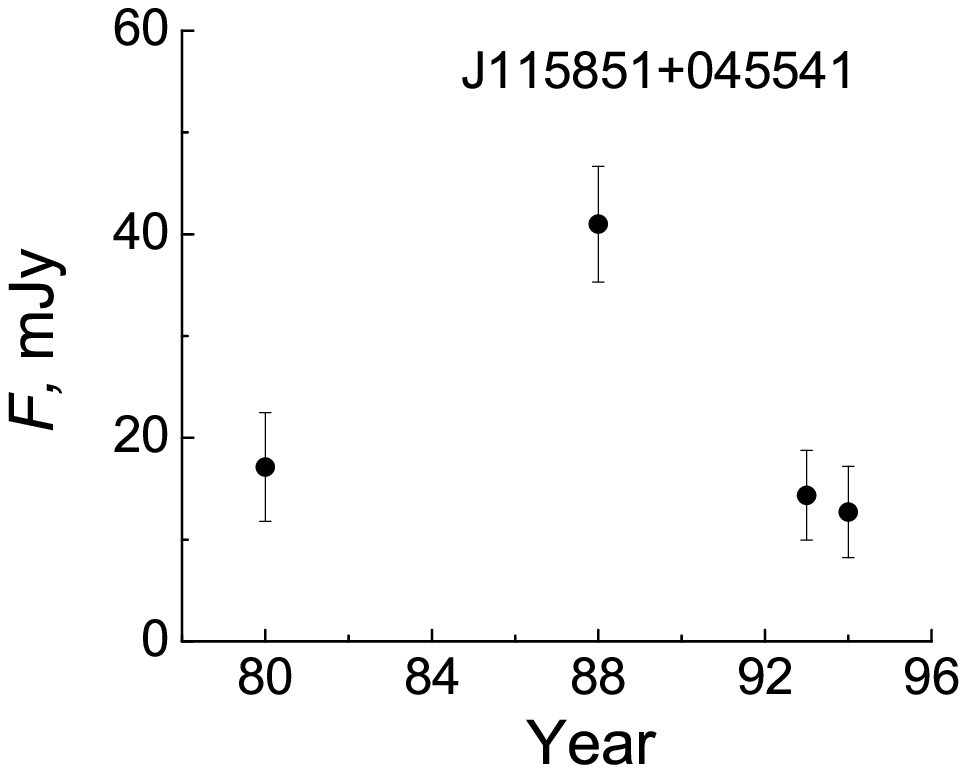}\hspace{21mm}
\includegraphics[angle=0,width=0.31\textwidth,clip]{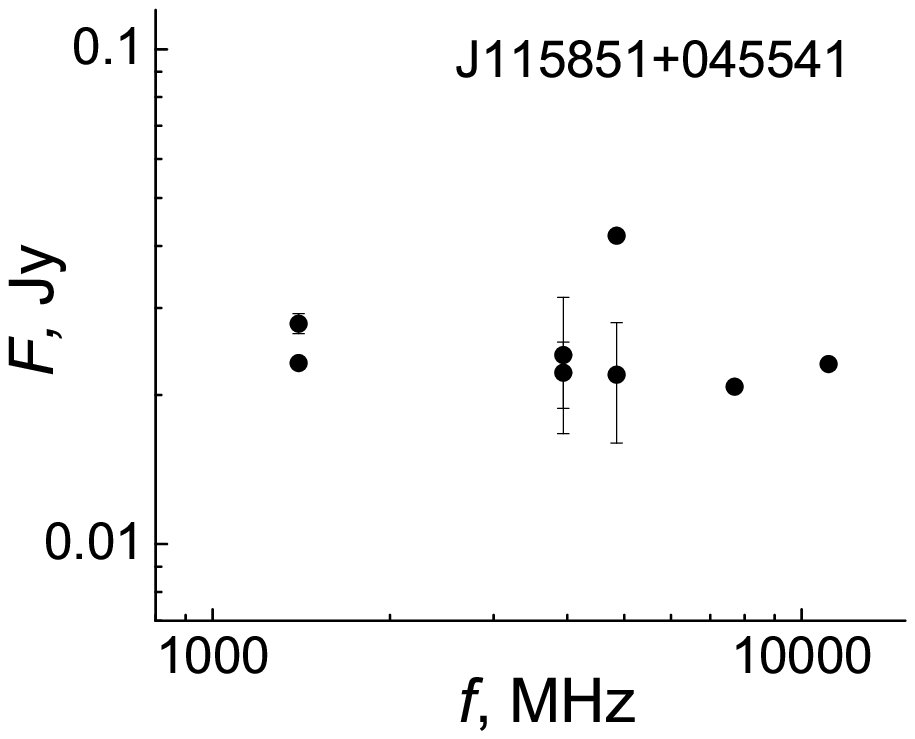}
} \hbox{
\includegraphics[angle=0,width=0.31\textwidth,clip]{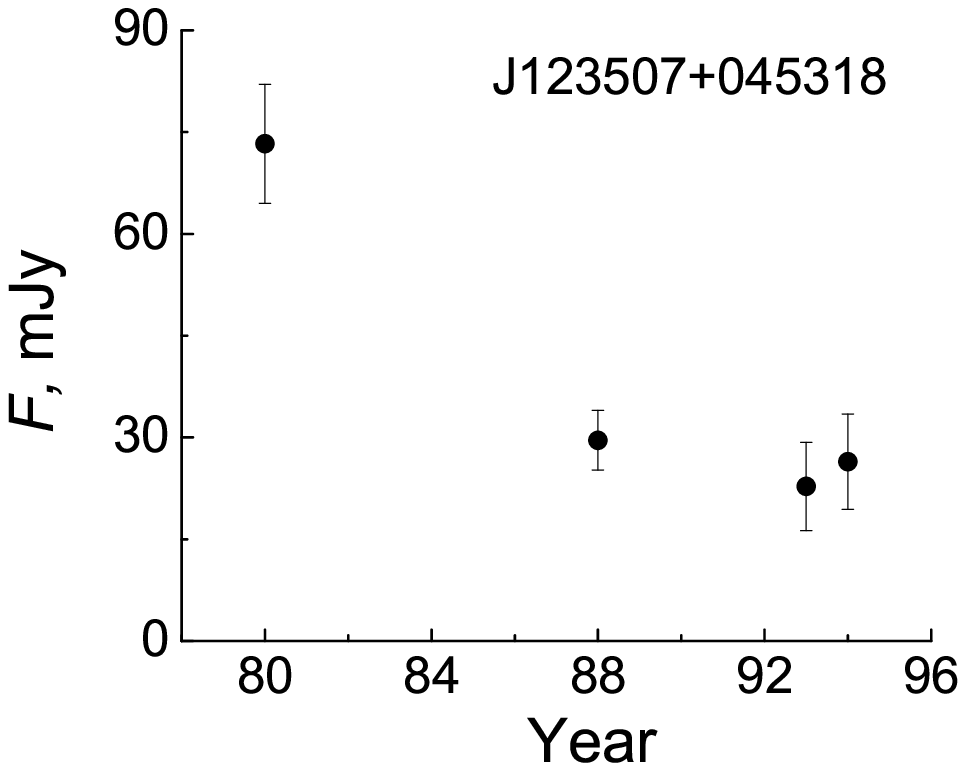}\hspace{21mm}
\includegraphics[angle=0,width=0.31\textwidth,clip]{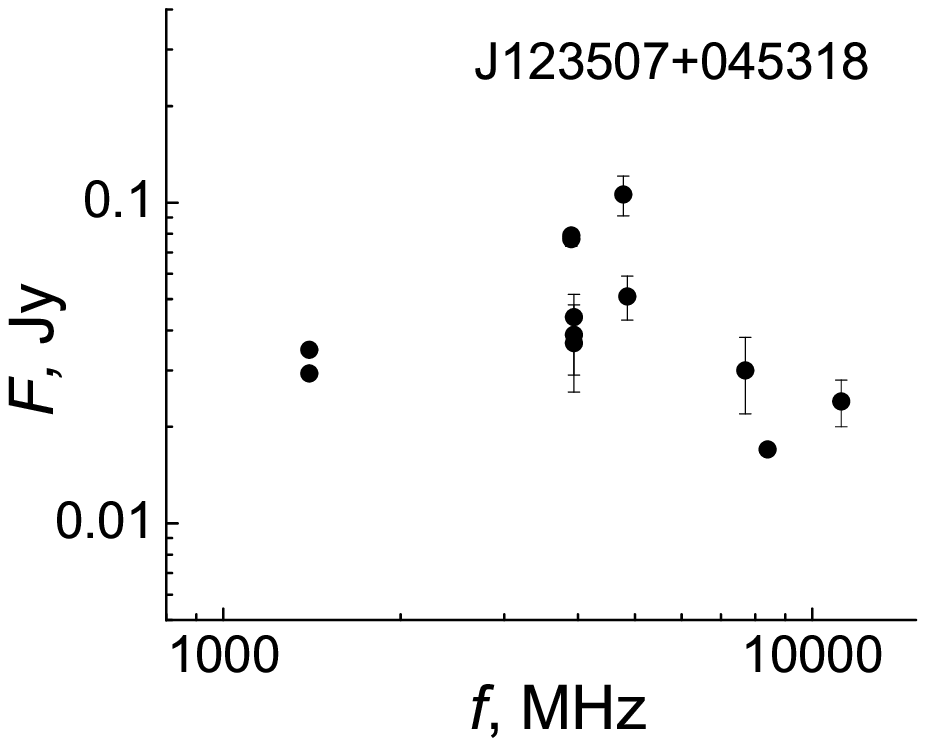}
} \hbox{
\includegraphics[angle=0,width=0.31\textwidth,clip]{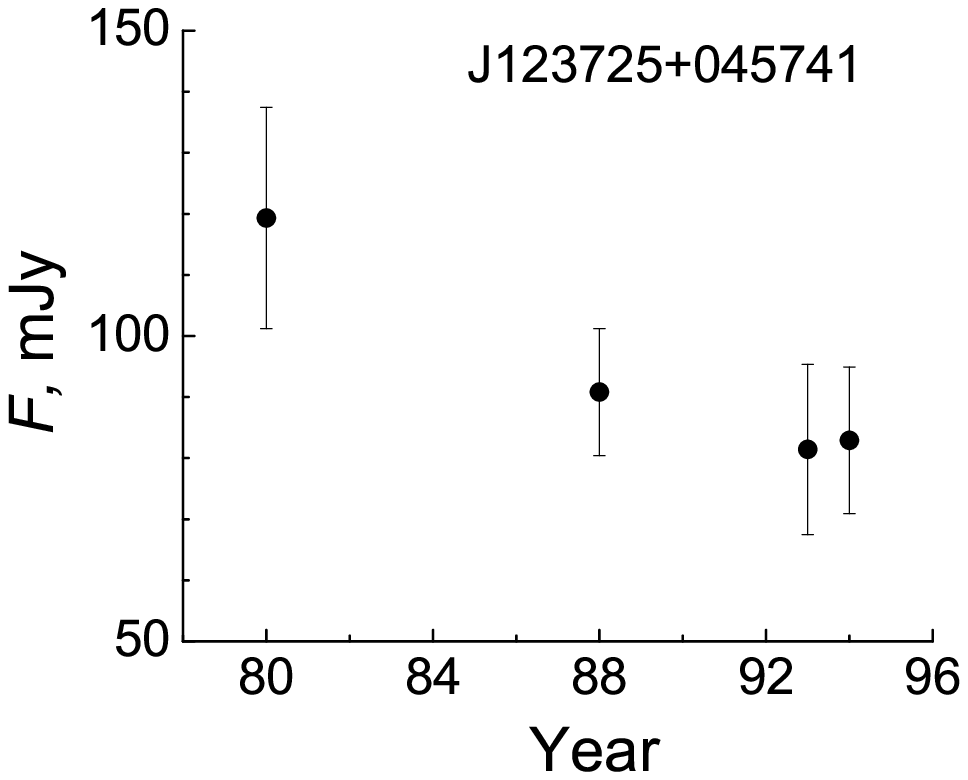}\hspace{20.5mm}
\includegraphics[angle=0,width=0.31\textwidth,clip]{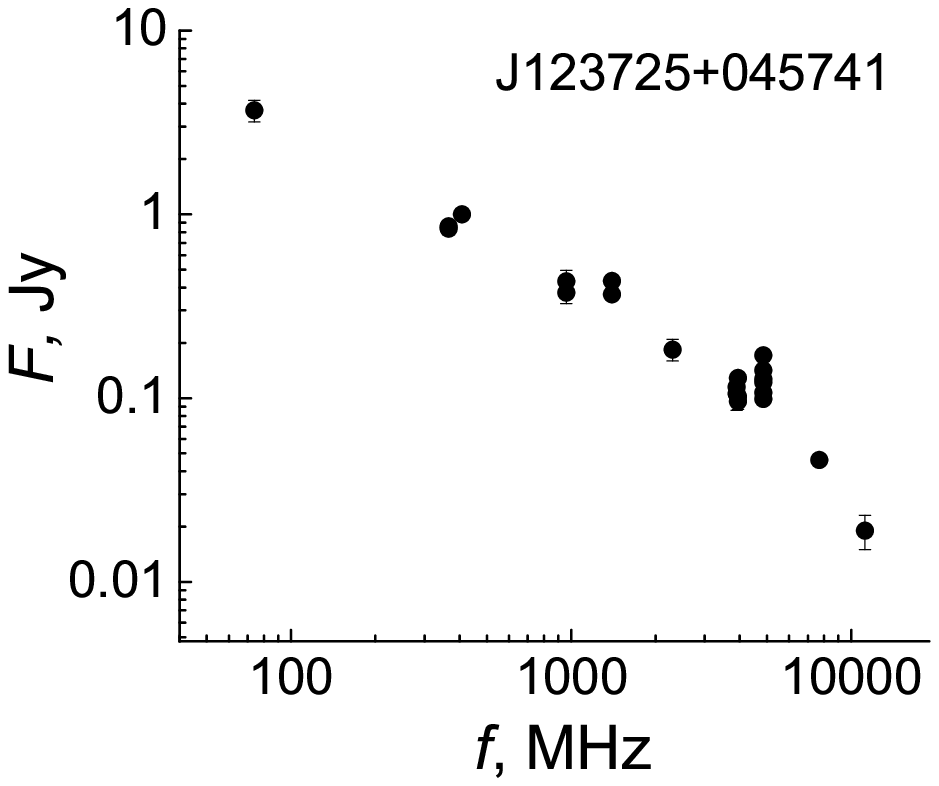}}}}
 \caption{(Contd.)}
\end{figure*}

\addtocounter{figure}{-1}
\begin{figure*}
\setcaptionmargin{5mm} \onelinecaptionstrue \centerline{ \vbox{
\hbox{
\includegraphics[angle=0,width=0.31\textwidth,clip]{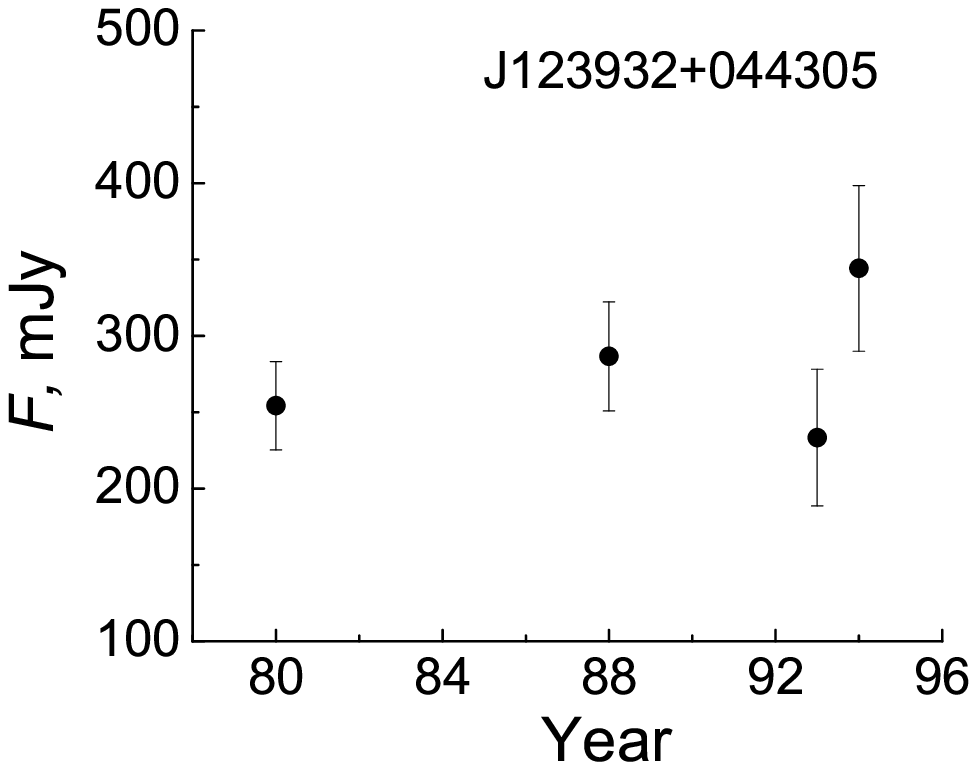}\hspace{20mm}
\includegraphics[angle=0,width=0.31\textwidth,clip]{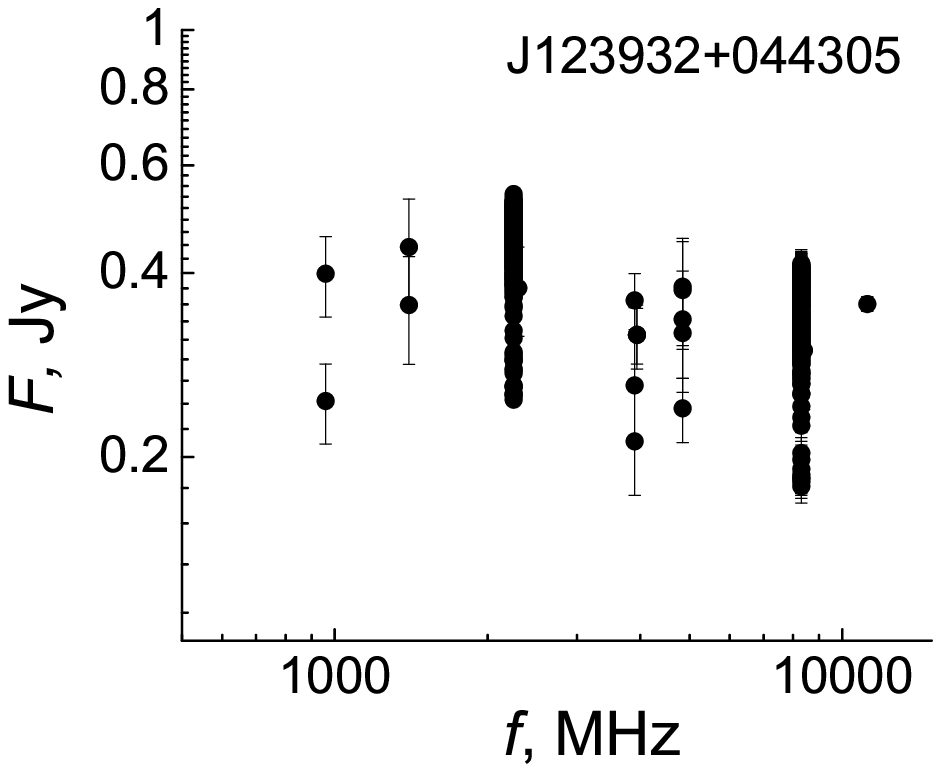}
} \hbox{
\includegraphics[angle=0,width=0.31\textwidth,clip]{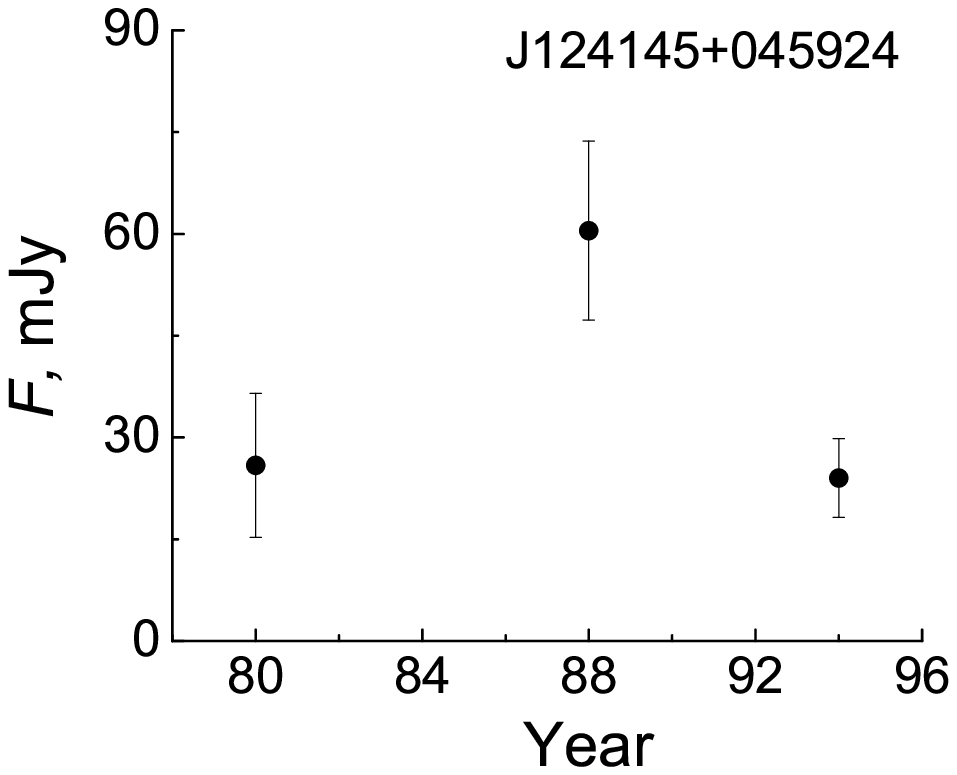}\hspace{21mm}
\includegraphics[angle=0,width=0.31\textwidth,clip]{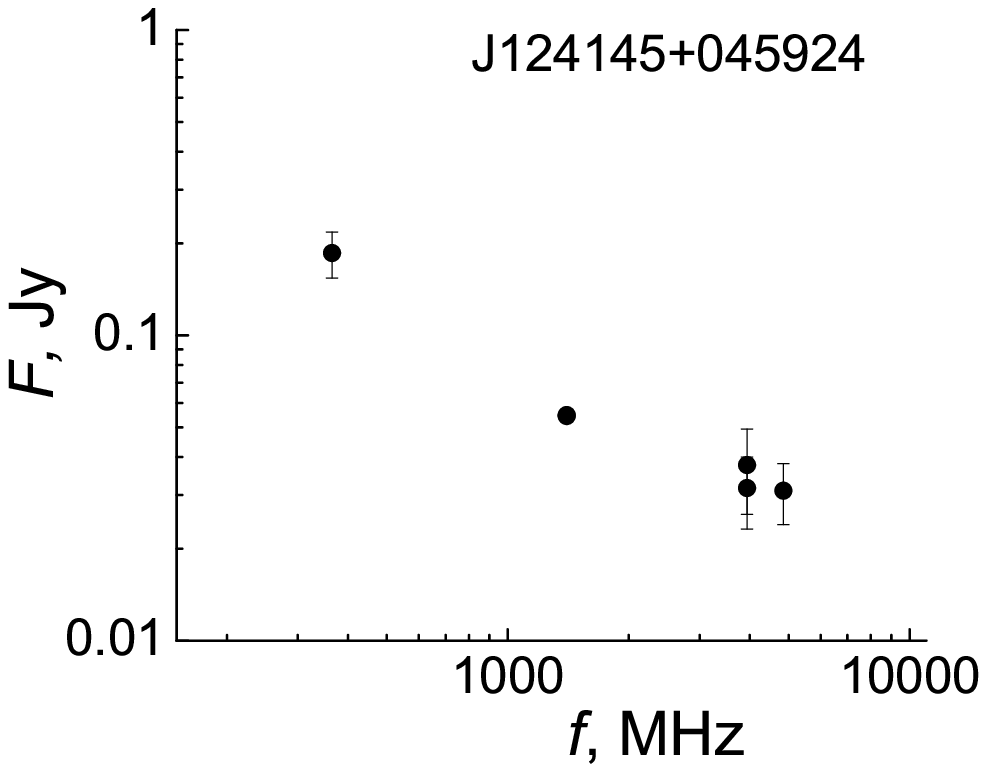}
} \hbox{
\includegraphics[angle=0,width=0.31\textwidth,clip]{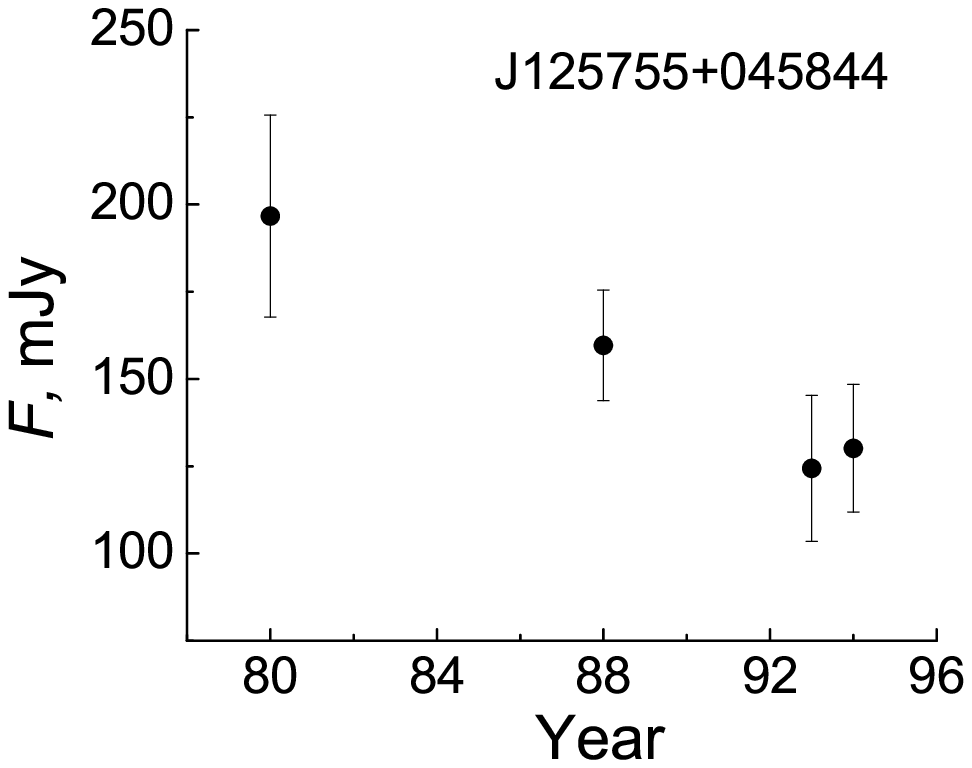}\hspace{20.5mm}
\includegraphics[angle=0,width=0.31\textwidth,clip]{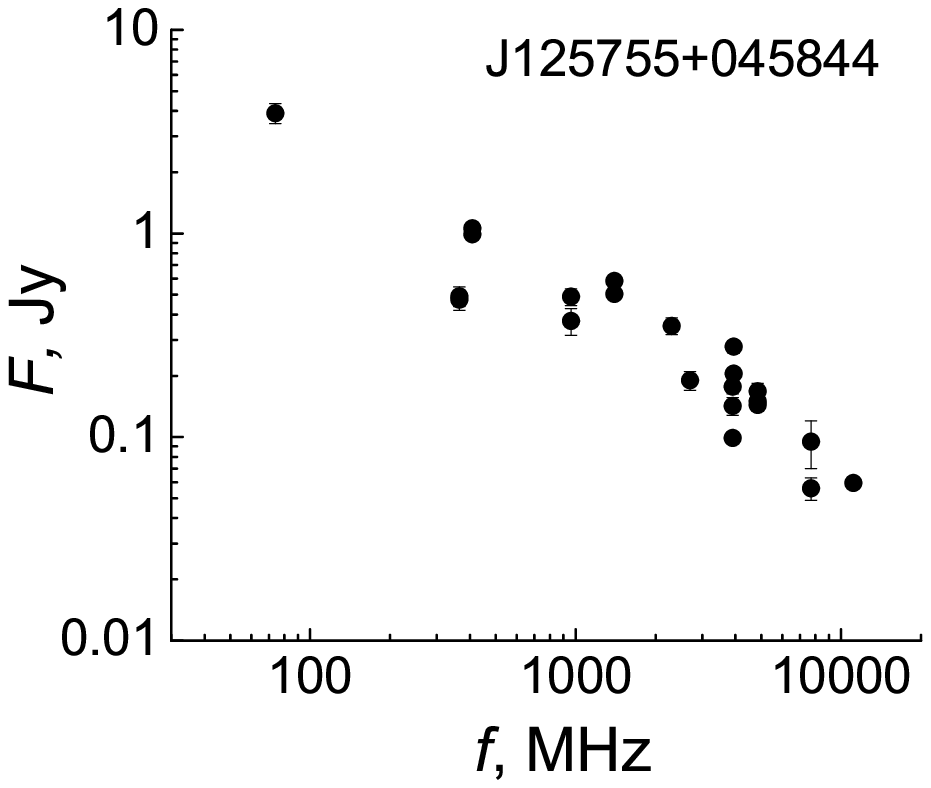}
} \hbox{
\includegraphics[angle=0,width=0.31\textwidth,clip]{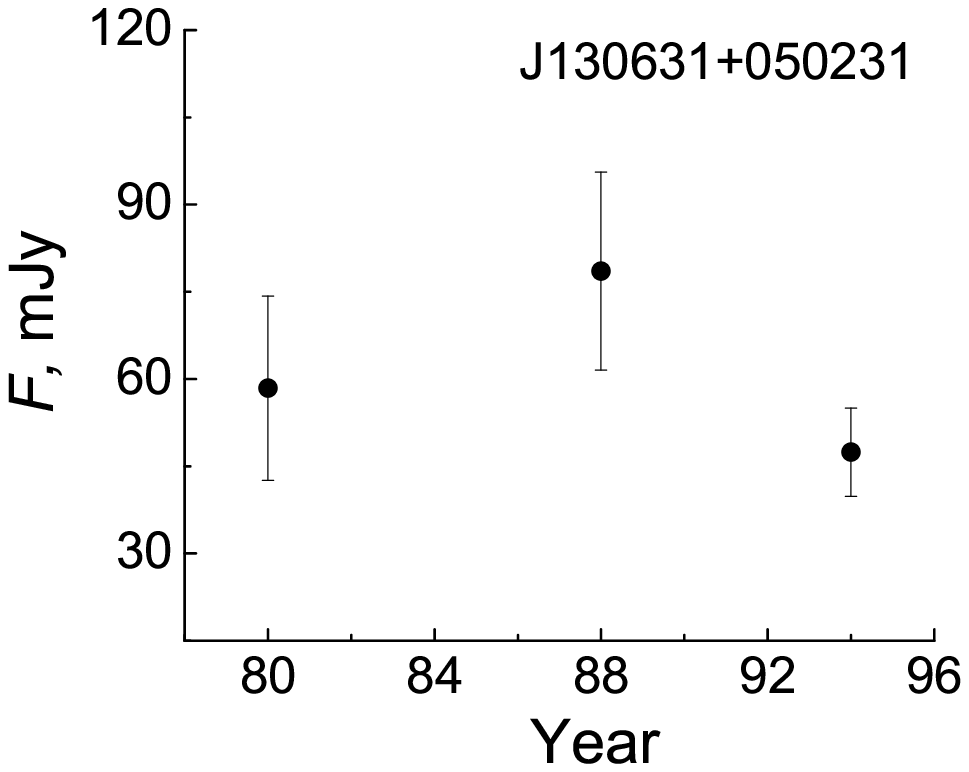}\hspace{21mm}
\includegraphics[angle=0,width=0.31\textwidth,clip]{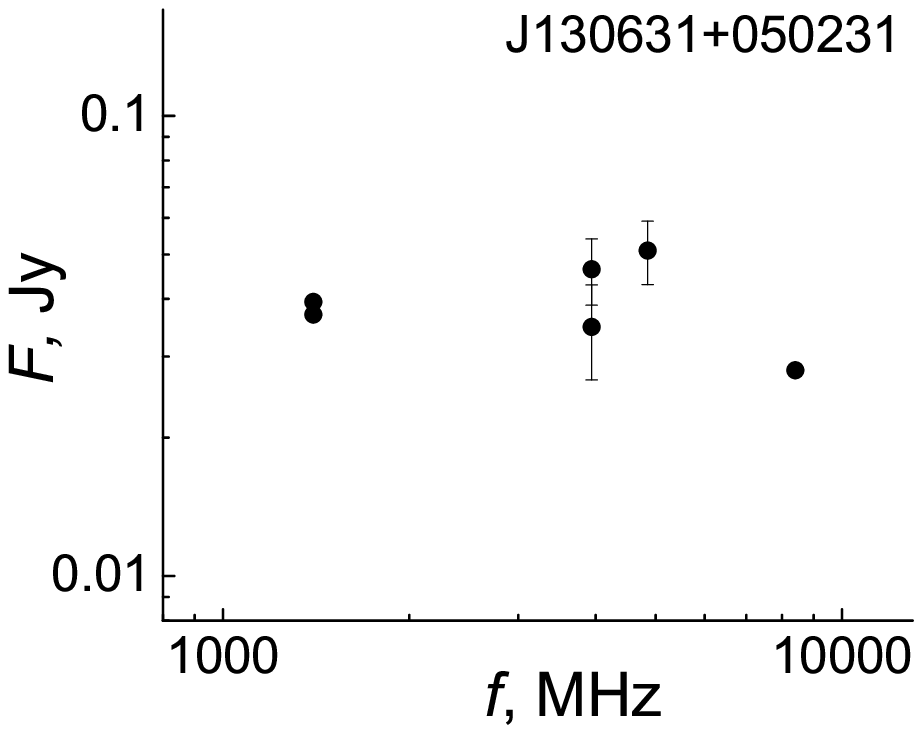}
} \hbox{
\includegraphics[angle=0,width=0.31\textwidth,clip]{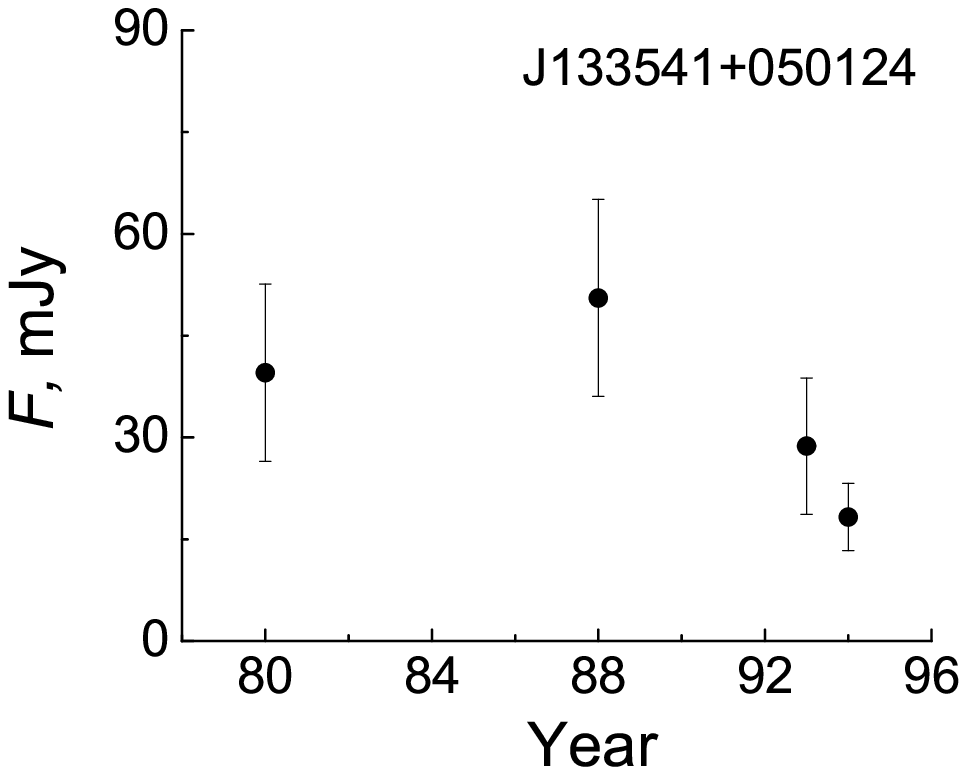}\hspace{21mm}
\includegraphics[angle=0,width=0.31\textwidth,clip]{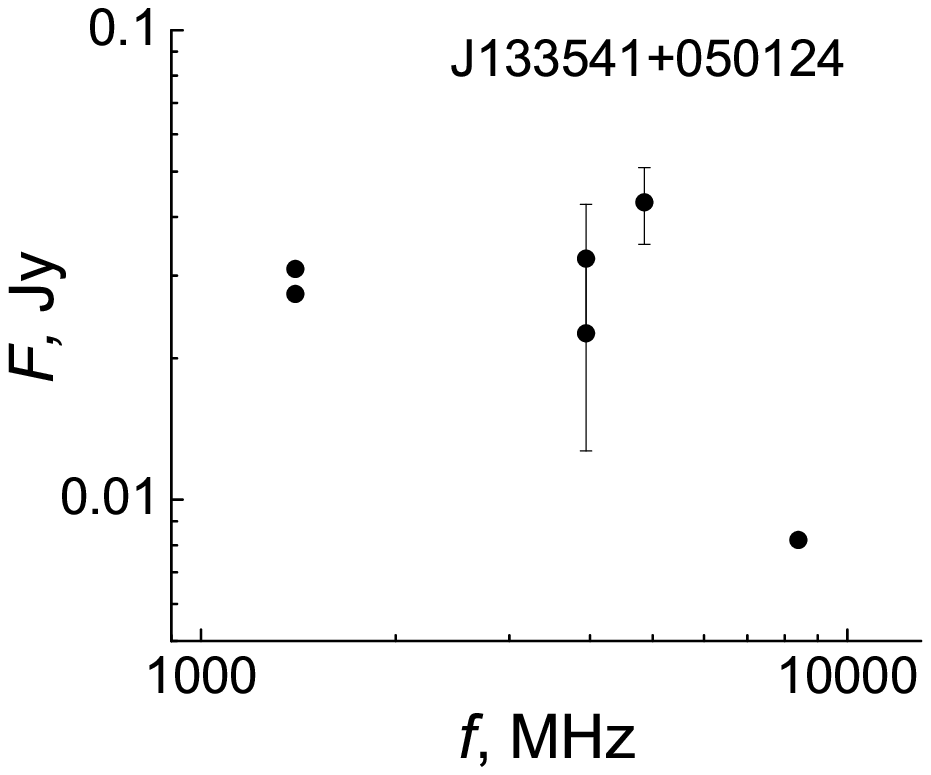}}}}
 \caption{(Contd.)}
\end{figure*}

\addtocounter{figure}{-1}
\begin{figure*}
\setcaptionmargin{5mm} \onelinecaptionstrue \centerline{ \vbox{
\hbox{
\includegraphics[angle=0,width=0.31\textwidth,clip]{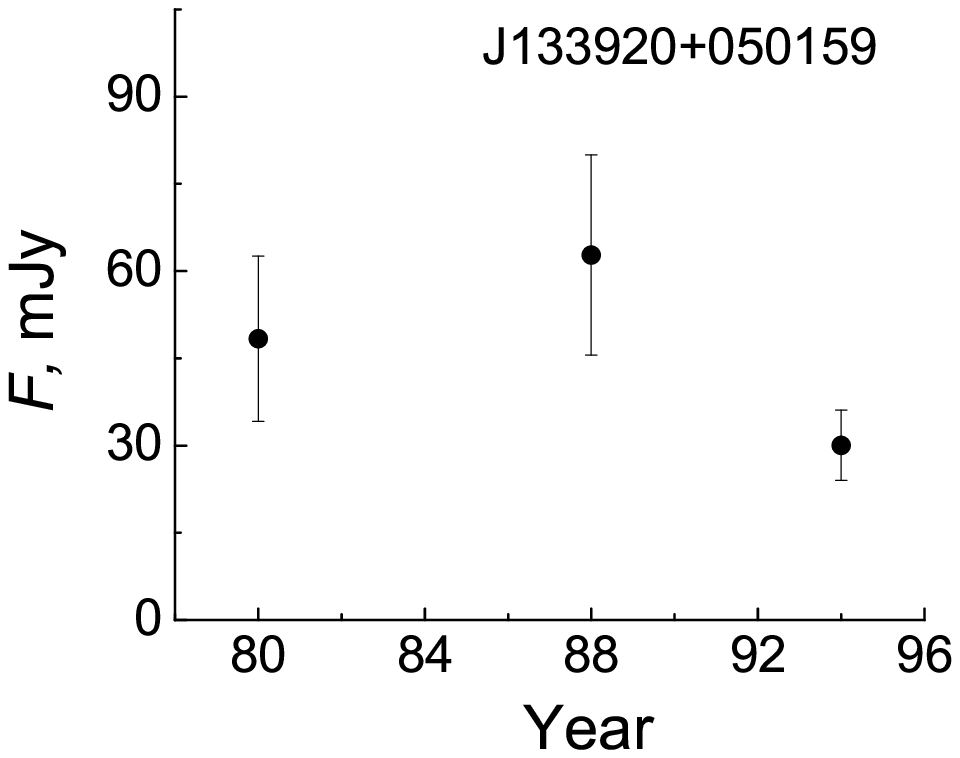}\hspace{20mm}
\includegraphics[angle=0,width=0.31\textwidth,clip]{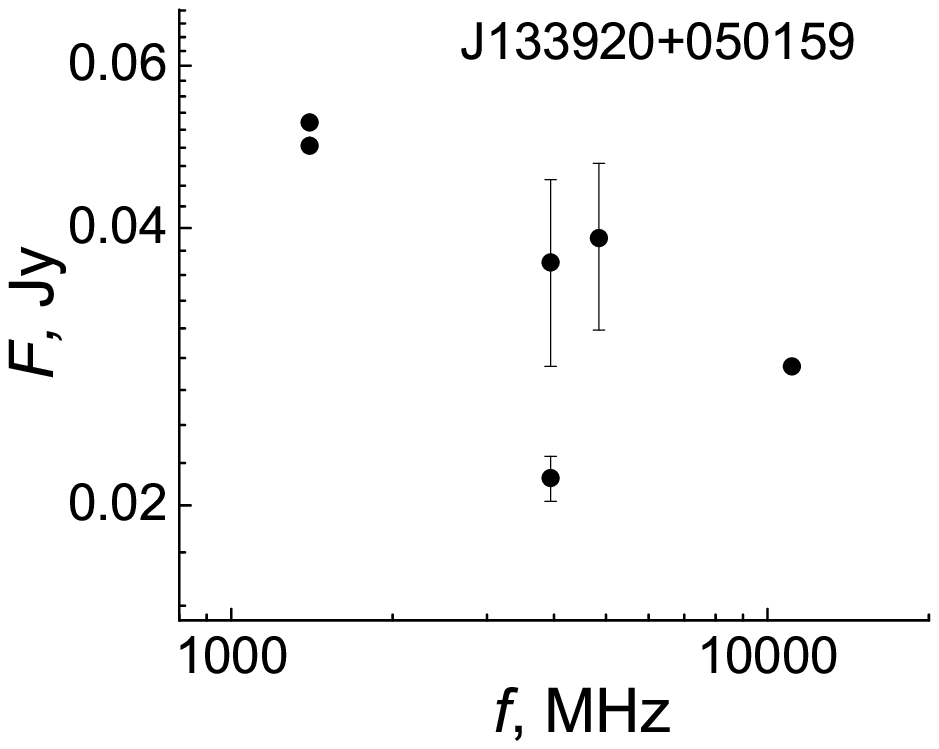}
} \hbox{
\includegraphics[angle=0,width=0.31\textwidth,clip]{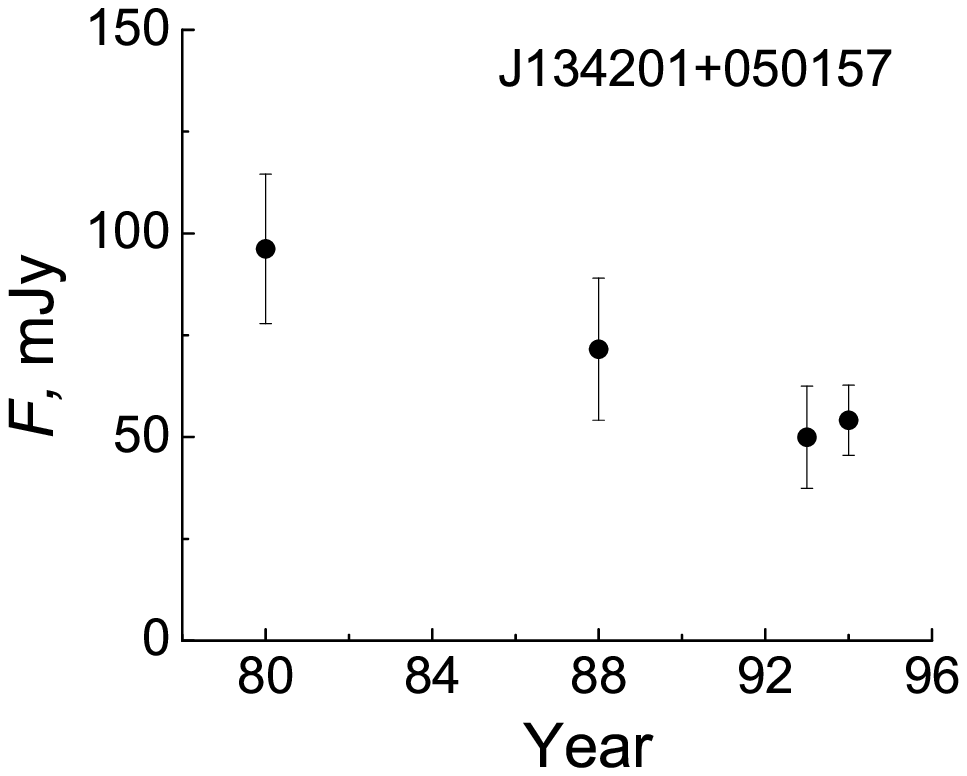}\hspace{21mm}
\includegraphics[angle=0,width=0.31\textwidth,clip]{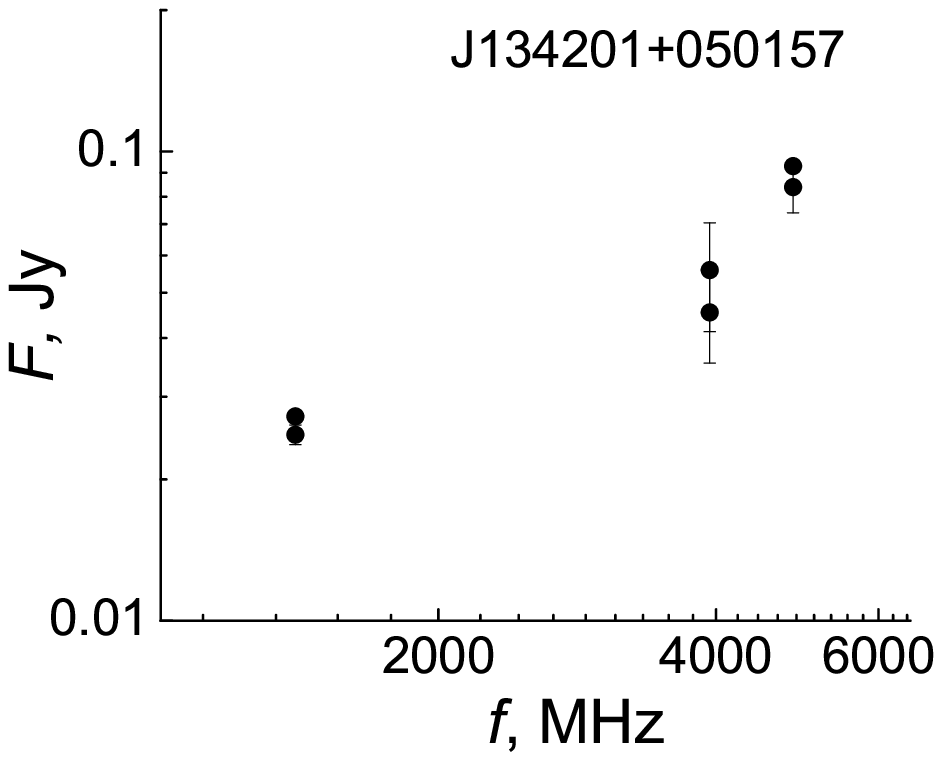}
} \hbox{
\includegraphics[angle=0,width=0.31\textwidth,clip]{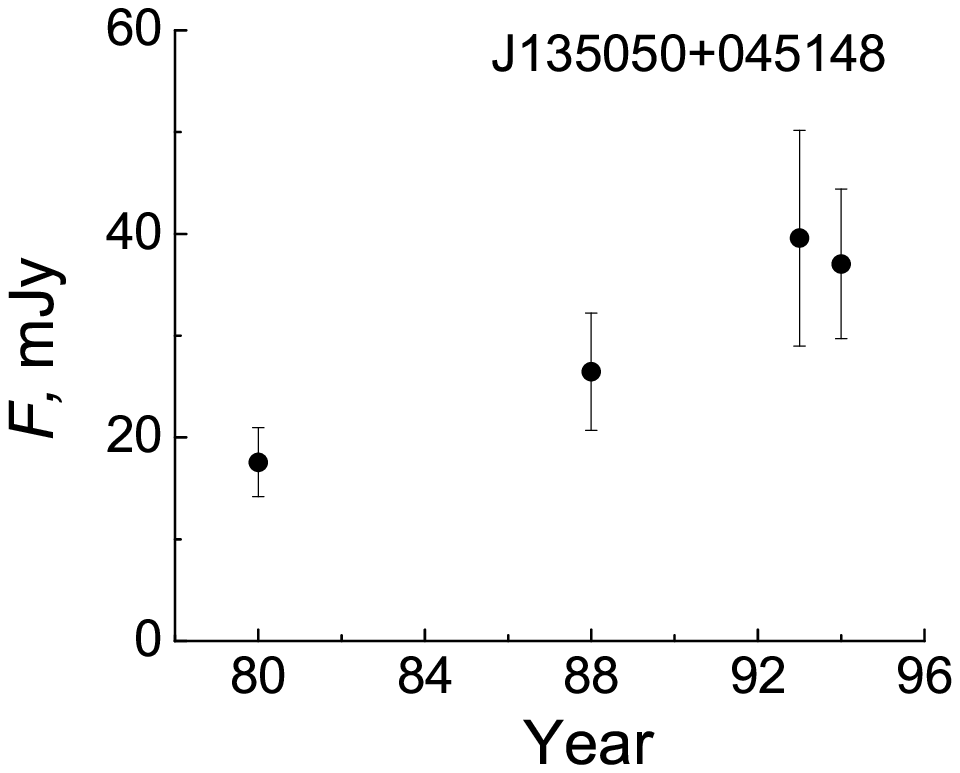}\hspace{21mm}
\includegraphics[angle=0,width=0.31\textwidth,clip]{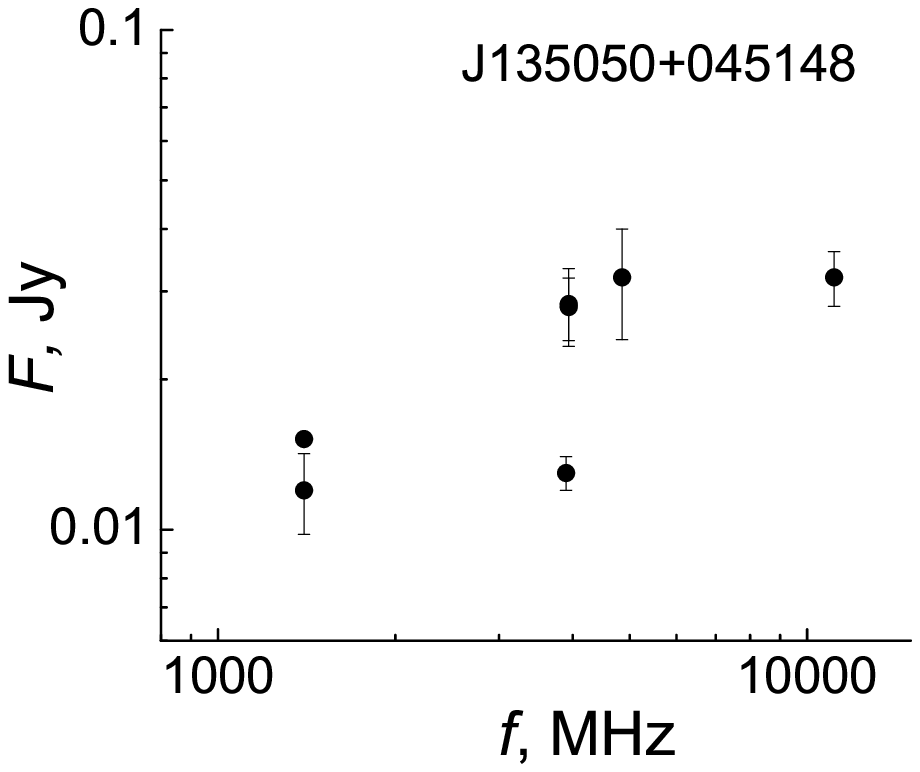}
} \hbox{
\includegraphics[angle=0,width=0.31\textwidth,clip]{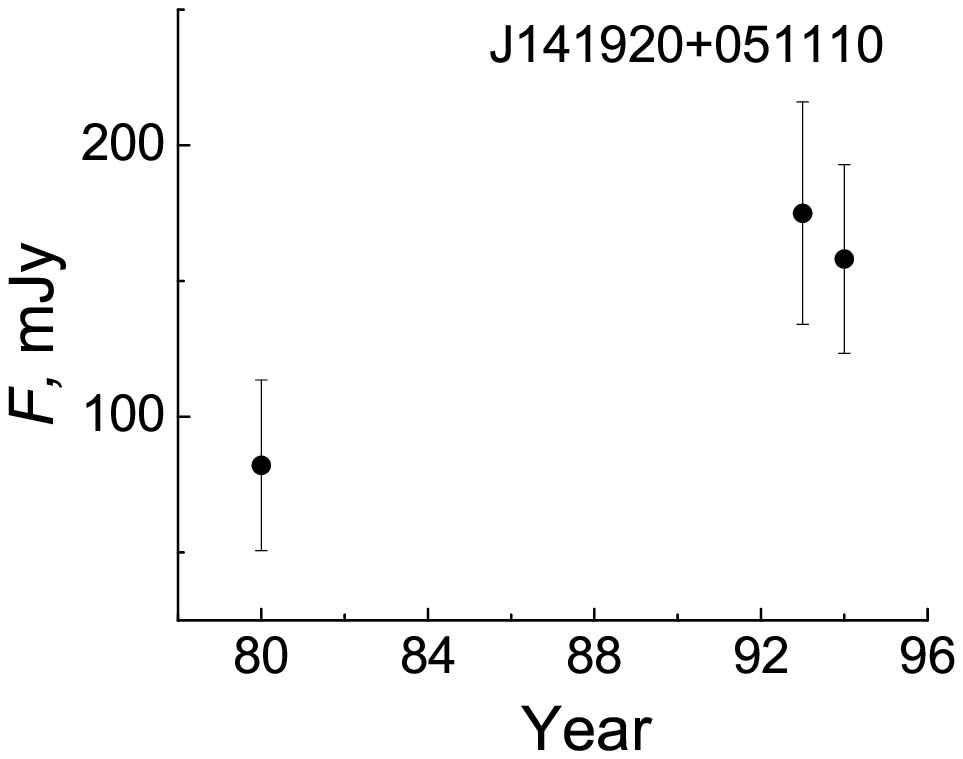}\hspace{21mm}
\includegraphics[angle=0,width=0.31\textwidth,clip]{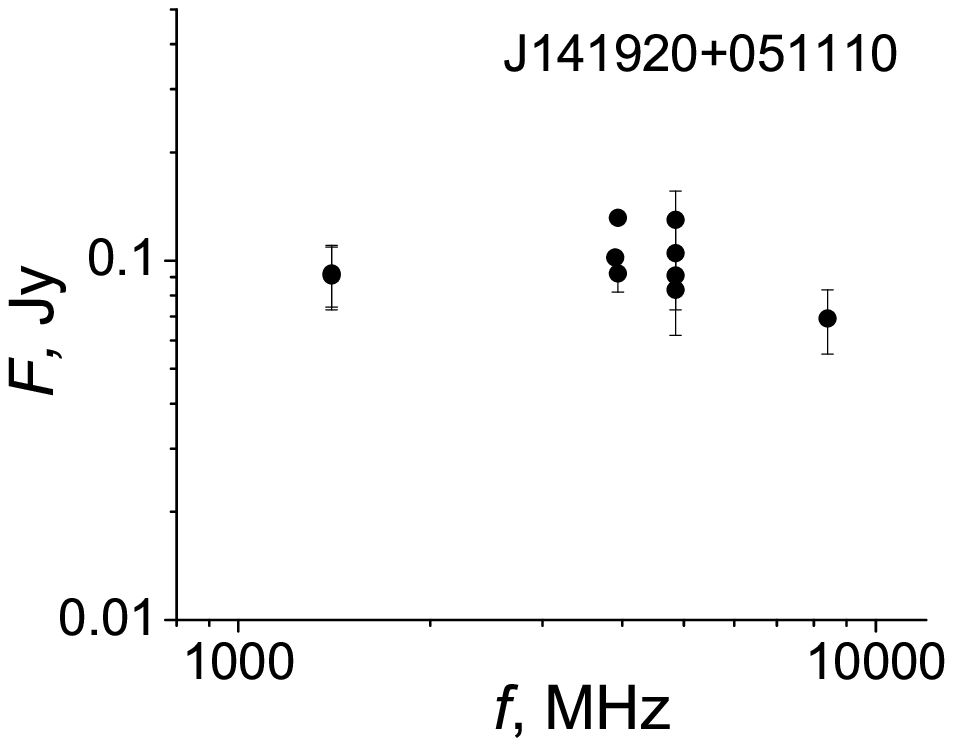}
} \hbox{
\includegraphics[angle=0,width=0.31\textwidth,clip]{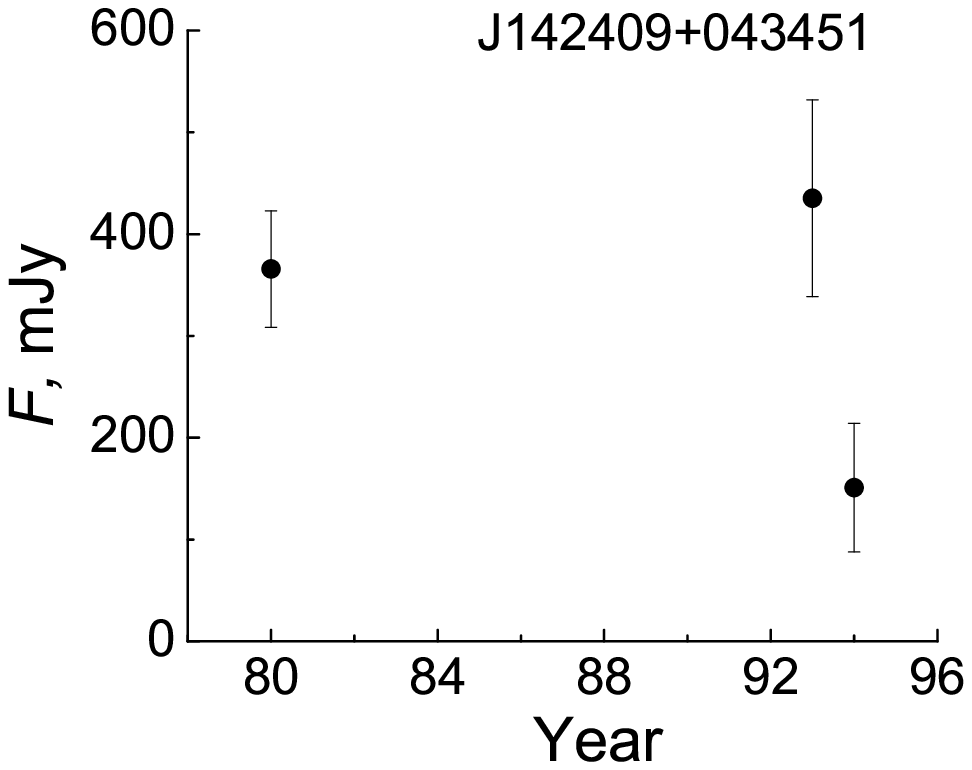}\hspace{22mm}
\includegraphics[angle=0,width=0.31\textwidth,clip]{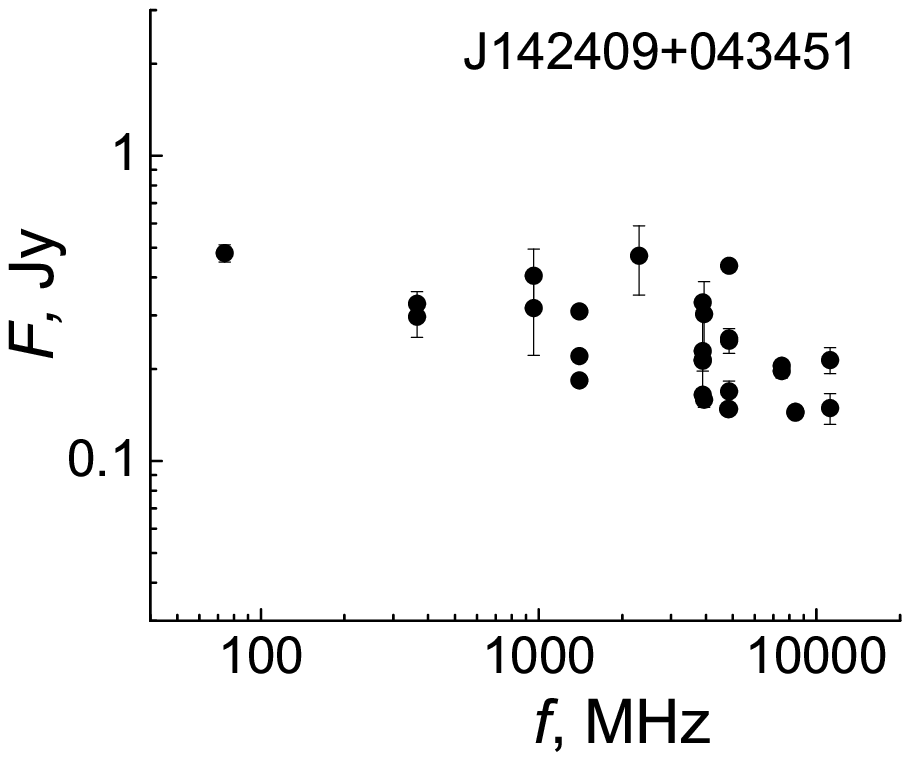}}}}
 \caption{(Contd.)}
\end{figure*}

\addtocounter{figure}{-1}
\begin{figure*}
\setcaptionmargin{5mm} \onelinecaptionstrue \centerline{ \vbox{
\hbox{
\includegraphics[angle=0,width=0.31\textwidth,clip]{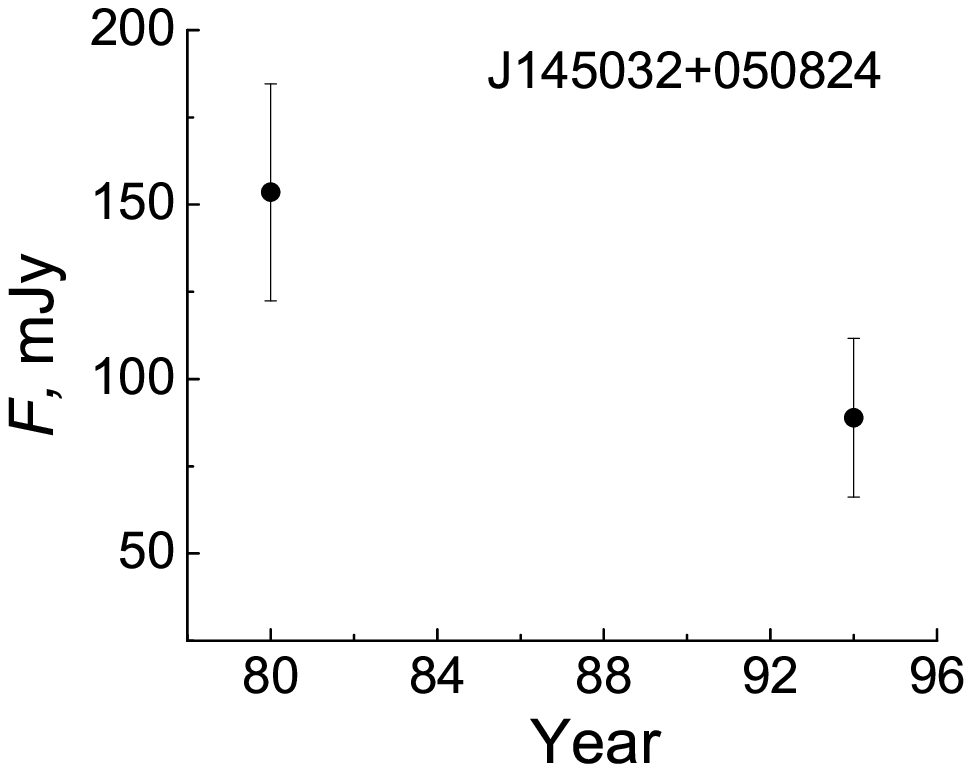}\hspace{20mm}
\includegraphics[angle=0,width=0.31\textwidth,clip]{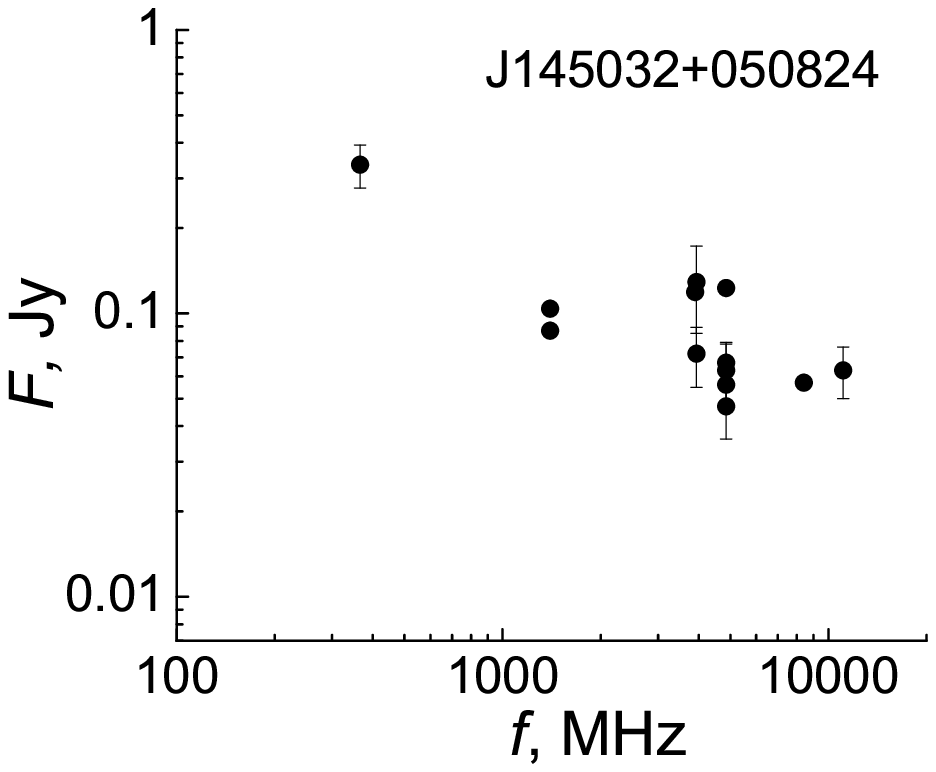}
} \hbox{
\includegraphics[angle=0,width=0.31\textwidth,clip]{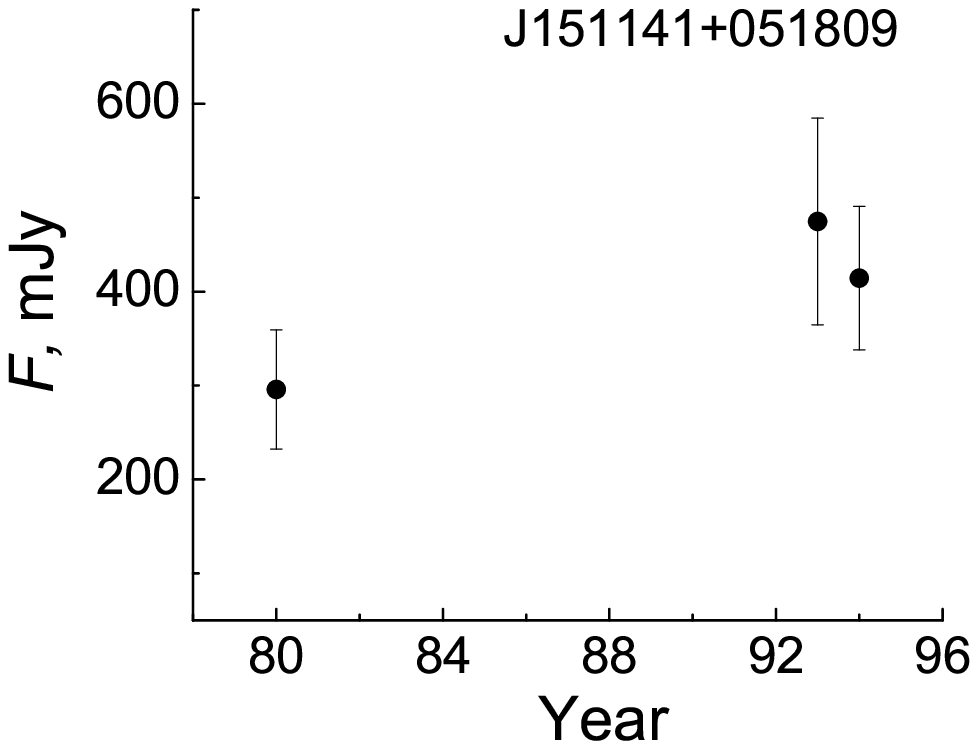}\hspace{21mm}
\includegraphics[angle=0,width=0.31\textwidth,clip]{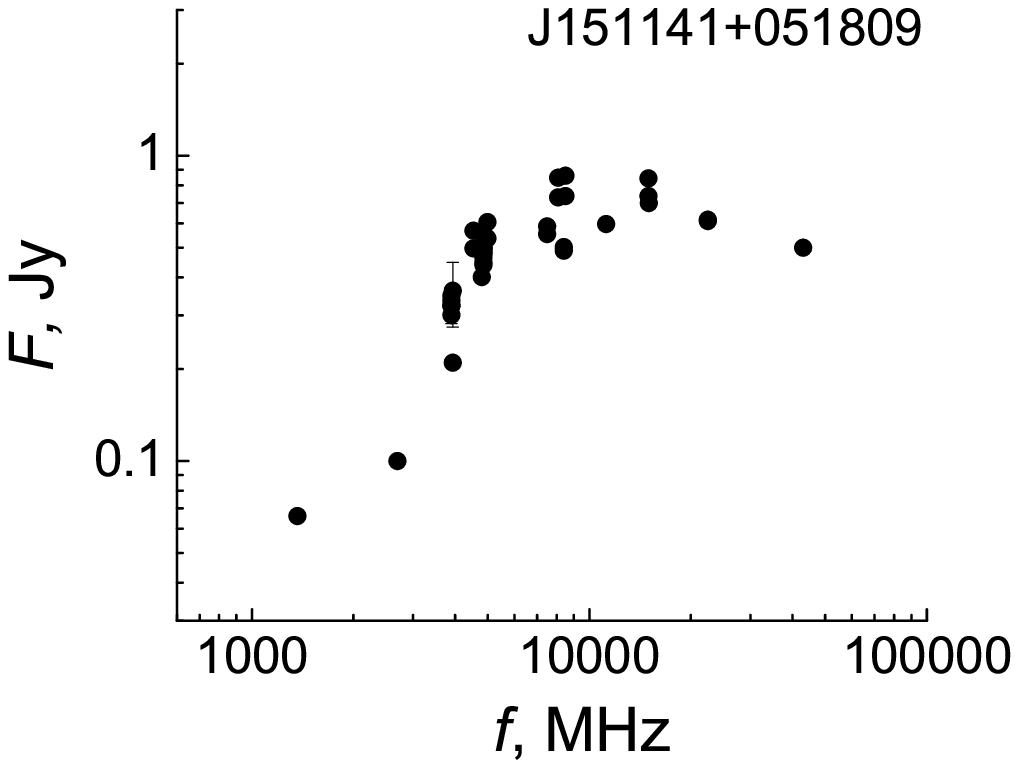}
} \hbox{
\includegraphics[angle=0,width=0.31\textwidth,clip]{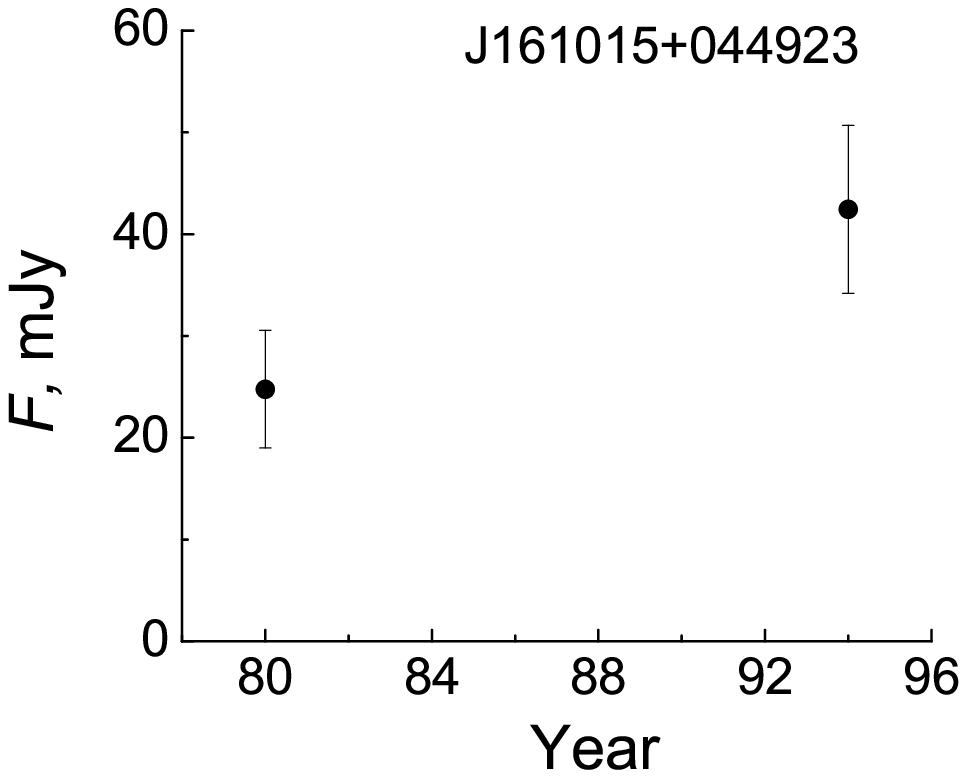}\hspace{21mm}
\includegraphics[angle=0,width=0.31\textwidth,clip]{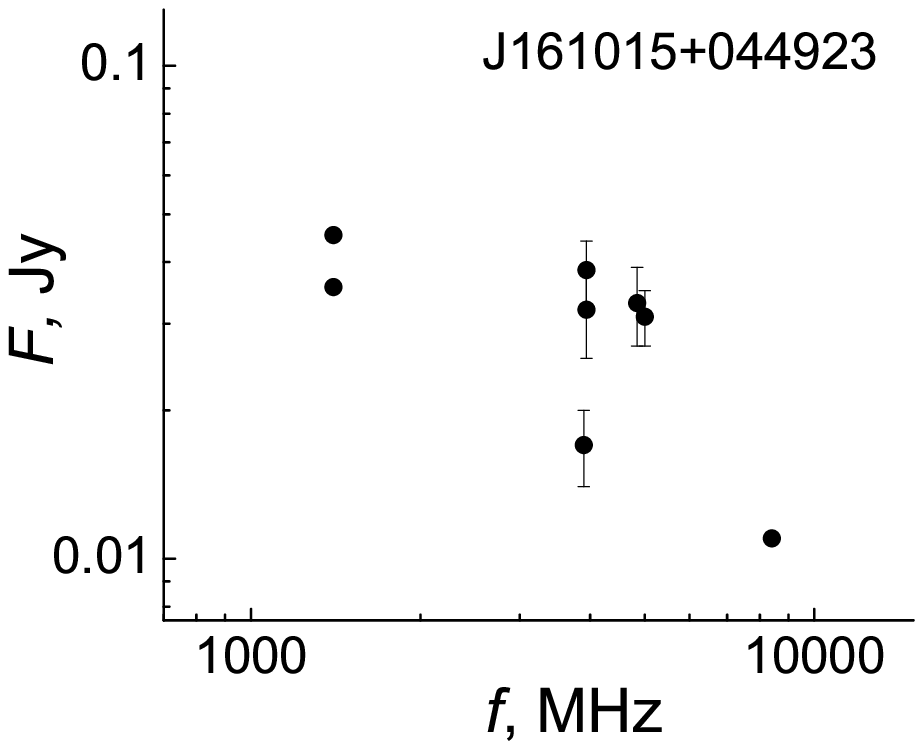}
} \hbox{
\includegraphics[angle=0,width=0.31\textwidth,clip]{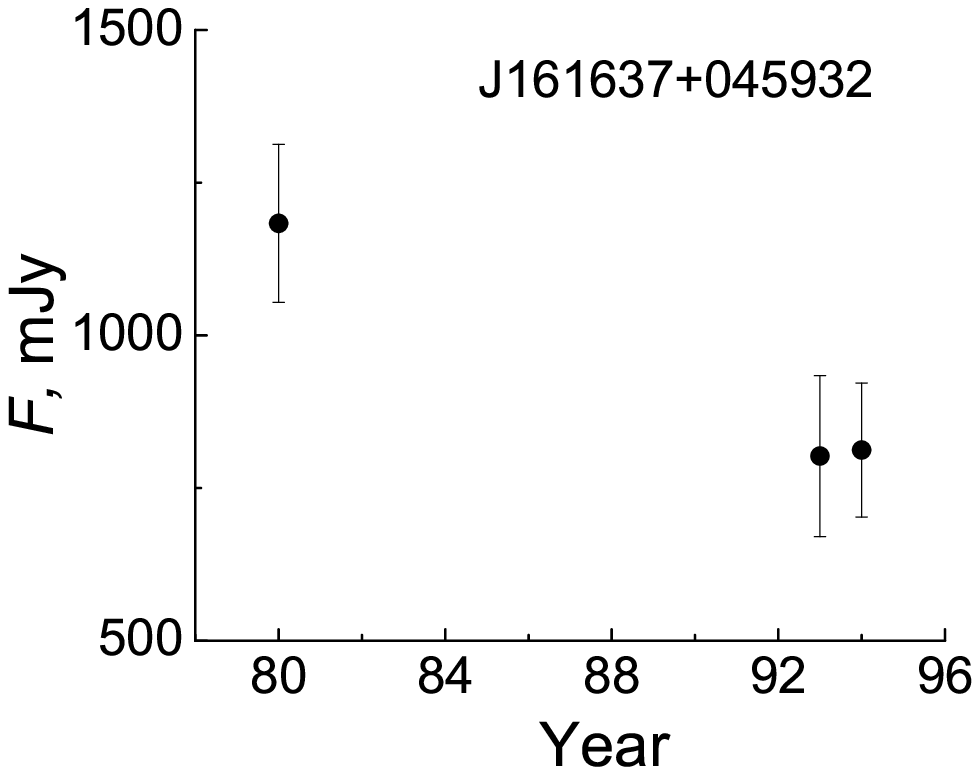}\hspace{21mm}
\includegraphics[angle=0,width=0.31\textwidth,clip]{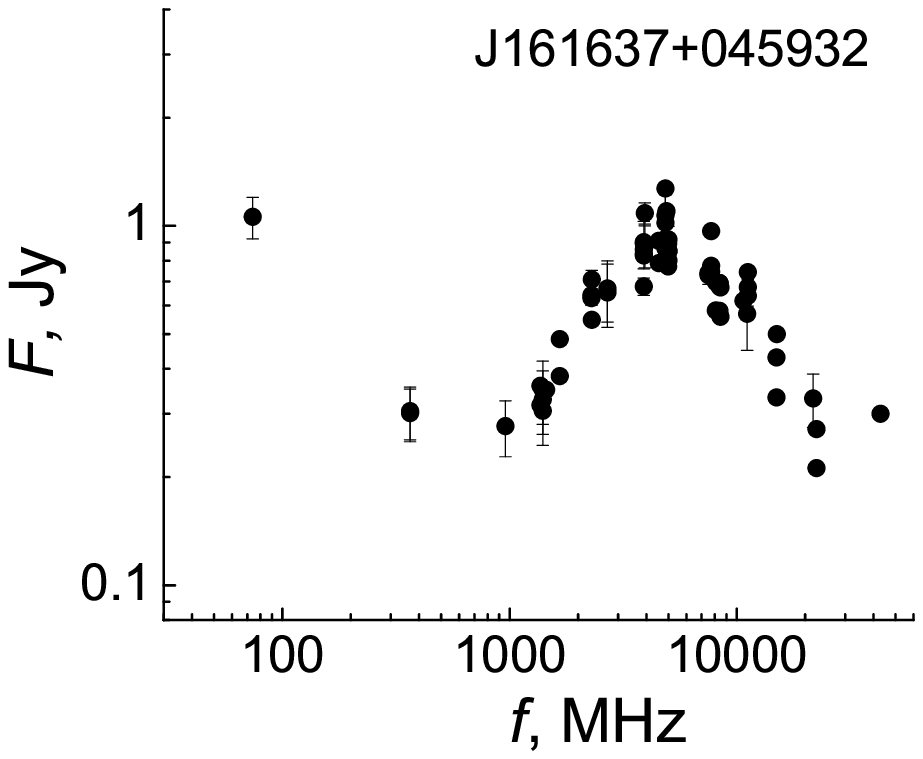}
} \hbox{
\includegraphics[angle=0,width=0.31\textwidth,clip]{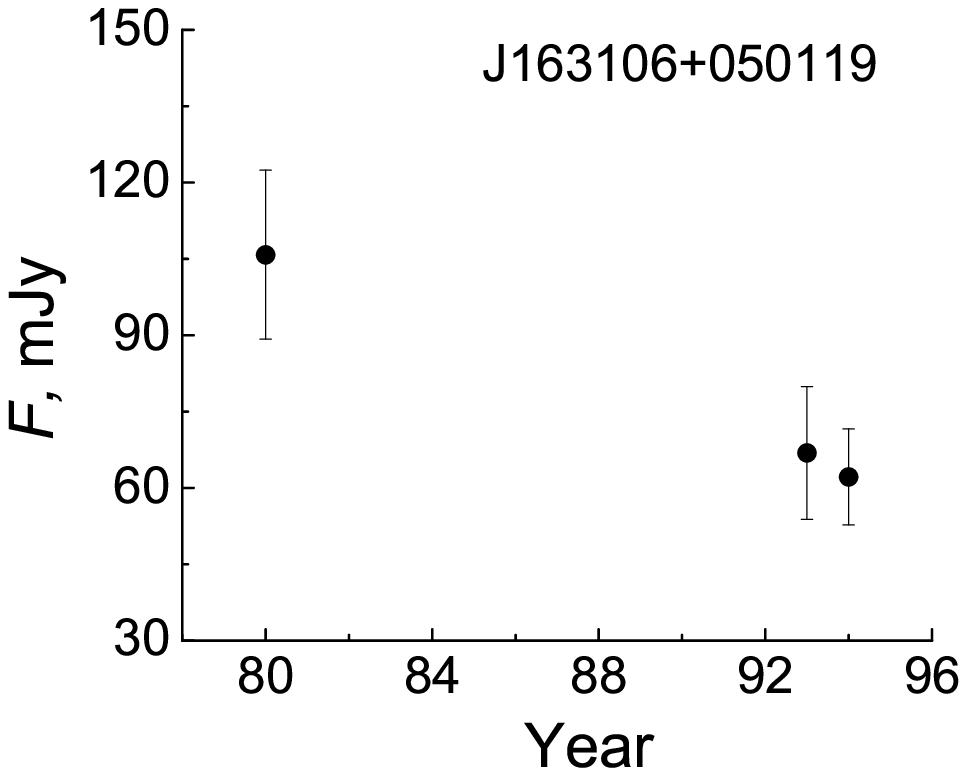}\hspace{22mm}
\includegraphics[angle=0,width=0.31\textwidth,clip]{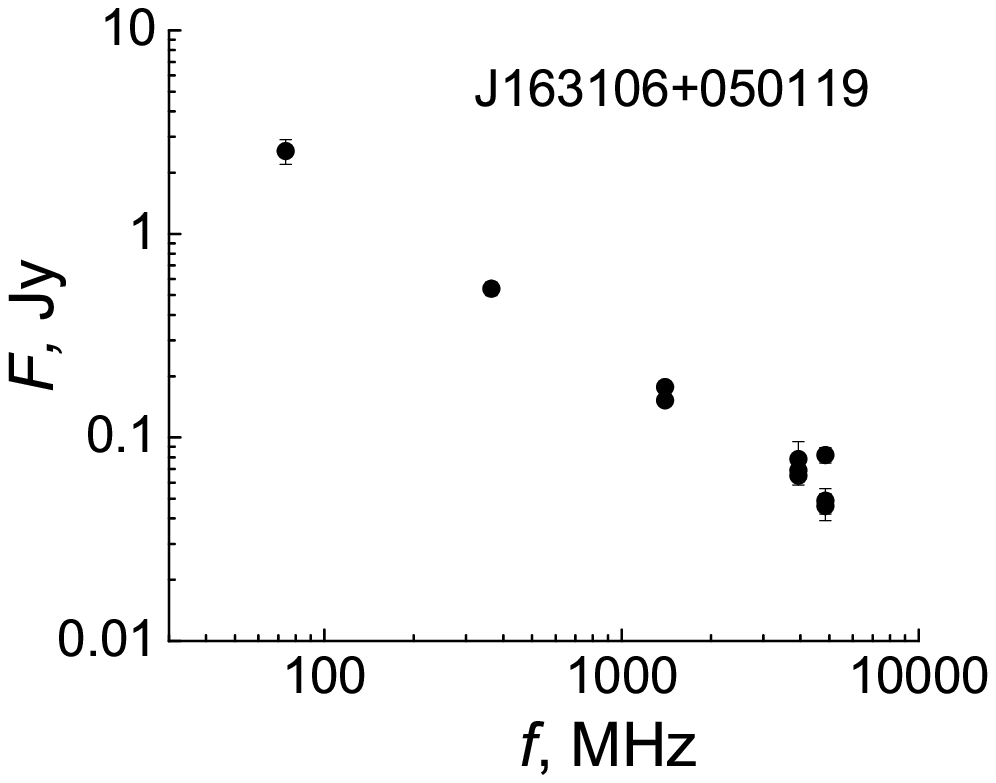}}}}
 \caption{(Contd.)}
\end{figure*}

\addtocounter{figure}{-1}
\begin{figure*}
\setcaptionmargin{5mm} \onelinecaptionstrue \centerline{ \vbox{
\hbox{
\includegraphics[angle=0,width=0.31\textwidth,clip]{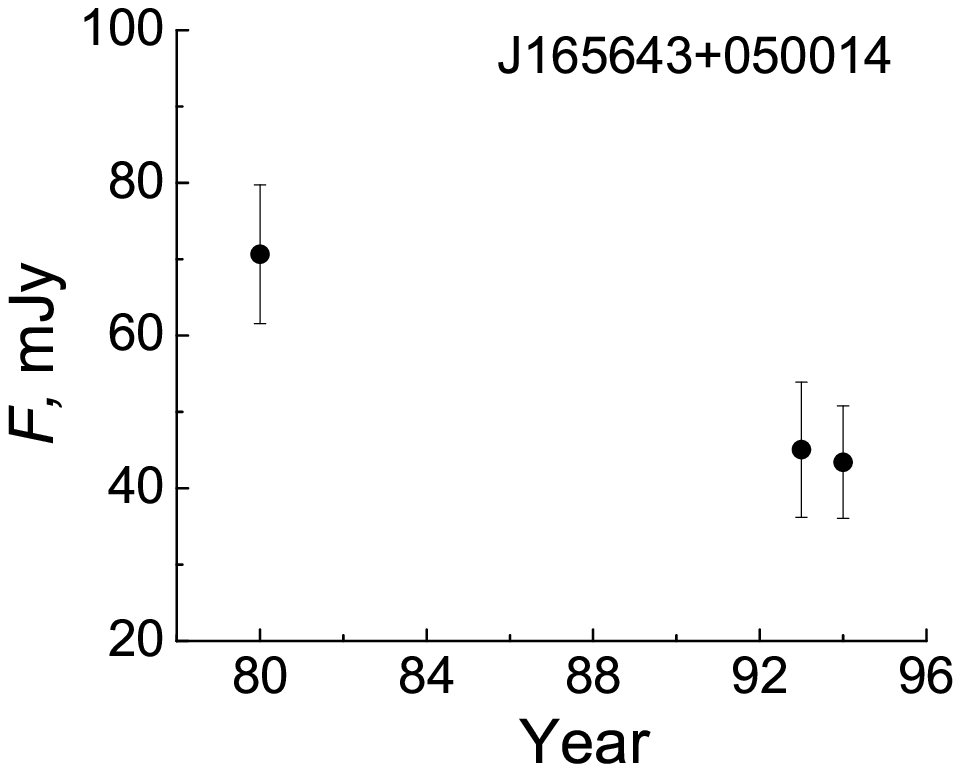}\hspace{20mm}
\includegraphics[angle=0,width=0.31\textwidth,clip]{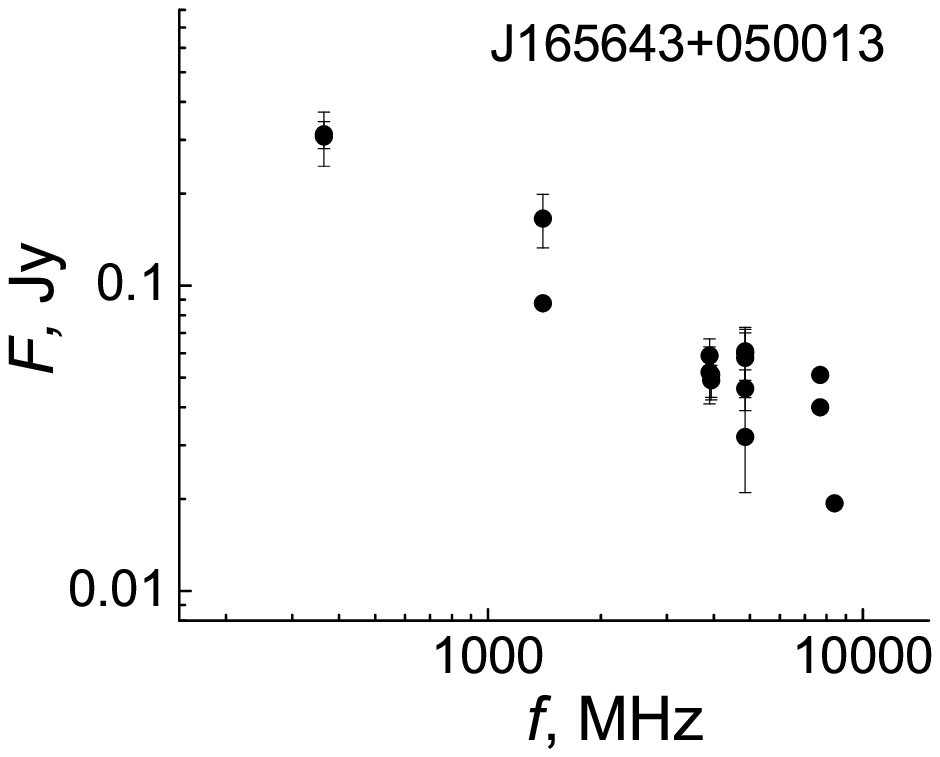}}}}
\caption{(Contd.)}
\end{figure*}

\begin{figure*}[]
\onelinecaptionsfalse \centerline{ \vbox{\vspace{10mm} \hbox{
\includegraphics[angle=0,width=0.31\textwidth,clip]{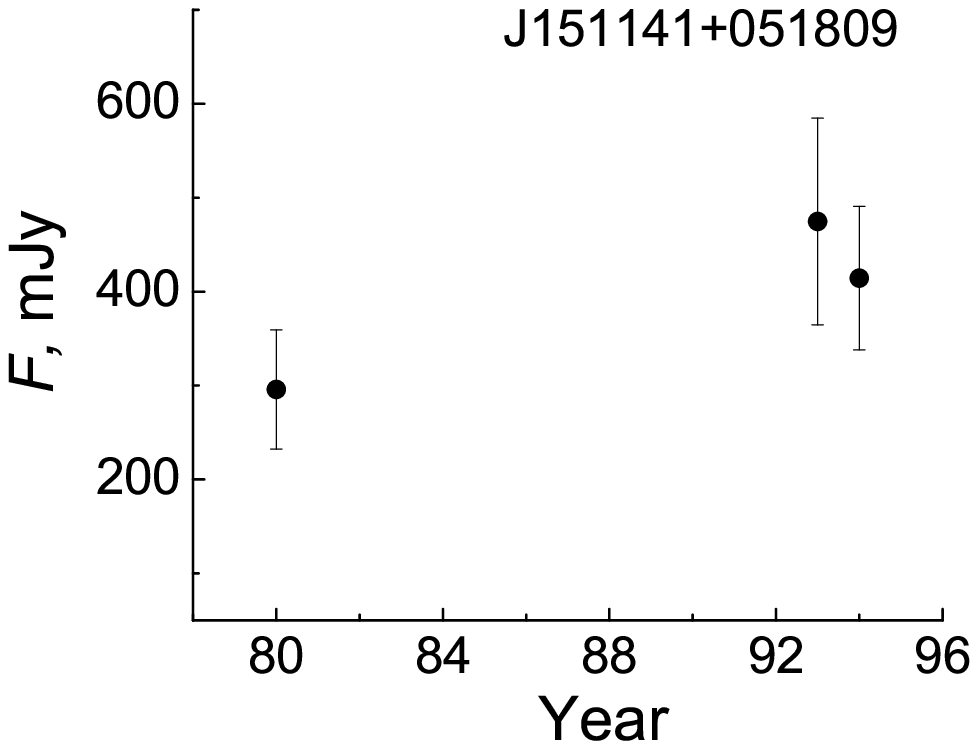}\hspace{24mm}
\includegraphics[angle=0,width=0.31\textwidth,clip]{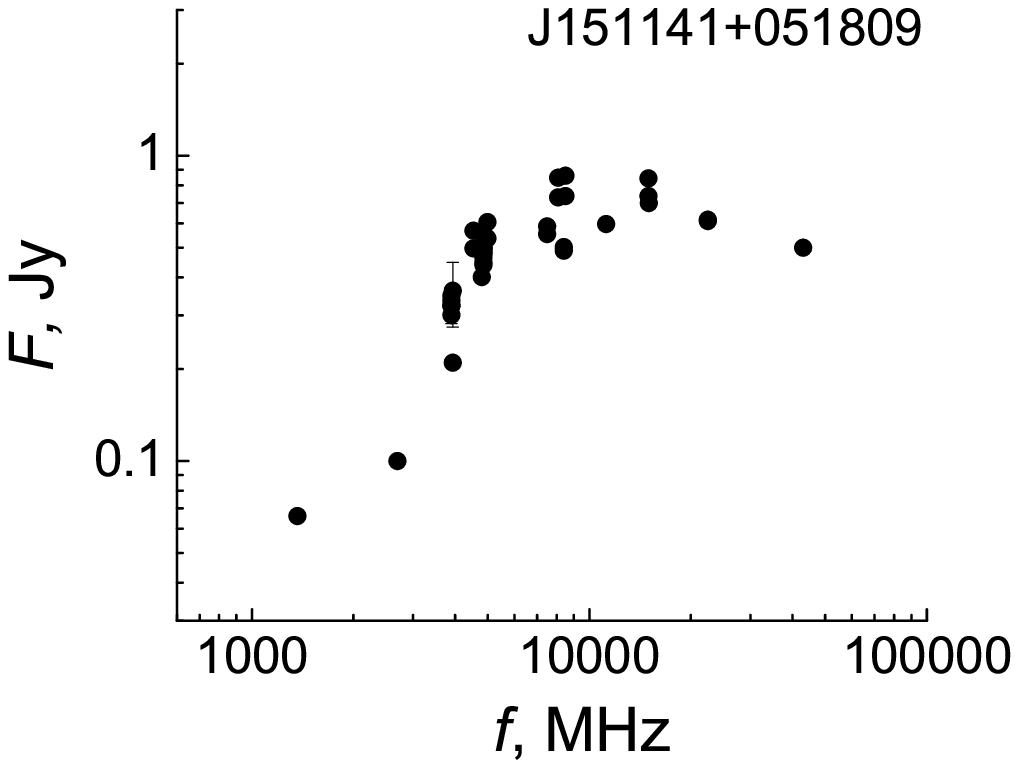}
} \hbox{
\includegraphics[angle=0,width=0.31\textwidth,clip]{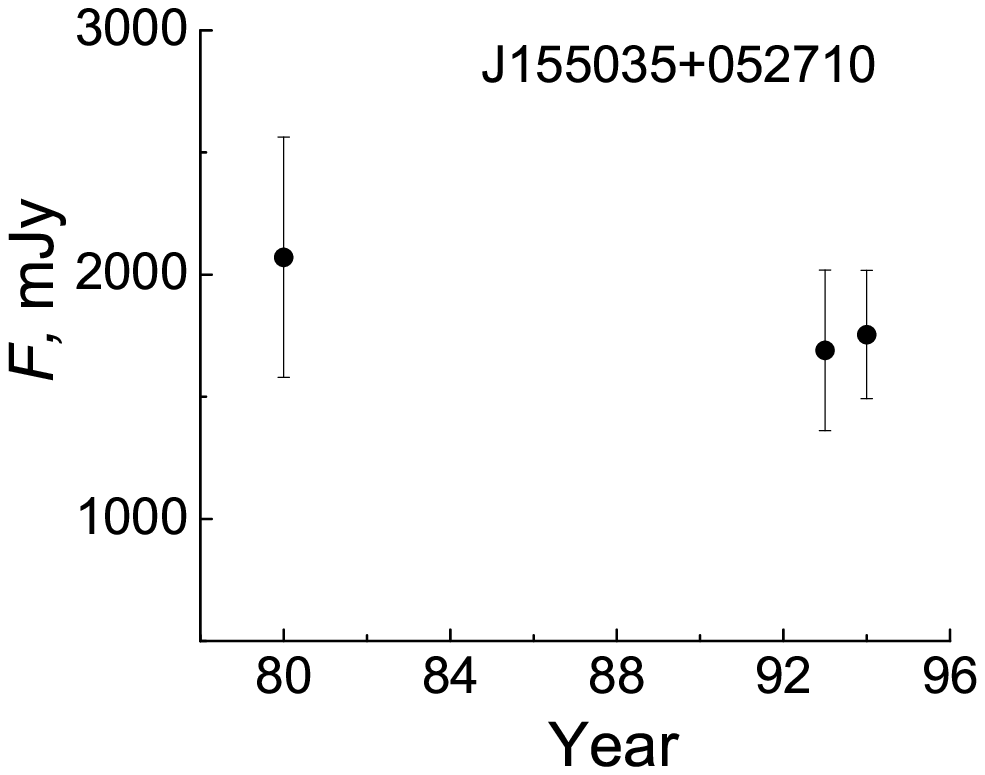}\hspace{24mm}
\includegraphics[angle=0,width=0.31\textwidth,clip]{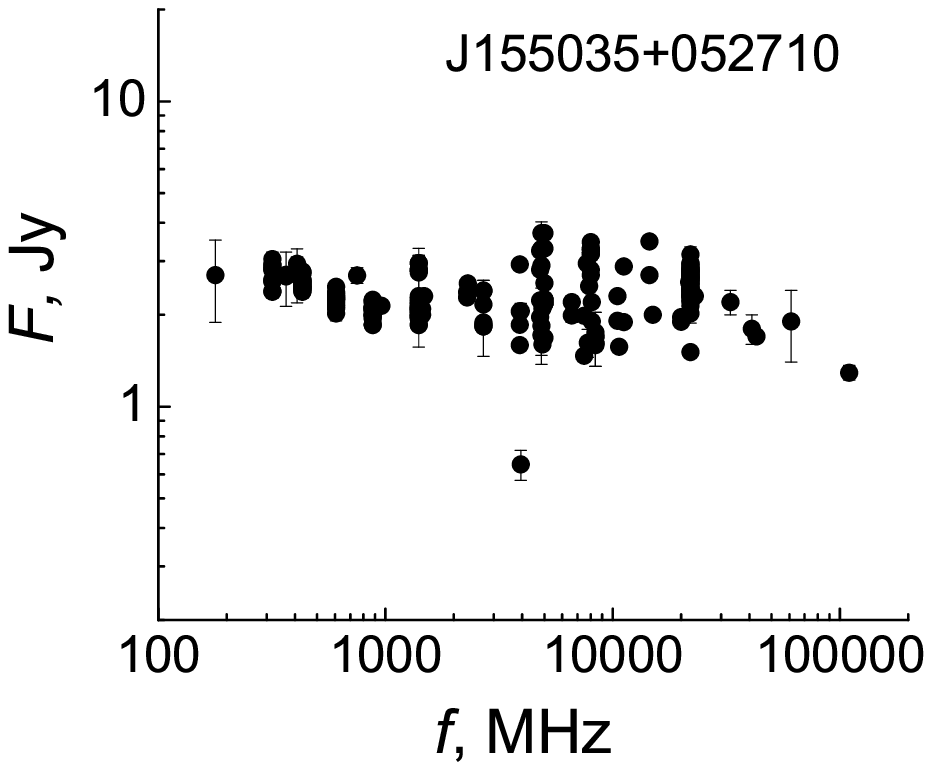}
} \hbox{
\includegraphics[angle=0,width=0.31\textwidth,clip]{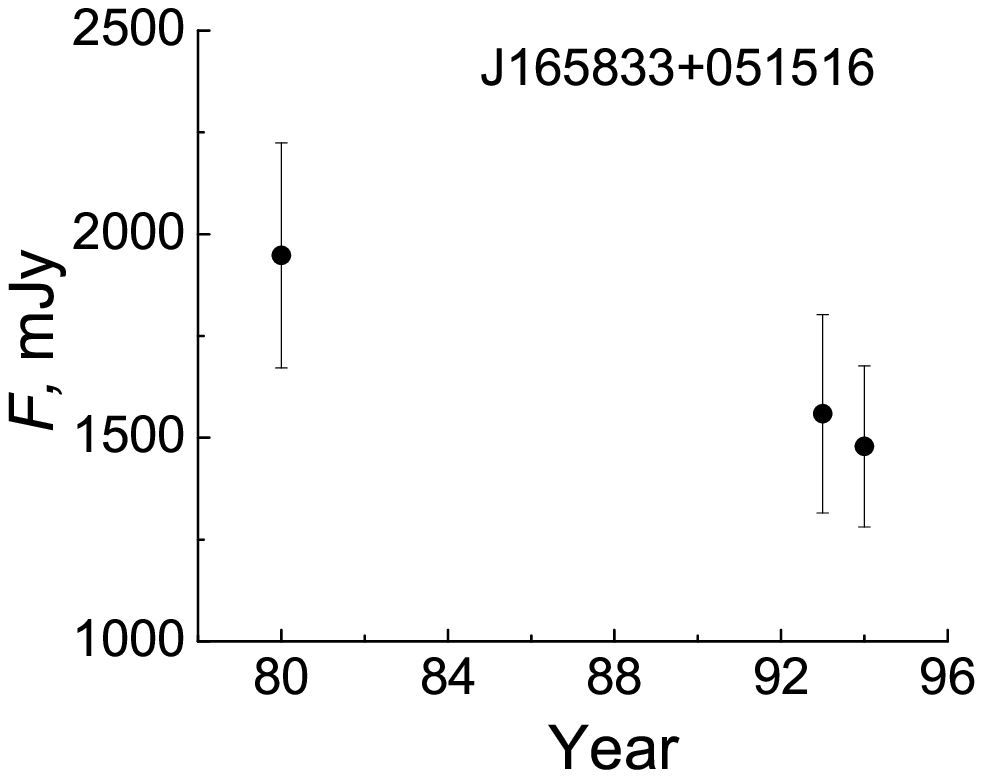}\hspace{24mm}
\includegraphics[angle=0,width=0.31\textwidth,clip]{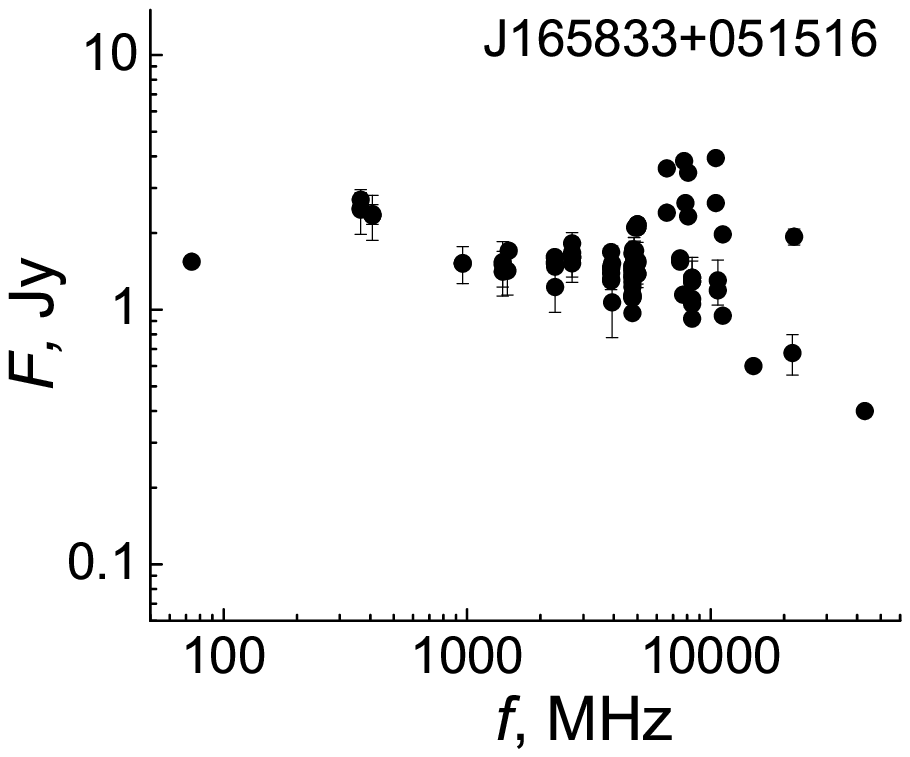}
} } } \setcaptionmargin{5mm} \captionstyle{normal}
 \caption{ Light
curves (the left panel) and spectra (the right panel) of variable
radio sources with flat spectra whose long-term variability
indices are negative, $V<0$, according to the data of our surveys.
} \label{fig10:Majorova_n} \vspace{10mm}
\end{figure*}

In conclusion, we summarize the main results of our search for
variable sources reported in this and our earlier
paper~\cite{maj:Majorova_n}. Out of the total number of sources
(about 280) studied in these papers, 55 proved to have positive
long-term variability indices for at least one pair of surveys. We
list these sources (with $V>0$) in Table~3. It includes objects
from Tables~1 and 2 of this paper and Tables~3 and 4
from~\cite{maj:Majorova_n}.

Among the 55 radio sources listed in Table~3, 15 and 9 objects
meet the condition of reliable (\mbox{$p > 0.98$}) and possible
($ 0.95 <  p < 0.98$) variability, respectively. Thirty-five out
of 55 objects have relative variability amplitudes $V_{\chi} > 0.2
$, and only six have $V_{\chi} < 0.1 $.

Column~1 gives the coordinates  (${\rm RA}_{2000}$, ${\rm
Dec}_{2000}$) of the objects, columns~2 and 3 give the long-term
variability index  $V$ and maximum $\chi^2$ probability $p$, and
column~4 gives the relative variability amplitude~$V_{\chi}$.
Column~5 gives the type of the host galaxy of the object in the
optical range (Type$_{\rm opt}$): QSO, qso---a quasar (lowercase
letters indicate that the type of the object was determined from
photometric data); BL\,Lac---BL~Lacertae type object; G,
g---\linebreak a galaxy, and Sy---a Seyfert galaxy. Column~6 gives
the redshift $Z$, and the ``+'' sign in column~7 indicates the
spectroscopic redshift. Column~8 gives the spectral index
($\alpha$), and column~9 gives the morphology (Mrph) of the radio
source. The uppercase letters denote the classification based on
FIRST maps, and the lowercase letters represent the classification
based on NVSS maps: C or c indicate a point source (Core); D or
d---a double source (Double)\footnote{DC is a double source with a
core (Double-Core), and DD is a doubled double structure
(Double-Double). The ``w'' symbol refers to the  ``winged'' or
``X-shaped'' morphology.}; CL---a core with components
(Core--Lobe)\footnote{The extra symbol ``S'' indicates
``S-shaped'' sources.}; CJ---a core with a jet (\mbox{Core--Jet}),
and T---a triple source (Triple). Column~10 gives the type of the
radio source (Type$_r$): FSRQ---a flat spectrum radio quasar;
FSRS---a flat spectrum radio source; GPS, MPS---a Gigahertz or
Megahertz peaked source, and FR\,I, FR\,II---the  Fanaroff and
Riley class~I or II objects~\cite{Fr:Majorova_n}. The letters
``o'' and ``r'' in column~11 indicate objects that are variable or
possibly variable (``?'') in the optical range and at radio
frequencies respectively. The ``\#'' symbol  \pagebreak

\onecolumngrid
\setcaptionwidth{\linewidth}%
\setcaptionmargin{0mm} %
\onelinecaptionstrue \captionstyle{normal}
\medskip
\begin{longtable}{c|c|c|c|c|c|c|r|c|l|c|c}
\caption{Full list of RCR objects with $V > 0$ obtained in this
study and in our previous paper~\cite{maj:Majorova_n} and the
properties of these objects}
\label{data:Majorova_n}\\
\hline
RCR & $V$  & $p$  & $V_\chi$ & Type$_{\rm opt}$ & $Z$ & $l_{Z}$  & \multicolumn{1}{c|}{$\alpha$}   & Mrph &  \multicolumn{1}{c|}{Type$_r$} & Var & Notes \\
~~~${\rm RA}_{2000}$~~~~~~${\rm Dec}_{2000}$  & & & & & &  &   &             &                   &             & \\
\hline
(1)&(2)&(3)&(4)&(5)&(6)&(7)&\multicolumn{1}{c|}{(8)}&(9)&\multicolumn{1}{c|}{(10)}&(11)&(12)\\
\endfirsthead
\caption{(Contd.) }\\
\hline
RCR & $V$  & $p$  & $V_\chi$ & Type$_{\rm opt}$ & $Z$ & $l_{Z}$  & \multicolumn{1}{c|}{$\alpha$}   & Mrph &  \multicolumn{1}{c|}{Type$_r$} & Var & Notes \\
~~~${\rm RA}_{2000}$~~~~~~${\rm Dec}_{2000}$  & & & & & &  &   &             &                   &             & \\
\hline
(1)&(2)&(3)&(4)&(5)&(6)&(7)&\multicolumn{1}{c|}{(8)}&(9)&\multicolumn{1}{c|}{(10)}&(11)&(12)\\
\hline
\endhead
\hline
\endfoot
\endlastfoot
\hline
J\,072919.57+044948.7~ &~$0.189$~ &~$0.966$~ &~$0.385$~ &~  g     ~&~$     $~ &~  ~&~$-0.67$ ~&~c    ~&~     ~&~     ~&~  \\
J\,073357.46+045614.1~ &~$0.199$~ &~$0.999$~ &~$0.388$~ &~QSO     ~&~$3.010$~ &~+ ~&~$ 0.12$ ~&~C    ~&~FSRQ ~&~o?   ~&~  \\
J\,074239.65+050704.3~ &~$0.077$~ &~$0.855$~ &~$0.180$~ &~Sy2     ~&~$0.160$~ &~+ ~&~$-0.85$ ~&~d?   ~&~FRII?~&~o    ~&~\#\\
J\,075314.02+045129.4~ &~$0.025$~ &~$0.796$~ &~$0.057$~ &~  G     ~&~$0.450$~ &~  ~&~$-0.35$ ~&~C    ~&~FSRS ~&~     ~&~  \\
J\,080757.60+043234.6~ &~$0.275$~ &~$0.995$~ &~$0.594$~ &~QSO     ~&~$2.877$~ &~+ ~&~$-0.30$ ~&~C    ~&~FSRQ ~&~o    ~&~  \\
J\,081218.14+050755.5~ &~$0.097$~ &~$0.906$~ &~$0.194$~ &~ qso    ~&~$     $~ &~  ~&~$-0.75$ ~&~C    ~&~     ~&~     ~&~  \\
J\,081626.62+045852.8~ &~$0.106$~ &~$0.990$~ &~$0.337$~ &~  G     ~&~$0.080$~ &~  ~&~$-0.61$ ~&~CL, S~&~GPS? ~&~o    ~&~  \\
J\,083148.89+042938.5~ &~$0.130$~ &~$0.984$~ &~$0.367$~ &~BL\,Lac ~&~$0.174$~ &~+ ~&~$ 0.04$ ~&~CJ   ~&~FSRS ~&~o, r ~&~  \\
J\,091636.22+044132.0~ &~$0.263$~ &~$0.965$~ &~$0.453$~ &~G       ~&~$0.184$~ &~+ ~&~$-0.84$ ~&~T    ~&~MPS? ~&~o    ~&~  \\
J\,095218.73+050559.3~ &~$0.201$~ &~$0.976$~ &~$0.403$~ &~QSO     ~&~$0.400$~ &~  ~&~$ 0.27$ ~&~C    ~&~     ~&~o    ~&~  \\
J\,100534.80+045119.8~ &~$0.104$~ &~$0.981$~ &~$0.377$~ &~g       ~&~$     $~ &~  ~&~$-0.56$ ~&~D?   ~&~FRII?~&~     ~&~  \\
J\,101515.33+045305.6~ &~$0.061$~ &~$0.864$~ &~$0.168$~ &~g       ~&~$     $~ &~  ~&~$-1.04$ ~&~C    ~&~     ~&~     ~&~\#\\
J\,101603.12+051303.6~ &~$0.147$~ &~$0.987$~ &~$0.334$~ &~QSO     ~&~$1.702$~ &~+ ~&~$ 0.04$ ~&~C    ~&~FSRQ ~&~r    ~&~  \\
J\,103846.84+051229.6~ &~$0.050$~ &~$0.839$~ &~$0.121$~ &~QSO     ~&~$0.473$~ &~+ ~&~$ 0.24$ ~&~T    ~&~FSRQ ~&~o?   ~&~  \\
J\,103938.62+051031.1~ &~$0.264$~ &~$0.984$~ &~$0.480$~ &~  G     ~&~$0.068$~ &~+ ~&~$-0.74$ ~&~D, w ~&~FRII ~&~o    ~&~\#\\
J\,104117.65+045306.4~ &~$0.157$~ &~$0.994$~ &~$0.414$~ &~  G     ~&~$0.068$~ &~+ ~&~$-0.82$ ~&~T?   ~&~     ~&~o    ~&~  \\
J\,104527.19+045118.7~ &~$0.043$~ &~$0.808$~ &~$0.202$~ &~g       ~&~$     $~ &~  ~&~$-0.85$ ~&~C    ~&~     ~&~     ~&~  \\
J\,104551.72+045553.9~ &~$0.035$~ &~$0.691$~ &~$0.083$~ &~g       ~&~$     $~ &~  ~&~$-0.99$ ~&~D?   ~&~     ~&~     ~&~\#\\
J\,105253.05+045735.3~ &~$0.039$~ &~$0.895$~ &~$0.149$~ &~g       ~&~$     $~ &~  ~&~$-0.28$ ~&~D?   ~&~GPS? ~&~     ~&~  \\
J\,105719.26+045545.4~ &~$0.200$~ &~$0.990$~ &~$0.427$~ &~QSO     ~&~$1.334$~ &~+ ~&~$-0.20$ ~&~C    ~&~GPS? ~&~o?   ~&~  \\
J\,110246.51+045916.7~ &~$0.082$~ &~$0.925$~ &~$0.228$~ &~  G     ~&~$0.630$~ &~  ~&~$-0.81$ ~&~D, w ~&~FRII ~&~     ~&~\#\\
J\,112437.45+045618.8~ &~$0.057$~ &~$0.895$~ &~$0.164$~ &~ Sy2    ~&~$0.283$~ &~+ ~&~$-0.87$ ~&~D    ~&~FRII ~&~o?   ~&~\#\\
J\,113156.47+045549.3~ &~$0.050$~ &~$0.842$~ &~$0.110$~ &~  G     ~&~$0.844$~ &~+ ~&~$-0.77$ ~&~C    ~&~     ~&~r    ~&~  \\
J\,114521.30+045526.7~ &~$0.018$~ &~$0.729$~ &~$0.061$~ &~QSO     ~&~$1.339$~ &~+ ~&~$-0.33$ ~&~D    ~&~FSRQ ~&~o?   ~&~  \\
J\,114631.64+045818.2~ &~$0.056$~ &~$0.924$~ &~$0.165$~ &~QSO     ~&~$     $~ &~  ~&~$-0.21$ ~&~C?   ~&~GPS? ~&~     ~&~  \\
J\,115248.33+050057.2~ &~$0.171$~ &~$0.974$~ &~$0.513$~ &~qso     ~&~$     $~ &~  ~&~$-0.89$ ~&~D?   ~&~     ~&~     ~&~  \\
J\,115336.08+045505.2~ &~$0.175$~ &~$0.989$~ &~$0.393$~ &~  G     ~&~$0.313$~ &~+ ~&~$ 0.78$ ~&~C    ~&~GPS? ~&~     ~&~  \\
J\,115851.23+045541.9~ &~$0.335$~ &~$1.0  $~ &~$0.882$~ &~QSO     ~&~$     $~ &~  ~&~$-0.11$ ~&~C    ~&~FSRQ?~&~o?   ~&~  \\
J\,121328.89+050009.9~ &~$0.076$~ &~$0.772$~ &~$0.176$~ &~  G     ~&~$0.700$~ &~  ~&~$-1.08$ ~&~DC   ~&~FRII ~&~     ~&~\#\\
J\,121852.16+051449.4~ &~$0.007$~ &~$0.681$~ &~$0.175$~ &~  G     ~&~$0.075$~ &~+ ~&~$-0.67$ ~&~D    ~&~     ~&~o    ~&~\#\\
J\,123507.25+045318.7~ &~$0.376$~ &~$1.0  $~ &~$0.753$~ &~  g?    ~&~$     $~ &~  ~&~$-0.06$ ~&~C    ~&~GPS? ~&~     ~&~  \\
J\,123723.63+045741.6~ &~$0.032$~ &~$0.828$~ &~$0.089$~ &~  g     ~&~$     $~ &~  ~&~$-1.16$ ~&~C    ~&~     ~&~     ~&~  \\
J\,123932.78+044305.3~ &~$0.027$~ &~$0.844$~ &~$0.135$~ &~QSO     ~&~$1.762$~ &~+ ~&~$-0.13$ ~&~C    ~&~FSRQ ~&~o, r ~&~  \\
J\,124145.15+045924.5~ &~$0.224$~ &~$0.961$~ &~$0.491$~ &~  G     ~&~$     $~ &~  ~&~$-0.56$ ~&~D    ~&~     ~&~     ~&~  \\
J\,125755.32+045917.6~ &~$0.072$~ &~$0.939$~ &~$0.184$~ &~  G     ~&~$0.240$~ &~  ~&~$-1.01$ ~&~DD   ~&~FRII ~&~o?   ~&~  \\
J\,130631.65+050231.3~ &~$0.056$~ &~$0.764$~ &~$0.159$~ &~ qso    ~&~$     $~ &~  ~&~$-0.25$ ~&~C    ~&~     ~&~     ~&~  \\
J\,132448.14+045758.8~ &~$0.088$~ &~$0.762$~ &~$0.165$~ &~ qso    ~&~$     $~ &~  ~&~$-1.03$ ~&~D    ~&~FRII ~&~     ~&~\#\\
J\,133541.21+050124.9~ &~$0.216$~ &~$0.933$~ &~$0.519$~ &~  G     ~&~$0.770$~ &~+ ~&~$-0.04$ ~&~C    ~&~     ~&~     ~&~  \\
J\,133920.76+050159.3~ &~$0.116$~ &~$0.871$~ &~$0.307$~ &~QSO     ~&~$1.358$~ &~+ ~&~$-0.27$ ~&~C    ~&~     ~&~o    ~&~  \\
J\,134201.37+050157.5~ &~$0.108$~ &~$0.931$~ &~$0.279$~ &~QSO     ~&~$3.166$~ &~+ ~&~$ 0.85$ ~&~C    ~&~     ~&~o    ~&~  \\
J\,134243.57+050431.5~ &~$0.008$~ &~$0.601$~ &~$0.026$~ &~Sy1     ~&~$0.136$~ &~+ ~&~$-0.65$ ~&~CL, S~&~~FRI ~&~o    ~&~\#\\
J\,135050.06+045148.9~ &~$0.173$~ &~$0.989$~ &~$0.481$~ &~QSO     ~&~$1.800$~ &~  ~&~$ 0.44$ ~&~C    ~&~GPS? ~&~o    ~&~  \\
J\,135137.56+043542.0~ &~$0.084$~ &~$0.960$~ &~$0.294$~ &~  g     ~&~$     $~ &~  ~&~$-0.89$ ~&~C    ~&~     ~&~     ~&~\#\\
J\,140730.77+044934.9~ &~$0.007$~ &~$0.675$~ &~$0.071$~ &~QSO     ~&~$1.756$~ &~+ ~&~$-0.75$ ~&~D?   ~&~     ~&~o?   ~&~\#\\
J\,141920.56+051110.6~ &~$0.104$~ &~$0.943$~ &~$0.203$~ &~QSO     ~&~$0.787$~ &~+ ~&~$-0.27$ ~&~C    ~&~GPS? ~&~o?   ~&~  \\
J\,142104.21+050845.0~ &~$0.092$~ &~$0.928$~ &~$0.280$~ &~  G     ~&~$0.455$~ &~+ ~&~$-0.69$ ~&~C    ~&~     ~&~o?   ~&~\#\\
J\,142409.47+043451.7~ &~$0.231$~ &~$0.988$~ &~$0.484$~ &~BL\,Lac ~&~$0.665$~ &~+ ~&~$-0.20$ ~&~CJ   ~&~FSRS ~&~o, r ~&~  \\
J\,145032.99+050824.6~ &~$0.082$~ &~$0.925$~ &~$0.235$~ &~QSO     ~&~$1.635$~ &~+ ~&~$-0.32$ ~&~D    ~&~FRII ~&~     ~&~  \\
J\,151141.19+051809.4~ &~$0.019$~ &~$0.725$~ &~$0.133$~ &~Sy1     ~&~$0.084$~ &~+ ~&~$ 1.11$ ~&~C    ~&~FRSQ ~&~o    ~&~  \\
J\,155148.09+045930.5~ &~$0.125$~ &~$0.961$~ &~$0.275$~ &~  G     ~&~$     $~ &~  ~&~$-1.17$ ~&~D    ~&~FRII ~&~     ~&~\#\\
J\,161015.24+044923.5~ &~$0.080$~ &~$0.919$~ &~$0.223$~ &~qso     ~&~$     $~ &~  ~&~$-0.20$ ~&~CJ   ~&~     ~&~o, r ~&~  \\
J\,161637.49+045932.8~ &~$0.067$~ &~$0.984$~ &~$0.214$~ &~QSO     ~&~$3.217$~ &~+ ~&~$ 0.22$ ~&~C    ~&~GPS  ~&~     ~&~  \\
J\,163106.83+050119.2~ &~$0.110$~ &~$0.972$~ &~$0.241$~ &~  g     ~&~$     $~ &~  ~&~$-0.90$ ~&~d    ~&~     ~&~     ~&~  \\
J\,165643.94+050014.2~ &~$0.096$~ &~$0.988$~ &~$0.269$~ &~  g?    ~&~$     $~ &~  ~&~$-0.72$ ~&~d?   ~&~     ~&~     ~&~  \\
\hline
\end{longtable}
\twocolumngrid

\noindent in column~12 indicates
possibly variable objects that we found in our earlier
study~\cite{maj:Majorova_n}.

We thus obtained a sample of  \mbox{RCR} objects that can be considered to be very
likely variable on time scales of  \mbox {6--7}~years. The sample contains about  10\%
of all the RCR catalog objects considered.

 \section{CONCLUSIONS}

To reveal the variable sources from the data of the ``Cold''
surveys, we used the technique developed in our earlier
paper~\cite{maj:Majorova_n}, where we derived the calibration
curves and performed a detailed analysis and estimation of the
relative root mean square errors for each survey. We searched for
variable objects in a sample of radio sources of the  RCR
catalog~\cite{so2:Majorova_n} containing about 200 objects with
flux density data available at three or more frequencies.

To reveal variable sources among the RCR objects of the considered sample,
we estimated the long-term variability index $V$, the
relative variability amplitude $V_{\chi}$, the $\chi^2$ probability $p$, and
the parameters  $V_{R}$  and $V_{F}$.

Out of about 200 RCR catalog objects considered in this study, 41
proved to have positive long-term variability indices, suggesting
their possible variability. Almost half of these sources are
bright objects with flux densities above  100~mJy, and about one
third of them are faint radio sources with $F \le 40$~mJy. The
distribution of spectral indices for this sample has two maxima:
at $\alpha=-0.75$ and $\alpha=-0.15$. However, most of the objects
with $V>0$ have spectral indices \mbox {$\alpha > -0.5$}.

It can be concluded, based on  the criteria proposed
in~\cite{kest:Majorova_n,fan:Majorova_n,sei:Majorova_n,g1:Majorova_n},
that of the 41 objects with positive long-term variability
indices, 15 can be considered to be reliably variable radio
sources. These sources have  $\chi^2$ probabilities  $p > 0.98$,
three of these objects have probabilities  \mbox {$ p \ge 0.999$},
six sources are possibly variable with the probabilities \mbox {$
0.95 <  p < 0.98$}, and 20 sources have variability probabilities
in the \mbox {$0.73 \le p < 0.95$} interval.

Of the 21 most likely variable sources with\linebreak \mbox {$p >
0.95$}, one third have flux densities above 150~mJy, seven sources
have flux densities in  the interval \mbox{$40<  F  < 100$}~mJy,
and seven sources have flux densities in the \mbox{$20<  F  <
40$}~mJy interval. Most of these sources are objects with nearly
flat spectra, and two objects have inverted spectra. One third of
the sources have spectral indices $\alpha < -0.67$ (steep
spectra).

Twenty-four of the 41 objects are variable or possibly variable in
the optical range, and five objects are known variable radio
sources. We obtained the light curves and spectra for the radio
sources with positive long-term variability indices and for a
number of ``nonvariable'' objects.

We thus studied about 280 radio sources of the RCR catalog. The
results of our search for variable sources performed within the
framework of this study and our earlier
paper~\cite{maj:Majorova_n} lead us to point out that 55~radio
sources have positive long-term variability indices. Fifteen of
these sources meet the criterion of reliable variability (their
$\chi^2$ probabilities are \mbox{$p > 0.98$}), and nine objects
meet the criterion of possible variability ($0.95 < p < 0.98$).
The variability probabilities of the remaining sources lie in the
\mbox{$0.6 < p < 0.95$} interval. Thirty-five of the  55 objects
have relative variability amplitudes \mbox {$V_{\chi} > 0.2 $}.
Thus about 10\% of all the RCR objects that we analyzed proved to
be variable.

Out of 24 most likely variable sources, two are BL\,Lac objects,
and the remaining ones are quasars (nine objects) and galaxies
(10~objects).

Conspicuous is the fact that the \mbox {FIRST survey} flux densities of 14 out of 55 sources
exceed the flux densities from the \mbox {NVSS survey} in what is yet another indicator
of the possible variability of the sources considered. Almost half of the objects are
flat spectrum sources, and about ten objects are  GPS and MPS sources
with the spectra peaking at GHz or MHz frequencies. Twenty-six objects out of 55 are variable
or possibly variable in the optical range and five are known variable radio sources.

\begin{acknowledgments}

This work was supported in part by the\linebreak Russian
Foundation for Basic Research (grants\linebreak
no.~\mbox{11-02-12036}, 11-02-00489, and \mbox{12-07-00503})  and
the Ministry of Education and Science\linebreak of the Russian
Federation (state contracts\linebreak no.~16.552.11.7028 and
16.518.11.7054). This research has made use of the VizieR
catalogue access tool and the SIMBAD database, operated at CDS,
Strasbourg, France, as well as NASA/IPAC Extragalactic Database
(NED) which is operated by the Jet Propulsion Laboratory,
California Institute of Technology, under contract with the
National Aeronautics and Space Administration.

\end{acknowledgments}

\end{document}